\documentclass[pldi-cameraready,10pt]{sigplanconf-pldi16}

\usepackage{amssymb,amsmath}
\usepackage{amsthm}
\usepackage{stmaryrd}
\usepackage{hyperref}
\usepackage{graphicx}
\usepackage{color}
\usepackage{fancyvrb}
\usepackage{tikz}
\usepackage{balance}
\usepackage{bm}
\usepackage{mathpartir}
\usepackage{wrapfig}
\usepackage{endnotes}

\usepackage[firstpage]{draftwatermark}
\SetWatermarkText{\hspace*{8in}\raisebox{4.7in}{\includegraphics[scale=0.09]{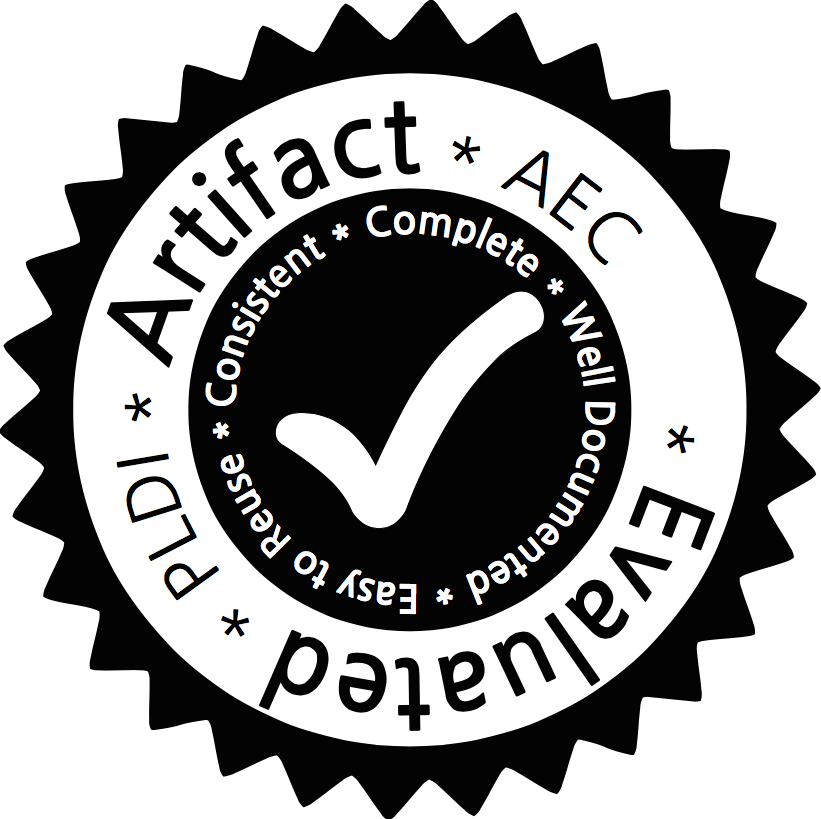}}}
\SetWatermarkAngle{0}

\newcommand{\snsLoc}{6,000}
\newcommand{\littleExamplesAll}{68}
\newcommand{\littleLoc}{2,000}
\newcommand{\preludeLoc}{500}
\newcommand{\numUsers}{25}

\newcommand{\pct}[1]{{#1}\%}

\newcommand{\numShapes}{3,772}

\newcommand{\numZones}{14,106}
\newcommand{\numZonesInactive}{991}
\newcommand{\pctZonesInactive}{\pct{7}}
\newcommand{\numZonesActive}{13,115}
\newcommand{\pctZonesActive}{\pct{93}}
\newcommand{\numZonesActiveOne}{4,856}
\newcommand{\pctZonesActiveOne}{\pct{34}}
\newcommand{\numZonesActiveMultiple}{8,259}
\newcommand{\pctZonesActiveMultiple}{\pct{59}}
\newcommand{\avgNumOfChoices}{3.83}

\newcommand{\numEquations}{28,222}
\newcommand{\numUniqueEquations}{4,574}
\newcommand{\numOutsideFragment}{919}
\newcommand{\pctOutsideFragment}{\pct{20}}
\newcommand{\numInsideFragment}{3,655}
\newcommand{\pctInsideFragment}{\pct{80}}
\newcommand{\numNoSolutionForOne}{194}
\newcommand{\pctNoSolutionForOne}{\pct{4}}
\newcommand{\numSolutionForOne}{3,461}

\newcommand{\numNoSolutionForOneHundred}{438}
\newcommand{\pctNoSolutionForOneHundred}{\pct{10}}
\newcommand{\numSolutionForOneHundred}{3,023}
\newcommand{\pctSolutionForOneHundred}{\pct{66}}

\newcommand{\minParseTime}{\millis{9}}
\newcommand{\maxParseTime}{\millis{520}}
\newcommand{\avgParseTime}{\millis{77}}
\newcommand{\medParseTime}{\millis{53}}
\newcommand{\minEvalTime}{\millis{$<${1}}}
\newcommand{\maxEvalTime}{\millis{165}}
\newcommand{\avgEvalTime}{\millis{12}}
\newcommand{\medEvalTime}{\millis{5}}
\newcommand{\minPrepareTime}{\millis{1}}
\newcommand{\maxPrepareTime}{\millis{6,789}}
\newcommand{\avgPrepareTime}{\millis{200}}
\newcommand{\medPrepareTime}{\millis{13}}
\newcommand{\minSolveTime}{\millis{$<${1}}}
\newcommand{\maxSolveTime}{\millis{14}}
\newcommand{\avgSolveTime}{\millis{$<${1}}}
\newcommand{\medSolveTime}{\millis{$<${1}}}

\def\parahead#1{\paragraph{\textbf{#1.}}}
\def\paraheadQuotes#1{\paragraph{\textbf{``#1.''}}}

\newcommand{\ie}{{\emph{i.e.}}}
\newcommand{\eg}{{\emph{e.g.}}}
\newcommand{\etc}{{\emph{etc.}}}
\newcommand{\cf}{{\emph{cf.}}}

\newcommand{\sns}{\ensuremath{\textsc{Sketch-n-Sketch}}}
\newcommand{\little}{\texttt{little}}
\newcommand{\snsItals}{\textrm{Sketch-n-Sketch}}
\newcommand{\littleItals}{\textrm{Little}}

\newcommand{\sineWaveBoxes}{\texttt{sineWaveOfBoxes}}

\newcommand{\powerpoint}{{PowerPoint}}
\newcommand{\illustrator}{{Illustrator}}

\newcommand{\myurl}[1]{\url{#1}}

\newcommand{\defeq}{\circeq}
\newcommand{\set}[1]{\{{#1}\}}

\newcommand{\setComp}[2]
           {\ensuremath{\{\miniSepThree #1 \hspace{0.02in}
                            \mid\hspace{0.02in} #2 \miniSepThree\}}}

\newcommand{\blindUrl}
  {\url{http://ravichugh.github.io/sketch-n-sketch}}
\newcommand{\theAppendix}
  {Supplementary Appendices~\cite{sns-supplementary}}
\newcommand{\theAppendixCite}
  {\cite{sns-supplementary}}
\newcommand{\example}[1]{{#1}}
\newcommand{\millis}[1]{{#1}\ ms}

\newcommand{\codeSize}{\small}

\newcommand{\sep}{\hspace{0.06in}}
\newcommand{\sepPremise}{\hspace{0.20in}}
\newcommand{\hsepRule}{\hspace{0.20in}}
\newcommand{\vsepRuleHeight}{0.12in}
\newcommand{\vsepRule}{\vspace{\vsepRuleHeight}}
\newcommand{\miniSepOne}{\hspace{0.01in}}
\newcommand{\miniSepTwo}{\hspace{0.02in}}
\newcommand{\miniSepThree}{\hspace{0.03in}}

\newcommand{\varNum}{n}
\newcommand{\varNumK}{k}
\newcommand{\varNumAnnotated}{N}
\newcommand{\varNumAnn}{\ensuremath{\beta}}
\newcommand{\varStr}{s}
\newcommand{\varBool}{b}
\newcommand{\varVal}{v}
\newcommand{\varValW}{w}

\newcommand{\varExp}{e}
\newcommand{\varLoc}{\ell}
\newcommand{\varLocs}{\ensuremath{\mathcal{L}}}

\newcommand{\varLocAnn}{\ensuremath{\alpha}}
\newcommand{\varTrace}{t}
\newcommand{\varVar}{x}
\newcommand{\varPat}{p}

\newcommand{\varOp}{op}
\newcommand{\varValCtx}{V}

\newcommand{\varSubst}{\rho}
\newcommand{\varAttrSubst}{\theta}
\newcommand{\varZoneSubst}{\gamma}

\newcommand{\varTrigger}{\tau}
\newcommand{\locName}[1]{\ensuremath{\mathit{#1}}}
\newcommand{\varZone}{\zeta}

\newcommand{\ttlcurly}{\ensuremath{\texttt{\char`\{}}\hspace{0.02in}}
\newcommand{\ttrcurly}{\ensuremath{\hspace{0.02in}\texttt{\char`\}}}}
\newcommand{\ttlparen}{\ensuremath{\texttt{(}}}
\newcommand{\ttrparen}{\ensuremath{\texttt{)}}}
\newcommand{\ttlbrack}{\ensuremath{\texttt{[}}}
\newcommand{\ttrbrack}{\ensuremath{\texttt{]}}}
\newcommand{\ttpipe}{\ensuremath{\texttt{|}}}
\newcommand{\ttdash}{\ensuremath{\texttt{-}}}

\newcommand{\num}[1]{\ensuremath{\texttt{#1}}}
\newcommand{\op}[1]{\ensuremath{\texttt{#1}}}

\newcommand{\inv}[2]{\ensuremath{\helperop{Inv}(#1)(#2)}}
\newcommand{\invL}[3]{\ensuremath{\helperop{InvL}(#1,#2)(#3)}}
\newcommand{\invR}[3]{\ensuremath{\helperop{InvR}(#1,#2)(#3)}}

\newcommand{\parens}[1]{\ensuremath{\ttlparen{#1}\ttrparen}}
\newcommand{\bracks}[1]{\ensuremath{\ttlbrack{#1}\ttrbrack}}
\newcommand{\spaceOne}[1]{\ensuremath{\miniSepOne{#1}\miniSepOne}}
\newcommand{\expFun}[2]{\ensuremath{\expTwoThings{\lambda}{#1}{#2}}}
\newcommand{\expNil}{\ensuremath{\ttlbrack\ttrbrack}}
\newcommand{\expCons}[2]{\ensuremath{\bracks{\spaceOne{#1}\ttpipe\spaceOne{#2}}}}
\newcommand{\expList}[3]{\bracks{\spaceOne{\threeThings{#1}{#2}{#3}}}}
\newcommand{\expApp}[2]{\ensuremath{\parens{{#1}\miniSepThree{ }{#2}}}}
\newcommand{\expAppTwo}[3]{\ensuremath{\expApp{#1}{\twoThings{#2}{#3}}}}
\newcommand{\expAppMulti}[3]{\ensuremath{\expApp{#1}{\threeThings{#2}{\cdots}{#3}}}}
\newcommand{\expStr}[1]
  {\ensuremath{\texttt{'{#1}'}}}
\newcommand{\expTrue}{\ensuremath{\texttt{true}}}
\newcommand{\expFalse}{\ensuremath{\texttt{false}}}

\newcommand{\expPair}[2]{\bracks{\spaceOne{\twoThings{#1}{#2}}}}
\newcommand{\expTriple}[3]{\bracks{\spaceOne{\threeThings{#1}{#2}{#3}}}}
\newcommand{\expListFour}[4]{\bracks{\spaceOne{\fourThings{#1}{#2}{#3}{#4}}}}

\newcommand{\labeledNum}[2]{\ensuremath{{#1}^{#2}}}
\newcommand{\numTrace}[2]{\labeledNum{#1}{#2}}

\newcommand{\ttNumLoc}[2]{\labeledNum{\num{#1}}{#2}}

\newcommand{\annFreeze}{\texttt{!}}
\newcommand{\annThaw}{\texttt{?}}
\newcommand{\numTriple}[3]{({#1},{#2},{#3})}
\newcommand{\numRange}[2]
  {\ttlcurly{#1}\miniSepTwo\ttdash\miniSepTwo{#2}\ttrcurly}
\newcommand{\numWithRange}[3]{{#1}\miniSepOne\numRange{#2}{#3}}

\newcommand{\expThaw}[1]{\ensuremath{{#1}\annThaw}}
\newcommand{\expLet}[3]{\expThreeThings{\texttt{let}}{#1}{#2}{#3}}
\newcommand{\expLetRec}[3]{\expThreeThings{\texttt{letrec}}{#1}{#2}{#3}}
\newcommand{\expDef}[3]{\twoThings{\expTwoThings{\texttt{def}}{#1}{#2}}{#3}}
\newcommand{\expDefRec}[3]{\twoThings{\expTwoThings{\texttt{defrec}}{#1}{#2}}{#3}}
\newcommand{\expIte}[3]{\expThreeThings{\texttt{if}}{#1}{#2}{#3}}
\newcommand{\expCase}[4]{\expFourThings{\texttt{case}}{#1}{#2}{#3}{#4}}
\newcommand{\expCaseTwo}[3]{\expThreeThings{\texttt{case}}{#1}{#2}{#3}}
\newcommand{\expBranch}[2]{\expOneThing{#1}{#2}}

\newcommand{\twoThings}[2]{\ensuremath{{#1}\miniSepThree{#2}}}
\newcommand{\threeThings}[3]{\ensuremath{{#1}\miniSepThree\twoThings{#2}{#3}}}
\newcommand{\fourThings}[4]{\ensuremath{{#1}\miniSepThree\threeThings{#2}{#3}{#4}}}
\newcommand{\fiveThings}[5]{\ensuremath{{#1}\miniSepThree\fourThings{#2}{#3}{#4}{#5}}}

\newcommand{\expOneThing}[2]{\ensuremath{\parens{\twoThings{#1}{#2}}}}
\newcommand{\expTwoThings}[3]{\ensuremath{\parens{\threeThings{#1}{#2}{#3}}}}
\newcommand{\expThreeThings}[4]{\ensuremath{\parens{\fourThings{#1}{#2}{#3}{#4}}}}
\newcommand{\expFourThings}[5]{\ensuremath{\parens{\fiveThings{#1}{#2}{#3}{#4}{#5}}}}

\newcommand{\figSyntaxLineBreak}{\\[2pt]}
\newcommand{\figSyntaxSpaceNextCategory}{\\[6pt]}
\newcommand{\figSyntaxSpaceItem}{\sep\mid\sep}
\newcommand{\figSyntaxSpaceItemNarrow}{\mid}
\newcommand{\figSyntaxEnd}{\end{array}$}

\newcommand{\figSyntaxBegin}{$\begin{array}{rcl}}
\newcommand{\figSyntaxRowLabel}[2]{{#2}&\miniSepThree{::=}\miniSepThree&}
\newcommand{\figSyntaxRowLabelSet}[2]{{#2}&\miniSepThree{\in}\miniSepThree&}
\newcommand{\figSyntaxRow}{&\miniSepThree\mid\miniSepThree&}

\newcommand{\relDescription}[1]{\ensuremath{\textrm{\textbf{#1}}}}

\newcommand{\judgementHeadThree}[3]
  {\ensuremath{\relDescription{#1}\ \textrm{{#2}}\hfill\fbox{#3}}}
\newcommand{\judgementHeadNameOnly}[1]
  {\ensuremath{\relDescription{#1}\hfill}}

\newcommand{\ruleName}[1]{\mbox{\textsc{\begin{normalsize}#1\end{normalsize}}}}
\newcommand{\ruleNameFig}[1]{\textsc{\begin{scriptsize}[#1]\end{scriptsize}}}

\newcommand{\tti}{\ensuremath{\texttt{i}}}

\newcommand{\reducesToMulti}[2]{{#1}\Downarrow{#2}}

\newcommand{\interp}[1]{\ensuremath{\llbracket\spaceOne{#1}\rrbracket}}
\newcommand{\hole}{\bullet}

\newcommand{\substBracks}[1]{[\miniSepOne{#1}\miniSepOne]}
\newcommand{\catSubsts}[2]{{#1}\miniSepTwo{#2}}

\newcommand{\substPlus}{\oplus}
\newcommand{\substOne}[2]{[{#2}/{#1}]}
\newcommand{\subst}[3]{{#1}\substOne{#2}{#3}}
\newcommand{\aSubst}[2]{{#1}\mapsto{#2}}
\newcommand{\aSubstOne}[2]
  {\substBracks{\aSubst{#1}{#2}}}
\newcommand{\aSubstTwo}[4]
  {\substBracks{\aSubst{#1}{#2},\aSubst{#3}{#4}}}

\newcommand{\relConsistent}[2]{\ensuremath{{#1}\sim{#2}}}

\newcommand{\manip}{\ensuremath{\leadsto}}

\newcommand{\helperop}[1]{\ensuremath{\mathsf{#1}}}
\newcommand{\locsOf}[1]{\ensuremath{\helperop{Locs}(#1)}}
\newcommand{\myEquation}[2]{\ensuremath{{#1}={#2}}}

\newcommand{\inferPlausible}[2]{\ensuremath{\helperop{SynthesizePlausible}(#1,#2)}}
\newcommand{\solve}[4]{\ensuremath{\helperop{Solve}(#1,#2,\myEquation{#3}{#4})}}
\newcommand{\solveOne}[4]{\ensuremath{\helperop{SolveOne}(#1,#2,\myEquation{#3}{#4})}}
\newcommand{\solvea}[4]{\ensuremath{\helperop{Solve_A}(#1,#2,\myEquation{#3}{#4})}}
\newcommand{\solveb}[4]{\ensuremath{\helperop{Solve_B}(#1,#2,\myEquation{#3}{#4})}}
\newcommand{\countPlus}[3]{\ensuremath{\helperop{WalkPlus}(#1,#2,#3)}}
\newcommand{\solveAndUpdate}[4]
  {(\aSubst{#2}{\solveOne{#1}{#2}{#3}{#4}})}

\newcommand{\mmVar}[1]{\ensuremath{\mathit{#1}}} 
\newcommand{\attrVal}[2]{#1[#2]}
\newcommand{\zone}[1]{\textsc{#1}}
\newcommand{\numOf}[1]{\ensuremath{\helperop{Num}(#1)}}
\newcommand{\traceOf}[1]{\ensuremath{\helperop{Tr}(#1)}}
\newcommand{\makeTrigger}[4]{\ensuremath{\helperop{ComputeTrigger}(#1,#2,#3,#4)}}
\newcommand{\makeTriggerSimple}[3]{\ensuremath{\helperop{ComputeTrigger}(#1,#2,#3)}}
\newcommand{\variance}[2]{{#1}\miniSepOne#2}
\newcommand{\dxPos}{\ensuremath{\variance{+}{\mmVar{dx}}}}
\newcommand{\dxNeg}{\ensuremath{\variance{-}{\mmVar{dx}}}}
\newcommand{\dyPos}{\ensuremath{\variance{+}{\mmVar{dy}}}}
\newcommand{\dyNeg}{\ensuremath{\variance{-}{\mmVar{dy}}}}

\newcommand{\mouseTrigger}[1]
  {\ensuremath{\lambda(\mmVar{dx},\mmVar{dy}).\ {#1}}}
\newcommand{\mouseTriggerMulti}[2]
  {\ensuremath{\lambda({#1}).\ {#2}}}

\newcommand{\svgRHS}[1]{\ensuremath{``{#1}"}}
\newcommand{\toSvg}{\ensuremath{\hookrightarrow}}
\newcommand{\toSvgSub}[1]{\ensuremath{\hookrightarrow_{#1}}}
\newcommand{\svgStr}[1]{\ensuremath{\texttt{'}{#1\texttt{'}}}}

\begin{document}

\setlength{\pdfpageheight}{\paperheight}
\setlength{\pdfpagewidth}{\paperwidth}

\conferenceinfo{PLDI '16}{June 13--17, 2016, Santa Barbara, CA, USA}
\copyrightyear{2016}
\copyrightdata{978-1-4503-4261-2/16/06}
\copyrightdoi{2908080.2908103}

\title{Programmatic and Direct Manipulation, Together at Last}

\authorinfo{Ravi Chugh \and Brian Hempel \and Mitchell Spradlin \and Jacob Albers}
           {University of Chicago, USA}
           {\{rchugh, brianhempel, mhs, jacobalbers\}\ @\ uchicago.edu}

\maketitle

\begin{abstract}

Direct manipulation interfaces and programmatic systems have
distinct and complementary strengths. The former provide intuitive,
immediate visual feedback and enable rapid prototyping, whereas the
latter enable complex, reusable abstractions. Unfortunately, existing
systems typically force users into just one of these two interaction
modes.

We present a system called \sns{} that integrates programmatic and
direct manipulation for the particular domain of
Scalable Vector Graphics (SVG).
%
In \sns{}, the user writes a program to generate an output SVG canvas.
Then the user may directly manipulate the canvas while the
system immediately infers a program update in order to match the
changes to the output, a workflow we call \emph{live synchronization}.
To achieve this,
we propose (i) a technique called \emph{trace-based program synthesis} that
takes program execution history into account in order to constrain the search
space and (ii) heuristics for dealing with ambiguities.
Based on our experience with examples spanning \littleLoc{} lines of code
and from the results of a preliminary user study,
we believe that \sns{} provides a novel workflow that can augment
traditional programming systems. Our approach may serve as the basis for
live synchronization in other application domains, as well as a starting
point for yet more ambitious ways of combining programmatic and direct
manipulation.

\end{abstract}



\section{Introduction}
\label{sec:intro}

\newcommand{\varConstraints}{\ensuremath{\mathcal{C}}}
\newcommand{\comma}{\ensuremath{,\ }}
\newcommand{\keyIdea}[2]{\parahead{Key Idea {#1}: {#2}}}

Direct manipulation user interfaces~\cite{Shneiderman1983}
such as Adobe \illustrator{}, Microsoft \powerpoint{}, and GIMP~\cite{GIMP}
and programmatic systems
such as Processing~\cite{Processing,ProcessingJs}
each have distinct strengths. The former
provide intuitive, immediate visual feedback and enable rapid
prototyping, whereas the latter allow for designing and reusing
complex abstractions.
Neither mode of interaction is preferable for all tasks.

\parahead{Motivation}

\setlength{\intextsep}{6pt}%
\setlength{\columnsep}{10pt}%
\begin{wrapfigure}{r}{0pt}
\includegraphics[scale=0.20]{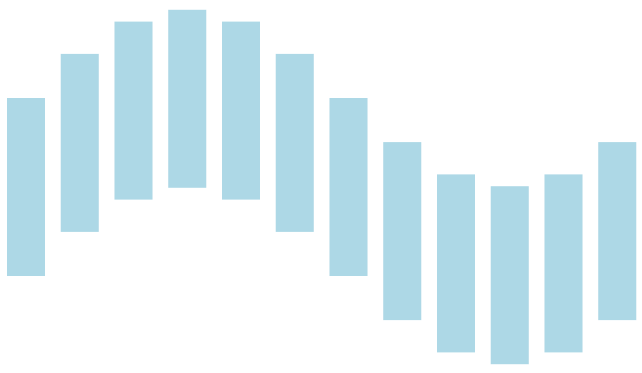}
\end{wrapfigure}

Imagine designing a series of identical shapes, laid out along the
contours of a sine wave. Designing the first ``prototype'' shape using
a visual, direct manipulation tool like \illustrator{} or
\powerpoint{} works well. But after copying the shape,
pasting it multiple times becomes tedious, and achieving the
sinusoidal pattern requires painstaking effort because built-in features
(\eg{} rulers, snapping to other shapes, and uniform spacing) are
unlikely to help. Worse yet are the edits
required to change high-level parameters of this design (\eg{} the
number of shapes, the spacing between them, or the amplitude of the
sine wave). The user will want to scrap the previous
effort and start from scratch.
On the other hand, by using a programmatic system, it is easy to churn
out variations of the high-level pattern by changing parameters and
re-running the program. But the disconnect between the software
artifact and its output can limit the pace of design, especially when
the appropriate parameters in the program are difficult to identify.

In a recent position paper, we identified several
domains for both visual and textual data where users
are forced to make the unfortunate choice between programmatic or direct
manipulation tools~\cite{prodirectvision}.
We proposed the notion that software systems ought to smoothly
integrate these two modes of use, a
combination we call \emph{prodirect manipulation}.
Ideally, a user could manipulate the output
of a program and the system would simultaneously update the program
to keep it synchronized with the changes.
In this paper, we present the first realization of that goal.

\parahead{Challenges}
Suppose that a program $\varExp$ generates $k$ output values
and that a {user action} changes the first ${j}$ values:
$
\set{
\varVal_{1} \manip \varVal_1' \comma \ldots \comma 
\varVal_{j} \manip \varVal_j' \comma
\varVal_{j+1} \comma \ldots \comma
\varVal_{k}
}
$.
\noindent
To synthesize a program $\varExp'$ that generates the updated output,
three primary challenges must be addressed:
(1) The meaning of the user's updates must be defined in a way
that constrains the program synthesis search space;
(2) The search algorithm must run quickly and find program updates
that are easy for the user to understand; and
(3) When multiple candidate solutions exist,
the ambiguity must be dealt with in a way that facilitates the
responsive, interactive workflow characteristic of direct
manipulation interfaces.

\keyIdea{1}{Trace-Based Program Synthesis}
To address challenge (1),
we propose a technique called \emph{trace-based program synthesis}
that comprises two components.
First, we instrument the evaluation of the program $\varExp$ so that
each of the $k$ values $\varVal_i$ it produces
comes with a trace $\varTrace_i$, which forms a set of constraints
$
\set{
\varVal_1=\varTrace_1 \comma \ldots \comma \varVal_k=\varTrace_k
}
$ relating the program to its output.
Second, to reconcile $\varExp$ with the $j$ updated
values, we specify that an ideal candidate update would be a program $\varExp'$
whose output satisfies the system of constraints
$
\varConstraints \defeq
\set{
\varVal'_{1}=\varTrace_{1} \comma \ldots \comma
\varVal'_{j}=\varTrace_{j} \comma
\varVal_{j+1}=\varTrace_{j+1} \comma \ldots \comma
\varVal_{k}=\varTrace_{k}
}
$.

\keyIdea{2}{Small Updates}
To address challenge (2),
our design principle is that only ``small''
program updates may be inferred; ``large'' changes
may require user intervention and are thus less conducive
to our goal of immediate synchronization.
In particular, we attempt only to change value literals
in the program, represented as a \emph{substitution} $\varSubst$
that maps locations in the abstract syntax tree to new values.

\keyIdea{3}{Heuristics for Disambiguation}

Even with the seemingly modest goal of inferring small updates, there
is still the problem of dealing with multiple solutions.
To address challenge (3), we first decompose $\varConstraints$
into $k$ separate constraints
$\varConstraints_1 \defeq \set{ \varVal'_{1}=\varTrace_{1} }$
through
$\varConstraints_k \defeq \set{ \varVal'_{k}=\varTrace_{k} }$
and then use heuristics to force each
$\varConstraints_i$ to have at most one solution $\varSubst_i$.
This enables us to define a \emph{trigger} function
$\mouseTriggerMulti{\varVal'_1 \comma \ldots \comma \varVal'_j}{\varSubst}$
that, given the concrete values changed by the user, computes a
solution $\varSubst_i$ for each constraint $\varConstraints_i$
and combines them into a single substitution
$\varSubst \defeq \catSubsts{\varSubst_1}{\catSubsts{\cdots}{\varSubst_k}}$.
This substitution is applied to the original program in real-time during
the user action, resulting in a new program $\varSubst\varExp$ that is
evaluated and ready for subsequent user manipulation.

\parahead{Contributions and Outline}

These three key ideas are general and may be developed for several settings.
In this paper, we focus on instantiating our approach for the
specific domain of Scalable Vector Graphics (SVG).
In particular:

\begin{itemize}

\item
We propose a framework for inferring program updates
called \emph{trace-based synthesis}, which may apply broadly
to domains where users wish to (indirectly) manipulate programs by
(directly) manipulating their outputs.
(\autoref{sec:sync})

\item
We define heuristics for disambiguating between possible
updates and define \emph{triggers} that keep a program
synchronized with changes to its SVG output in real-time.
(\autoref{sec:prodirect})

\item
We implement our ideas in a Web-based system called \sns{}
and evaluate its interactivity.
(\autoref{sec:implementation})

\item
We use \sns{} to design many examples
difficult to develop or edit using existing tools.
(\autoref{sec:examples})

\end{itemize}

\noindent
Our implementation and examples,
as well as additional tutorial documentation and videos,
are publicly available at \blindUrl{}.

\section{Overview}
\label{sec:overview}

\begin{figure}[t]
\begin{center}

\textbf{(A)} Excerpt from \texttt{prelude.little}

\begin{Verbatim}[commandchars=\\\{\},codes={\catcode`\$=3\catcode`\^=7\catcode`\_=8},fontsize=\codeSize]
  (defrec range ($\lambda$(i j)
    (if (> i j) nil
        (cons i (range (+ $\ttNumLoc{1}{\varLoc_1}$ i) j)))))
 
  (def zeroTo ($\lambda$n (range $\ttNumLoc{0}{\varLoc_0}$ (- n 1))))
\end{Verbatim}

\textbf{(B)} \sineWaveBoxes\texttt{.little}

\begin{Verbatim}[commandchars=\\\{\},codes={\catcode`\$=3\catcode`\^=7\catcode`\_=8},fontsize=\codeSize]
  (def [x0 y0 w h sep amp] [50 120 20 90 30 60])
  (def n 12!\{3-30\})
  (def boxi ($\lambda$i
    (let xi (+ x0 (* i sep))
    (let yi (- y0 (* amp (sin (* i (/ twoPi n)))))
      (rect 'lightblue' xi yi w h)))))
 
  (svg (map boxi (zeroTo n)))
\end{Verbatim}

\textbf{(C)}
Suppose the user clicks on the third box from the left
(colored darker in red for emphasis)
and drags it to a new position down and to the right
(colored lighter in gray):
\vspace{0.05in}

\includegraphics[scale=0.35]{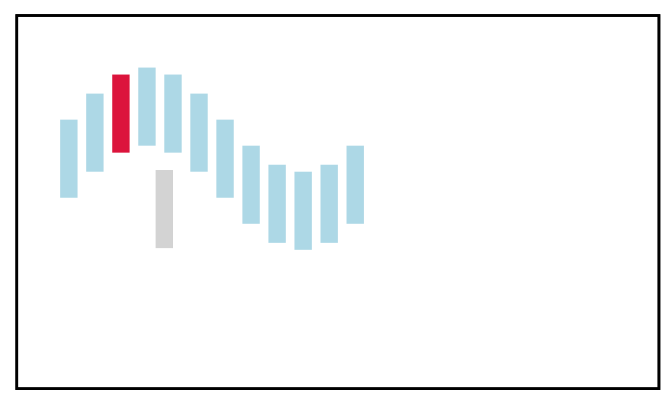}

\textbf{(D)}
\sns{} synthesizes four candidate updates to the program,
which have the following effects:
\vspace{0.05in}

\includegraphics[scale=0.35]{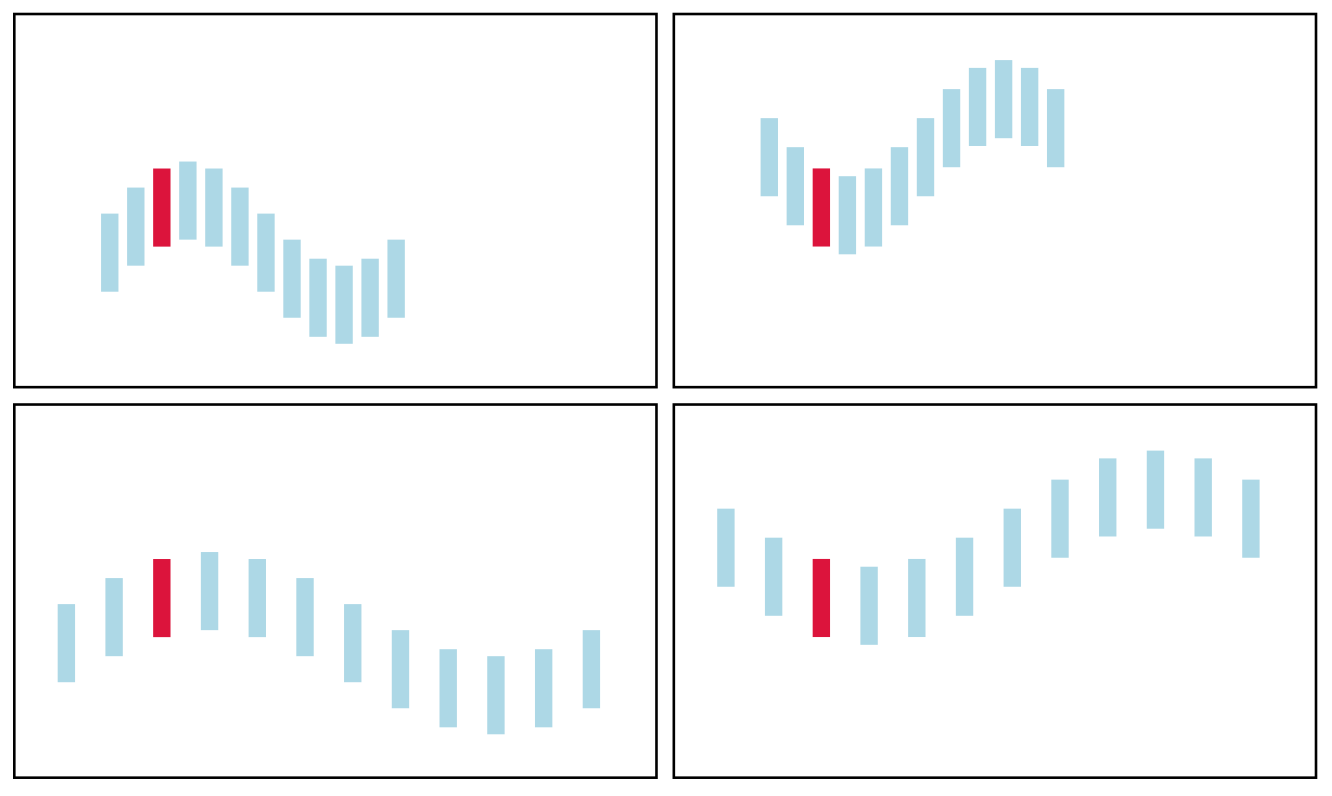}
\end{center}
\caption{Sine Wave of Boxes in \sns{}}
\label{fig:little-sine-wave}
\end{figure}

We will provide an overview of \sns{} by considering the
program in \autoref{fig:little-sine-wave} that generates
the sine wave box design in \autoref{sec:intro}.
Suppose the user clicks on the third box
and drags it to a new position down and to the right
(depicted in \autoref{fig:little-sine-wave}C).
In this section, we will describe how \sns{} synthesizes
four candidate updates to the program, each of which
computes the position of the third box to match the user's
direct manipulation but have different effects on the remaining boxes
(depicted in \autoref{fig:little-sine-wave}D).

\parahead{A \littleItals{} Programming Language}

The programming language in \sns{} is a core, untyped functional language
called \little{}, which includes
base values (floating-point numbers, booleans, and strings)
and lists (represented as cons-cells, or pairs).
For reference, we define the syntax of \little{} in
\autoref{fig:syntax}.
All expression and value forms are standard except for numbers.

Because numeric attributes are often directly manipulated in graphical
user interfaces, each numeric literal in \little{} comes with three
pieces of additional information: a \emph{location} $\varLoc$, which
is an integer inserted by the parser that identifies its position in
the abstract syntax tree (AST); an optional, user-specified \emph{location annotation}
$\varLocAnn$, where $\annFreeze$ is used to \emph{freeze} the number; and
an optional, user-specified \emph{range annotation} $\varNumAnn$,
where $\numRange{\varNum_1}{\varNum_2}$ denotes the
intended range of the number.
We discuss these annotations further later in this section.

\parahead{Representing SVG Values}

An \emph{SVG node} is represented as a list
\verb+[ svgNodeKind attributes children ]+
with three values
where the string \verb+svgNodeKind+ describes the kind of the node
(\eg{} \expStr{rect}, \expStr{circle}, or \expStr{line} for particular shapes,
or \expStr{svg} for a canvas, or collection, of shapes);
\verb+attributes+ is a list of attributes (\ie{} key-value pairs); and
\verb+children+ is a list of child nodes.
The intended result of a \little{} program is a node with
kind \expStr{svg}. The result of evaluating the program in
\autoref{fig:little-sine-wave}B takes the form below,
which gets translated directly to the SVG format~\cite{SVG} and
resembles the screenshot in \autoref{sec:intro} when rendered.
\begin{Verbatim}[commandchars=\\\{\},codes={\catcode`\$=3\catcode`\^=7\catcode`\_=8},fontsize=\codeSize]
 ['svg' []
   [ ['rect' [['x'  50] ['y' 120] ...] []]
     ['rect' [['x'  80] ['y'  90] ...] []]
     ['rect' [['x' 110] ['y'  68] ...] []] ... ] ]
\end{Verbatim}

\noindent
Our translation of key-value attributes provides a
thin wrapper over the target SVG format, allowing \little{}
programs to specify arbitrary XML attributes using strings.
We discuss the translation further in \theAppendix{}, but the details
are not needed in the rest of the paper.

\subsection{Locations and Traces}

To enable direct manipulation, the \little{} evaluator produces
run-time \emph{traces} to track the evaluation of numeric values
(for attributes such as \expStr{x}, \expStr{y}, \expStr{width},
and \expStr{height}).
There are two kinds of traces $\varTrace$ that are used to
infer program updates based on direct manipulation changes.

\parahead{Locations}

The simplest kind of trace records that a numeric literal $\varNum$ originates
from a particular source-code location $\varLoc$ in the AST
of the program.
For the program in \autoref{fig:little-sine-wave}B, the
\little{} parser annotates all numbers
\numTrace{\num{50}}{\varLoc_2},
\numTrace{\num{120}}{\varLoc_3},
\numTrace{\num{20}}{\varLoc_4},
\numTrace{\num{90}}{\varLoc_5}
\numTrace{\num{30}}{\varLoc_6},
\numTrace{\num{60}}{\varLoc_7}, and
\numTrace{\num{12}}{\varLoc_8}
with unique location identifiers.
These locations do not affect program evaluation or SVG rendering.
When the user manipulates a numeric value $\labeledNum{\varNum}{\varLoc}$
in the canvas (\eg{} by dragging or stretching a box), \sns{}
updates the value at program location $\varLoc$ in real-time.

In the rest of the paper,
when a number $\varNum$ is immediately bound to a variable $\mathtt{\varVar}$,
we choose the canonical name $\locName{\varVar}$ for the location
--- resulting in the value $\numTrace{\varNum}{\locName{\varVar}}$ ---
rather than one of the form $\varLoc_k$.
In the examples that follow,
we sometimes annotate numeric literals explicitly with locations
for explanatory purposes
even though the programmer does not write them ---
they are inserted implicitly by our parser.

\parahead{Expression Traces}

Smaller traces are combined into larger ones during the evaluation of
primitive operations, such as addition, multiplication, \etc{}
Whereas the \expStr{width} and \expStr{height} attribute
values originate from atomic AST locations, the \expStr{x}
value of each box is the result of evaluating
\texttt{xi} $\defeq$ \texttt{(+ x0 (* i sep))}
with different bindings for \tti{}.
Each run-time value that is bound to \tti{} is generated
by the function \verb+zeroTo+
(from a \verb+Prelude+ library included in every program),
which computes the list of integers from \verb+0+ to \verb+n-1+.

When evaluating a primitive operation, the run-time semantics
of \little{} records the structure of the expression along
with the resulting value.
Below are the values of \verb+xi+ for each of the first three boxes,
respectively, and their corresponding traces. Each value-trace
pair forms an equation:
\begin{align}
\num{50} &=
  {\expAppTwo{\op{+}}
    {\locName{x_0}}
    {\expAppTwo{\op{*}}{\varLoc_0}{\locName{sep}}}}
\label{eq:box1}
\\
\num{80} &=
  {\expAppTwo{\op{+}}
    {\locName{x_0}}
    {\expAppTwo{\op{*}}{\expAppTwo{\op{+}}{\varLoc_1}{\varLoc_0}}{\locName{sep}}}}
\label{eq:box2}
\\
\num{110} &=
  {\expAppTwo{\op{+}}
    {\locName{x_0}}
    {\expAppTwo{\op{*}}
      {\expAppTwo{\op{+}}{\varLoc_1}{\expAppTwo{\op{+}}{\varLoc_1}{\varLoc_0}}}
      {\locName{sep}}}}
\label{eq:box3}
\end{align}

\noindent
The locations $\varLoc_0$ and $\varLoc_1$ identify
literals \verb+0+ and \verb+1+, respectively, in
the \verb+Prelude+ function \verb+zeroTo+ that computes
increasing integers.
The value-trace equations for the remaining boxes are analogous.
These equations, together with the following substitution that
records location-value mappings from the source program,
relate the program to its output.
\begin{align*}
\varSubst_0 &\defeq
  \substBracks{
    \aSubst{\locName{x_0}}{\num{50}}
  , \aSubst{\locName{sep}}{\num{30}}
  , \aSubst{\varLoc_0}{\num{0}}
  , \aSubst{\varLoc_1}{\num{1}}
  , \ \ldots
  }
\end{align*}

\parahead{Dataflow-Only Traces}

The reduction rule \ruleName{E-Op-Num} (\autoref{fig:syntax}) builds expression
traces in parallel with the evaluation of primitive operations,
producing values $\labeledNum{\varNum}{\varTrace}$.
Traces $\varTrace$ record data flow but not control flow.
This design is based on the insight that programs generating output in
visual domains are often structured so that the control flow of the
program is similar across multiple runs.
We have not found this limitation to be a problem in practice for the
examples we have developed.
However, it may be useful to enrich traces with additional information
in subsequent work.

\subsection{Synthesizing Program Updates}
\label{sec:overview-sync}

The main idea behind \emph{trace-based program synthesis} is to use
value-trace equations in order to infer program updates that conform to
output values changed by the user.
In the setting of direct manipulation interfaces, changing attributes of
visual objects corresponds to changing the values on the left-hand
side of the equations above.
In our \sineWaveBoxes{} example, the result of dragging
the third box directly to the right so that its new \expStr{x} value
is \num{155}, is to replace \autoref{eq:box3} with the following equation:
\begin{align}
\num{155} &=
  {\expAppTwo{\op{+}}
    {\locName{x_0}}
    {\expAppTwo{\op{*}}
      {\expAppTwo{\op{+}}{\varLoc_1}{\expAppTwo{\op{+}}{\varLoc_1}{\varLoc_0}}}
      {\locName{sep}}}}
\tag{3'}
\label{eq:newthirdbox}
\end{align}
Our goal is to synthesize an updated program that satisfies
Equation 3'; satisfying all other (unchanged) equations would be ideal,
but this one is ``more important'' because it was induced by the user's
change.

\parahead{Local Updates}

We aim only to infer \emph{local updates}, which are
substitutions that map locations to updated numeric values.
We describe design and implementation decisions in
\autoref{sec:prodirect} and \autoref{sec:implementation}
that limit the cost of solving equations to ensure
responsive interaction with the user.
With our approach, \sns{} can infer four
substitutions based on \autoref{eq:newthirdbox}:
$\varSubst_1 \defeq \varSubst_0 \substBracks{\aSubst{\locName{x_0}}{\num{95}}}$,
$\varSubst_2 \defeq \varSubst_0 \substBracks{\aSubst{\locName{sep}}{\num{52.5}}}$,
$\varSubst_3 \defeq \varSubst_0 \substBracks{\aSubst{\varLoc_0}{\num{1.5}}}$, and
$\varSubst_4 \defeq \varSubst_0 \substBracks{\aSubst{\varLoc_1}{\num{1.75}}}$.

These substitutions, if applied to the program in
\autoref{fig:little-sine-wave}B, would
produce various effects.
The first option would translate all of the boxes in unison.
The second would increase the spacing between boxes.
The third and fourth would, respectively, translate all boxes
or the change the separation, but both would also change the
number of boxes because the constants \verb+0+ and \verb+1+
at locations $\varLoc_0$
and $\varLoc_1$ were used in the original
program to compute integer indices.
The user is unlikely to want either the third or fourth options,
since the list of integers
$\op{0}$ to $\op{n-1}$ specifies the number of boxes.
Moreover, the locations $\varLoc_0$ and $\varLoc_1$ appear in the
\verb+Prelude+, so changing their values would affect
the behavior of other programs!

\parahead{Frozen Constants}

By default, \sns{} will consider updating the value of
any program location used in a value-trace equation in order
to reconcile the program with the user's changes.
The user can direct the synthesis procedure not to change the
value of particular constants by \emph{freezing} them, denoted with
exclamation points (\eg{} \verb+3.14!+).
All numbers in \verb+Prelude+ are automatically frozen, so solutions
$\varSubst_3$ and $\varSubst_4$ are not actually considered by \sns{}.
Without freezing either \verb+x0+ or \verb+sep+, however,
the ambiguity between $\varSubst_1$ and $\varSubst_2$ remains.

\subsection{Heuristics}
\label{sec:overview-ambiguities}

Pausing and asking the user to choose between updating $\locName{x_0}$ or
$\locName{sep}$ would stymie the interactive, \emph{live synchronization}
we strive for.
Instead, we employ heuristics for automatically
resolving ambiguities that attempt to strike a
balance between interactivity and predictability.
Our key insight
is that the essence of a local update is the set of constants
$\varLocs$ (\ie{} program locations) that are changed and \emph{not}
necessarily their new values. Even if we arbitrarily decide that a
particular user action should cause a set $\varLocs_1$ of constants
to change rather than a set $\varLocs_2$, we can often assign a different
user action to manipulate the constants in $\varLocs_2$.

If the user drags the first box horizontally to a new \expStr{x} position
$\varNum$, the change would induce the value-trace equation
$\varNum=
  {\expAppTwo{\op{+}}
    {\locName{x_0}}
    {\expAppTwo{\op{*}}{\varLoc_0}{\locName{sep}}}}$
based on \autoref{eq:box1},
which can be solved by changing the value of either
$\locName{x_0}$ or $\locName{sep}$.
In preparation,
we arbitrarily choose to update $\locName{x_0}$ to $\varNum$
in order to solve the new equation if and when the user drags this box.
If the user drags the second box, again, either $\locName{x_0}$ or
$\locName{sep}$ could be changed in order to solve the equation
$\varNum=
  {\expAppTwo{\op{+}}
    {\locName{x_0}}
    {\expAppTwo{\op{*}}{\expAppTwo{\op{+}}{\varLoc_1}{\varLoc_0}}{\locName{sep}}}}$
based on \autoref{eq:box2}.
Because we already assigned $\locName{x_0}$ to vary if the user drags
the first box, we choose to vary $\locName{sep}$ if the user drags
the second box.
If the user drags the third box, again, either $\locName{x_0}$
or $\locName{sep}$ could be changed to solve the induced equation
$\varNum=
  {\expAppTwo{\op{+}}
      {\locName{x_0}}
      {\expAppTwo{\op{*}}
        {\expAppTwo{\op{+}}{\varLoc_1}{\expAppTwo{\op{+}}{\varLoc_1}{\varLoc_0}}}
        {\locName{sep}}}}$
based on \autoref{eq:box3}.
Because each location has
already been assigned to vary in response to some user action,
we arbitrarily choose $\locName{x_0}$.

We continue to assign program updates in this fashion
by trying to balance the number of times a location set is assigned to
some user action in the canvas.
When the user performs an action, we use the
pre-determined location set, together with the concrete
values from the mouse manipulation,
to compute a local update (\ie{} substitution), apply it
to the original program, run the resulting new program, and render the
new output canvas.

\subsection{Sliders}
\label{sec:overview-sliders}

There are several situations in which
it can be difficult or impossible to make
a desired change to the program via direct manipulation:
(1) when there is no natural, visual representation of the
program parameter of interest;
(2) when a visual representation may be hard to
directly manipulate, for example, because the shape is too small or
there are too many adjacent shapes; and
(3) when ambiguous updates are hard to resolve using
the built-in heuristics or by deciding which constants to freeze or
not.

\sns{} provides a feature that can help in all three situations.
If a number is annotated with a range, written
$\numWithRange{\labeledNum{\varNum}{\varLoc}}
              {\varNum_{\mmVar{min}}}
              {\varNum_{\mmVar{max}}}$,
then \sns{} will display a slider in the output pane
that can be used to manipulate the $\varNum$ value
between $\varNum_{\mmVar{min}}$ and $\varNum_{\mmVar{max}}$
(as opposed to having to edit the program).
Therefore,
sliders can provide control over otherwise hard-to-manipulate attributes.

\setlength{\intextsep}{6pt}%
\setlength{\columnsep}{10pt}%
\begin{wrapfigure}{r}{0pt}
\includegraphics[scale=0.40]{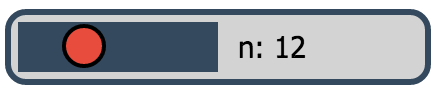}
\end{wrapfigure}
In the \sineWaveBoxes{} example, the number of boxes \verb+n+
is the only parameter that is hard to directly manipulate. The
heuristics choose to update \verb+n+ in response to certain
actions, but the effects are not intuitive.
Therefore, on line
2 of \autoref{fig:little-sine-wave}B, we freeze the
value of \verb+n+ (so that it will never change as a result
of directly manipulating the boxes) and declare the range
\verb+{3-30}+ so that we can easily adjust the number of boxes
using a slider.

\begin{figure}[t]

\centering

\judgementHeadNameOnly{Syntax of Expressions}

\vsepRule

\figSyntaxBegin

\figSyntaxRowLabel{Expressions}{\varExp}
  \varNumAnnotated
  \figSyntaxSpaceItem
  \varStr
  \figSyntaxSpaceItem
  \varBool
  \figSyntaxSpaceItem
  \expNil
  \figSyntaxSpaceItem
  \expCons{\varExp_1}{\varExp_2}
  \figSyntaxSpaceItem
  \varVar
\figSyntaxLineBreak
\figSyntaxRow
  \expFun{\varPat}{\varExp}
  \figSyntaxSpaceItem
  \expApp{\varExp_1}{\varExp_2}
  \figSyntaxSpaceItem
  \expAppMulti{\varOp_m}{\varExp_1}{\varExp_m}
\figSyntaxLineBreak
\figSyntaxRow
  \expLet{\varPat}{\varExp_1}{\varExp_2}
  \figSyntaxSpaceItem
  \expLetRec{\varPat}{\varExp_1}{\varExp_2}
\figSyntaxLineBreak
\figSyntaxRow
  \expCase{\varExp}{\expBranch{\varPat_1}{\varExp_1}}{\cdots}
                   {\expBranch{\varPat_m}{\varExp_m}}
\figSyntaxSpaceNextCategory
\figSyntaxRowLabel{Patterns}{\varPat}
  \varVar
  \figSyntaxSpaceItem
  \varNum
  \figSyntaxSpaceItem
  \varStr
  \figSyntaxSpaceItem
  \varBool
  \figSyntaxSpaceItem
  \expNil
  \figSyntaxSpaceItem
  \expCons{\varPat_1}{\varPat_2}
\figSyntaxSpaceNextCategory
\figSyntaxRowLabel{Numbers}{\varNumAnnotated}
  \numTriple{\labeledNum{\varNum}{\varLoc}}{\varLocAnn}{\varNumAnn}
  \hspace{0.13in}
  \varLocAnn \sep{::=}\sep
    \cdot
    \figSyntaxSpaceItemNarrow
    \annFreeze
  \hspace{0.13in}
  \varNumAnn \sep{::=}\sep
    \cdot
    \figSyntaxSpaceItemNarrow
    \numRange{\varNum_{1}}{\varNum_{2}}
\figSyntaxSpaceNextCategory
\figSyntaxRowLabelSet{}{\varOp_0}
  \set{\op{pi}, \ldots}
\figSyntaxLineBreak
\figSyntaxRowLabelSet{}{\varOp_1}
  \set{\op{not}, \op{cos}, \op{sin}, \op{arccos},
       \op{round}, \op{floor}, \ldots}
\figSyntaxLineBreak
\figSyntaxRowLabelSet{}{\varOp_2}
  \set{\op{+}, \op{-}, \op{*}, \op{/}, \op{<}, \op{=},
       \op{mod}, \op{pow}, \ldots}
\figSyntaxEnd

\vsepRule

\judgementHeadNameOnly{Syntax of Values}

\vsepRule

\figSyntaxBegin

\figSyntaxRowLabel{Values}{\varVal,\varValW}
  \labeledNum{\varNum}{\varTrace}
  \figSyntaxSpaceItem
  \varStr
  \figSyntaxSpaceItem
  \varBool
  \figSyntaxSpaceItem
  \expNil
  \figSyntaxSpaceItem
  \expCons{\varVal_1}{\varVal_2}
  \figSyntaxSpaceItem
  \expFun{\varPat}{\varExp}
\figSyntaxSpaceNextCategory
\figSyntaxRowLabel{Traces}{\varTrace}
  \varLoc
  \figSyntaxSpaceItem
  \expApp{\varOp_m}{\threeThings{\varTrace_1}{\cdots}{\varTrace_m}} 
\figSyntaxEnd

\vsepRule

\judgementHeadThree
  {Operational Semantics}{(excerpted)}
  {$\reducesToMulti{\varExp}{\varVal}$}

\vsepRule

$\inferrule*[right=\ruleNameFig{E-Op-Num}]
  {\varNum = \interp{\expAppMulti{\varOp_m}{\varNum_1}{\varNum_m}}
   \sepPremise
   \varTrace = \expAppMulti{\varOp_m}{\varTrace_1}{\varTrace_m}
  }
  {\reducesToMulti
     {\expAppMulti{\varOp_m}
                  {\labeledNum{\varNum_1}{\varTrace_1}}
                  {\labeledNum{\varNum_m}{\varTrace_m}}}
     {\labeledNum{\varNum}{\varTrace}}
     }
$

\vsepRule

\caption{Syntax and Semantics of \little{}}
\label{fig:syntax}
\end{figure}

\section{Trace-Based Program Synthesis}
\label{sec:sync}

In the previous section, we defined the semantics of \little{}
to produce run-time traces for numeric attributes. In this
section, we formulate \emph{trace-based program synthesis} as
a way to define the relationship between a program and updates
to its output, without regard to the particular domain.

\parahead{User Actions}

Suppose that a \little{} program $\varExp$ generates output
that contains $k$ numeric values.
With a single action, the user may
directly manipulate $j$ of the numeric values,
for some $1{\le}j{\le}k$.
The following table shows how, as alluded to in \autoref{sec:intro}, the $j$
updated values, together with the $k-j$ unchanged ones, form a system
of constraints that, ideally, an updated program $\varExp'$
would satisfy:

\begin{center}
$
\begin{array}{c|c|c|cc}
\textrm{Program} &
\textrm{Output} &
\textrm{Updates} &
\multicolumn{2}{c}{\textrm{Constraints}}
\\ \hline
&
\varValW_1 = \numTrace{\varNum_1}{\varTrace_1} &
\manip \varNum_1' &
\myEquation{\varNum_1'}{\varTrace_1} &
\textrm{(hard)}
\\
&
\cdots &
\cdots &
\cdots &
\textrm{(hard)}
\\
\varExp\ \ \Rightarrow &
\varValW_j = \numTrace{\varNum_j}{\varTrace_j} &
\manip \varNum_j' &
\myEquation{\varNum_j'}{\varTrace_j} &
\textrm{(hard)}
\\
&
\cdots &
&
\cdots &
\textrm{(soft)}
\\
&
\varValW_k = \numTrace{\varNum_k}{\varTrace_k} &
&
\myEquation{\varNum_k}{\varTrace_k} &
\textrm{(soft)}
\end{array}
$
\end{center}

\noindent
The user's changes may lead to an unsatisfiable
set of equations (when considering only local updates).
We treat equations induced by changes as ``hard''
constraints that a solution ought to satisfy, whereas the rest
are ``soft'' constraints that should be satisfied if possible.
This design principle prioritizes explicit changes made by
the user, which is the goal of our workflow.
Next, we will formally define what constitutes a valid solution
to a system of constraints.

\parahead{Contexts and Substitutions}

We define a \emph{value context} $\varValCtx$ below to be a value with
$m>0$ placeholders, or \emph{holes}, labeled $\hole_1$ through $\hole_m$.
We define the application of a value context to a list of values as
$
\varValCtx(\varVal_1,\cdots,\varVal_m) \ \defeq\
  \subst{\subst{\varValCtx}{\hole_1}{\varVal_1} \cdots}
        {\hole_m}{\varVal_m}
$.
A \emph{substitution} $\varSubst$ is a mapping from
program locations $\varLoc$ to numbers $\varNum$.
When applied to an expression, the bindings of a substitution are applied
from left-to-right.
Thus, the rightmost binding of any location takes precedence.
We use juxtaposition $\catSubsts{\varSubst}{\varSubst'}$ to denote concatenation,
and we write $\varSubst\substPlus(\aSubst{\varLoc}{\varNum})$ to
denote $\catSubsts{\varSubst}{\aSubstOne{\varLoc}{\varNum}}$.
We define \emph{value context similarity} below to relate values that
are structurally equal up to the values of numeric constants.
$$
\varValCtx \sep{::=}\sep
  \hole_i
  \figSyntaxSpaceItem
  \labeledNum{\varNum}{\varTrace}
  \figSyntaxSpaceItem
  \varStr
  \figSyntaxSpaceItem
  \varBool
  \figSyntaxSpaceItem
  \expNil
  \figSyntaxSpaceItem
  \expCons{\varValCtx_1}{\varValCtx_2}
  \figSyntaxSpaceItem
  \expFun{\varPat}{\varExp}
$$
\vspace{-0.19in} 
$$
\inferrule*
  { }
  {\relConsistent{\varValCtx_1}{\varValCtx_1}}
\hsepRule
\inferrule*
  { }
  {\relConsistent{\numTrace{\varNum_1}{\varTrace}}
                 {\numTrace{\varNum_2}{\varTrace}}}
\hsepRule
\inferrule*
  {\relConsistent{\varValCtx_1}{\varValCtx_1'} \sepPremise
   \relConsistent{\varValCtx_2}{\varValCtx_2'}}
  {\relConsistent{\expCons{\varValCtx_1}{\varValCtx_2}}
                 {\expCons{\varValCtx_1'}{\varValCtx_2'}}}
$$

\vspace{0.02pt} 

\parahead{Definition: Faithful Updates}

If

\begin{enumerate}
\setlength{\itemindent}{8pt}
\item[(a)]
$\reducesToMulti{\varExp}{\varVal}$, where
$\varVal = \varValCtx(\varValW_1\comma\ldots\comma\varValW_k)$; and
\item[(b)]
the user updates $\varValW_1\comma\ldots\comma\varValW_j$ to
$\varValW_1'\comma\ldots\comma\varValW_j'$,
\end{enumerate}

\noindent
then a substitution $\varSubst$ is \emph{faithful} if

\begin{enumerate}
\setlength{\itemindent}{8pt}
\item[(c)]
$\reducesToMulti{\varSubst\varExp}{\varVal'
   = \varValCtx'(\varValW_1''\comma\ldots\comma\varValW_k'')}$
where $\relConsistent{\varValCtx'}{\varValCtx}$; implies
\item[(d)]
$\varValW_i'' = \varValW_i'$ for all ${1}\le{i}\le{j}$.
\end{enumerate}

\noindent Premises (a) and (b) identify the list of $j$ values
manipulated by the user, and properties (c) and (d) capture the notion
that hard constraints induced by these changes should be satisfied
by the update $\varSubst$. The value similarity relation checks that
two value contexts are structurally equal but says
nothing about the soft constraints from the original program
(namely, it does not say $\varValW''_i=\varValW_i$ for all ${j}<{i}\le{k}$).
In a setting where multiple updates are synthesized, ranking
functions could be used to optimize for soft constraints.

It is important to note that our definition states
``(c) \emph{implies} (d)'' rather than the stronger property
``(c) \emph{and} (d)''
because the control flow may change and produce
$\varValCtx' \not\sim \varValCtx$.
We choose the weaker version because we do not intend to reason
about control flow either in traces or our synthesis algorithm
(\autoref{sec:implementation})
when considering how one program compares to another.
In other settings, it may be worthwhile to require the stronger version,
which would necessitate a richer trace language that records
control-flow information.

\parahead{Definition: Plausible Updates}
We define an alternative, weaker correctness criterion.
In particular,
we define a \emph{plausible update} to be one that satisfies \emph{some}
(\ie{} at least one) of the user's updates. Concretely, a plausible update
is defined just like a faithful one, except that the following condition
replaces (d) in the original definition:

\begin{enumerate}
\setlength{\itemindent}{8pt}
\item[(d')]
$\varValW_i'' = \varValW_i'$ for some ${1}\le{i}\le{j}$
\end{enumerate}

The general framework presented in this section
can be instantiated with solvers (which we will refer to as
$\helperop{Solve}$) that aim for different
points along this spectrum of faithful and plausible updates.

\section{Live Synchronization for SVG}
\label{sec:prodirect}

Given changes to the output of a program, in the previous section we
defined how value-trace equations can be used to specify candidate
program updates in order to reconcile the changes.
In this section, we describe how to compute program updates in
real-time for the specific domain of Scalable Vector Graphics (SVG).
First, we identify what constitutes a user action in this setting.
Second, we formulate how to compute \emph{triggers} that dictate
program updates based on such actions. For the latter, we propose
heuristics to automatically resolve ambiguities that result
from trace-based program synthesis problem instances.

In the following, we write $\attrVal{\varVal}{\expStr{k}}$ to refer to the
value of attribute \expStr{k} in the \little{} SVG value $\varVal$.
We also define the abbreviations
$\numOf{\numTrace{\varNum}{\varTrace}} \defeq \varNum$ and
$\traceOf{\numTrace{\varNum}{\varTrace}} \defeq \varTrace$.

\parahead{User Actions}

Consider a value $\mmVar{r}$ that represents a rectangle positioned at
$(\attrVal{\mmVar{r}}{\expStr{x}},\attrVal{\mmVar{r}}{\expStr{y}})=
 (\numTrace{\varNum_x}{\varTrace_x},\numTrace{\varNum_y}{\varTrace_y})$.
Suppose the user clicks the mouse button somewhere inside the borders of
$\mmVar{r}$ (rendered visually) and then drags the cursor $\mmVar{dx}$ pixels
in the $x$-direction and $\mmVar{dy}$ pixels in the $y$-direction.
As a result, the new desired position of $\mmVar{r}$ is
given by
$(\varNum_x', \varNum_y')=
 (\varNum_x+\mmVar{dx}, \varNum_y+\mmVar{dy})$.
Our goal is to reconcile this change to the position of $\mmVar{r}$
with the original program that generated it.
One option is to wait until the user finishes dragging the rectangle,
that is, when the user releases the mouse button.
At that point, we could invoke
$\helperop{Solve}(\set{\varNum_x'=\varTrace_x, \varNum_y'=\varTrace_y})$
to compute a set of substitutions.
Our goal with live synchronization, however, is to immediately a
apply program update during the user's actions.

\subsection{Mouse Triggers}

When the user clicks on a shape, we compute a \emph{mouse trigger}
$\varTrigger = \mouseTrigger{\varSubst}$, which is a function that,
based on the distance the mouse has moved,
returns a substitution to be immediately applied to the program.

For now, let us assume that all shapes are rectangles and that
user actions manipulate only their \expStr{x} and \expStr{y} attributes.
For every shape $\mmVar{r_i}$ in the canvas,
there are two steps to compute a mouse trigger.
First, for each attribute \expStr{x} and \expStr{y}, we
choose exactly one number (\ie{} location) in the program to
modify \emph{before} the user initiates any changes to
$(\attrVal{\mmVar{r_i}}{\expStr{x}},
  \attrVal{\mmVar{r_i}}{\expStr{y}})$.
The results of this step are two univariate equations to solve.
Second, we define a mouse trigger that invokes the solver
with each equation and then combines their resulting substitutions.
Once mouse triggers have been computed for all shapes, the editor is
prepared to respond to any user action with a local update to the
program.
We will now describe each step in detail.

\parahead{Shape Assignments}

Our task is to determine a \emph{shape assignment} $\varZoneSubst$
that maps each shape to an attribute assignment.
We define an \emph{attribute assignment} $\varAttrSubst$ to map
attribute names (\ie{} \little{} strings) to program locations.
We refer to the range of an attribute assignment as a \emph{location set}.

Let $\mmVar{box_i}$ refer to each rectangle from \sineWaveBoxes{}
in left-to-right order.
Using a procedure $\helperop{Locs}$
to collect all non-frozen locations that appear in a trace,
we see that the \expStr{x} and \expStr{y} attributes are each
computed using two locations:
$\locsOf{\traceOf{\attrVal{\mmVar{box_i}}{\expStr{x}}}} =
  \set{\locName{x_0},\locName{sep}}$ and
$\locsOf{\traceOf{\attrVal{\mmVar{box_i}}{\expStr{y}}}} =
  \set{\locName{y_0},\locName{amp}}$.
As a result, there are four possible attribute assignments for each shape:
\begin{align*}
\varAttrSubst_1 &\defeq
   \aSubstTwo{\expStr{x}}{\locName{x_0}}{\expStr{y}}{\locName{y_0}}
\\
\varAttrSubst_2 &\defeq
   \aSubstTwo{\expStr{x}}{\locName{x_0}}{\expStr{y}}{\locName{amp}}
\\
\varAttrSubst_3 &\defeq
   \aSubstTwo{\expStr{x}}{\locName{sep}}{\expStr{y}}{\locName{y_0}}
\\
\varAttrSubst_4 &\defeq
   \aSubstTwo{\expStr{x}}{\locName{sep}}{\expStr{y}}{\locName{amp}}
\end{align*}

\noindent
These assignments correspond to the four options
(top-left, top-right, bottom-left, and bottom-right, respectively)
depicted in \autoref{fig:little-sine-wave}D.

\parahead{``Fair'' and Other Heuristics}

As described in \autoref{sec:overview},
our default strategy is to choose an attribute assignment whose
range (\ie{} location set) has not yet been assigned to any other
shape in the output canvas.
When all possible assignments have been chosen an equal number of
times (\ie{} when they have been treated ``fairly''), then we
arbitrarily choose. As a result, we ``rotate'' through each of the
four attribute assignments, assigning
$\varZoneSubst(\mmVar{box_i})=\varAttrSubst_{j}$,
for all $i$ where $j= 1 + (i \bmod 4)$.

The fair heuristic will not always make choices that the user would prefer
best. However, we find that even simple heuristics such as this one
already enable a large degree of desirable interactivity. Therefore,
designing more sophisticated heuristics could be a fruitful avenue
for future work. In \theAppendix{}, we describe a second heuristic
that we have implemented, which ``biases'' towards program locations
that are used in relatively few run-time traces and, thus, have fewer
opportunities to be assigned to a zone.
We will not discuss this alternative in detail, because the vast majority
of the examples we have written to date, including the ones discussed in
this paper, work at least as well using the fair heuristic.

\parahead{Computing Triggers}

The next task is to prepare for when the user might click a
$\mmVar{box_i}$ in the output and drag it
$\mmVar{dx}$ pixels in the $x$-direction and
$\mmVar{dy}$ pixels in the $y$-direction.

Let $\varSubst_0$ be the mapping from locations to numbers in the
original \sineWaveBoxes{} program
and let $\varZoneSubst_0$ be the shape assignment
computed using the heuristics described above.
For each $\mmVar{box_i}$, we evaluate the helper procedure 
$
  \makeTriggerSimple
    {\varSubst_0}{\varZoneSubst_0}{\mmVar{box_i}}
$,
where $\helperop{SolveOne}$ is a solver
that is given exactly one univariate equation to solve:
\begin{align*}
& \makeTriggerSimple
    {\varSubst}{\varZoneSubst}{\varVal} \defeq
\\
& \hspace{0.20in}
  \lambda(\mmVar{dx},\mmVar{dy}).\ 
    \varSubst
    \substPlus
    \solveAndUpdate{\varSubst}{\varLoc_x}{\varNum_x'}{\varTrace_x}
\\
& \hspace{0.20in}
  \phantom{\lambda(\mmVar{dx},\mmVar{dy}).\ \varSubst}
    \substPlus
    \solveAndUpdate{\varSubst}{\varLoc_y}{\varNum_y'}{\varTrace_y}
\end{align*}
\vspace{-0.29in} 
\begin{align*}
\textrm{where }\ \
\numTrace{\varNum_x}{\varTrace_x} &= \attrVal{\varVal}{\expStr{x}}
&
\varNum_x' &\defeq \varNum_x + \mmVar{dx}
&
\varLoc_x &= \varZoneSubst(\varVal)(\expStr{x})
\\
\numTrace{\varNum_y}{\varTrace_y} &= \attrVal{\varVal}{\expStr{y}}
&
\varNum_y' &\defeq \varNum_y + \mmVar{dy}
&
\varLoc_y &= \varZoneSubst(\varVal)(\expStr{y})
\end{align*}
When the user drags some \mmVar{box_i},
its new attribute values $\varNum_x'$ and $\varNum_y'$
(directly manipulated by the user) are used
to solve the value-trace equations using the locations
assigned by $\varZoneSubst(\varVal)$.
This substitution is then applied to the original program,
the new program is run, and the new output is rendered
as the user moves the mouse.
When the user releases the mouse button,
we compute new shape assignments and mouse
triggers in anticipation of the next user action.

\parahead{Recap: Design Decisions}

There are two aspects of our approach that warrant emphasis.
The first that is we choose exactly one location to modify
per updated attribute, even though there may be additional solutions
(\ie{} local updates) that modify multiple locations.
For example, \autoref{eq:newthirdbox} can also be satisfied by
the substitution
$\varSubst_0 \substBracks{\aSubst{\locName{x_0}}{\num{55}}}
             \substBracks{\aSubst{\locName{sep}}{\num{20}}}$
(among many others).
By considering only ``small'' local updates, however, we reduce
the space of possible updates to synthesize and choose from.

The second is that our solutions are only plausible, not faithful,
because the same location may appear in multiple attributes being
directly manipulated (and, therefore, multiple equations).
For example, consider the box generated by the following expression:

\begin{Verbatim}[commandchars=\\\{\},codes={\catcode`\$=3\catcode`\^=7\catcode`\_=8},fontsize=\codeSize]
  (let xy 100 (rect 'red' xy xy ... ...) ...
\end{Verbatim}

\noindent
The attribute assignment
$\aSubstTwo{\expStr{x}}{\locName{xy}}{\expStr{y}}{\locName{xy}}$
is the only one to consider,
but the corresponding system of constraints on $\locName{xy}$
is overconstrained;
the new values computed by
$\solveOne{\varSubst}{\locName{xy}}{\mathtt{100}+\mmVar{dx}}{\locName{xy}}$
will differ from
$\solveOne{\varSubst}{\locName{xy}}{\mathtt{100}+\mmVar{dy}}{\locName{xy}}$
whenever $\mmVar{dx}\neq\mmVar{dy}$.
We could choose to apply an update only when the individual solutions agree,
or, more conservatively, disallow the shape from being manipulated at all.
Instead, we simply apply the individual substitutions in an arbitrary
(implementation-specific) order, which has the effect of satisfying at
least one of the constraints imposed by the user action.
This approach trades synthesizing only faithful updates in exchange
for additional opportunities to directly manipulate output values.

\subsection{Other Shapes and Zones}

For the purposes of presentation, so far we described a single type
of user action, namely, dragging the interior of a rectangle. In
practice, there are many other kinds of user actions.
For each kind of SVG shape, we define \emph{zones}
that identify and name particular visual areas of
a shape that can be directly manipulated by the user in order
to affect particular attributes.
The screenshot below depicts zones for several
kinds of shapes.

\begin{wrapfigure}{r}{0pt}
\includegraphics[scale=0.24]{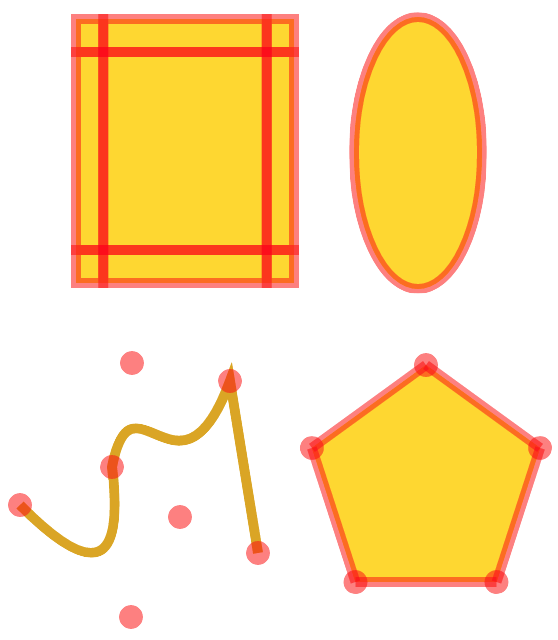}
\end{wrapfigure}
As we have described, dragging the \zone{Interior}
zone of a rectangle allows the user to manipulate its
\expStr{x} and \expStr{y} attributes.
Not all zones are tied to exactly two attributes, however.
For example, the rectangle \zone{RightEdge} zone is tied to one 
attribute (\expStr{width}) and the \zone{BotLeftCorner} zone is tied to
three (\expStr{x}, \expStr{width}, and \expStr{height}). 
Furthermore, not all attributes vary \emph{covariantly} with $\mmVar{dx}$
or $\mmVar{dy}$. For example, when the user manipulates the \zone{BotLeftCorner}
of a rectangle, the \expStr{width} attribute varies \emph{contravariantly} with
$\mmVar{dx}$ (and, at the same time, \expStr{x} varies covariantly).
Nevertheless, the approach we described for assigning triggers for
\zone{Interior} zones generalizes in 
a straightforward way to the remaining shapes and zones.
One slight change is that shape assignments are indexed by
shape and zone, for example,
$\varZoneSubst(\varVal)(\zone{Interior})(\expStr{x})$.
We provide more details in \theAppendix{}.

\section{Implementation}
\label{sec:implementation}

\newcommand{\tableTab}{\hspace{0.15in}}

We have implemented \sns{}
(available at {\blindUrl})
in approximately \snsLoc{} lines of
Elm~\cite{Elm} and JavaScript code.
When the user hovers over a zone, our implementation
displays a caption that indicates whether the zone is ``Inactive'' or
``Active'' and, for the latter, identifies the constants (\ie{} location set)
that will change if the user manipulates it. Furthermore, we highlight
these constants in yellow before the user begins manipulating the
zone; in green while they are being updated during manipulation; and
in red if the solver fails to compute a solution based on the user's
update. We use gray to highlight constants that contributed to an
attribute value but were not selected by the heuristics.

In the rest of this section, we describe the simple value-trace equation
solver that we currently use and we evaluate the overall
interactivity of our tool.
In \theAppendix{}, we describe additional features of
our implementation.

\subsection{Solving Value-Trace Equations}

The mouse triggers defined in the previous section
require a procedure
$\solveOne{\varSubst}{\varLoc}{\varNum'}{\varTrace}$
that, given the substitution $\varSubst$ from the previous
program and a location $\varLoc$, computes a new value for $\varLoc$
that satisfies the equation $\myEquation{\varNum'}{\varTrace}$.
Currently, we implement a simple solver that supports only
``single-occurrence'' equations, where the
location $\varLoc$ being solved for occurs exactly once.
Our top-down procedure
uses the inverses of primitive operations to recursively solve a
univariate equation in a syntax-directed manner
(see~\theAppendixCite{} for details).
Not all primitive operations have total inverses, so
$\helperop{SolveOne}$ sometimes fails to compute a solution.

As we will discuss below,
supporting this syntactic class of equations is already enough to
enable program synthesis for a variety of interesting examples.
Our solver is easy to implement and deploy in our Web-based setting
and fast enough to provide interactivity.
Future work, however, may incorporate more
powerful solvers (such as MATLAB or Z3~\cite{Z3})
while taking care to ensure that
synthesis is quick enough to incorporate into an interactive,
portable, direct manipulation editor.

\subsection{Interactivity}

The goal of \sns{} is to provide immediate, live synchronization
updates in response to direct manipulation changes. For a user
action to be ``successful'' requires
that the particular zone be Active,
that the solver computes an update in response to the mouse manipulation, and
that the resulting update is applied to the program and re-evaluated
within a short period of time.
We discuss each of these aspects in turn based on measurements
collected from \littleExamplesAll{} \little{} programs of
varying complexity, spanning more than \littleLoc{} lines of code in total.
Below, we discuss summary statistics across all examples;
for reference, detailed tables can be found in \theAppendixCite{}.

\subsubsection{Active Zones}

For any particular zone, our assignment algorithm may consider
zero, one, or more candidate location assignments based on the
traces of its attributes.
A zone is Inactive when there are zero candidates and is Active otherwise.
Across all of our examples, there were a total of \numShapes{} shapes
with \numZones{} zones,
of which \numZonesInactive{} (\pctZonesInactive{}) were Inactive and
\numZonesActive{} (\pctZonesActive{}) were Active.

\vspace{0.3em}
\begin{center}
\begin{tabular}{l r r}
\hline
\textbf{Zones} & \textbf{\numZones{}} \\\hline
Inactive & \numZonesInactive{} & \pctZonesInactive{} \\
Active & \numZonesActive{} & \\ 
\tableTab Unambiguous & \numZonesActiveOne{} & \pctZonesActiveOne{} \\
\tableTab Ambiguous & \numZonesActiveMultiple{} & \pctZonesActiveMultiple{}
\\\hline
\end{tabular}
\end{center}
\vspace{0.3em}

\parahead{Ambiguity}

Among Active zones,
\numZonesActiveOne{} (\pctZonesActiveOne{} of all zones)
had exactly one candidate location assignment and
\numZonesActiveMultiple{} (\pctZonesActiveMultiple{} of all zones)
had more than one
(\avgNumOfChoices{} candidates on average).
To provide responsive interaction, it is important to
deal with ambiguities because they are so frequent.
Our heuristics resolve ambiguities without user intervention.
It may be fruitful to explore other approaches, such as showing
multiple options for the user to choose from (particularly when there
are relatively few), or allowing the user to make multiple
user actions before attempting to infer an update.

\subsubsection{Solving Equations}

Next, we evaluate the solvability of equations that correspond
to Active zones.
Consider a program with initial location substitution $\varSubst$
and shape assignment $\varZoneSubst$,
and a shape $\varVal$ with an active zone $\varZone$.
For each attribute \expStr{k} that $\varZone{}$ controls,
$\varZoneSubst(\varVal)(\varZone)(\expStr{k})$
identifies a location $\varLoc$ to update in order to solve the equation
$\myEquation{\varNum+d}{\varTrace}$,
where $\numTrace{\varNum}{\varTrace}$ is the original value of
$\attrVal{\varVal}{\expStr{k}}$,
$\varLoc$ is one of the locations in $\varTrace$, and
$d$ is the change dictated by a user action.
Across all examples, there are \numEquations{} such
$(\varSubst,\varVal,\varZone,\varLoc,\varNum,\varTrace)$
tuples.
Because traces are often shared by multiple shapes and zones,
we filter out tuples that are identical modulo
$\varVal$ and $\varZone$, leaving \numUniqueEquations{}
unique $(\varSubst,\varLoc,\varNum,\varTrace)$ tuples.
In the following, we refer to each of these tuples as
a ``pre-equation.''

\vspace{0.3em}
\begin{center}
\begin{tabular}{l r r}
\hline
\textbf{Unique Pre-Equations} & \textbf{\numUniqueEquations{}} \\\hline
Outside Fragment & \numOutsideFragment{} & \pctOutsideFragment{} \\
Inside Fragment & \numInsideFragment{} & \\ 
\tableTab No Solution for $d=1$ & \numNoSolutionForOne{} & \pctNoSolutionForOne{} \\
\tableTab Solution for $d=1$ & \numSolutionForOne{} & \\ 
\tableTab \tableTab No Solution for $d=100$ & \numNoSolutionForOneHundred{} & \pctNoSolutionForOneHundred{} \\
\tableTab \tableTab Solution for $d=100$ & \numSolutionForOneHundred{} & \pctSolutionForOneHundred{}
\\\hline
\end{tabular}
\end{center}
\vspace{0.3em}

\parahead{Syntactic Fragment}

The majority of pre-equations
(\numInsideFragment{}, which constitutes \pctInsideFragment{})
fall into the syntactic fragment handled by our solver.
We paid little attention to the structure of traces when writing
examples, so we have been surprised that this number is so high.
We fully expected to incorporate a more
full-featured solver early in our work, but we have
been able to leave this to future work without severely hampering the
examples we have written so far.

The remaining \numOutsideFragment{} (\pctOutsideFragment{})
pre-equations fall outside the fragment and are guaranteed not to
be solvable.
Our current attribute assignment algorithm does not take this into
consideration and will sometimes assign such pre-equations to a
zone.
It would be worthwhile to avoid making such choices in the future.

\parahead{Solvability}
\label{sec:interactivity-solve}

For each pre-equation $(\varSubst,\varLoc,\varNum,\varTrace)$, we would
like to know whether the solver can compute an update if the user
manipulates the given attribute to be $\varNum+d$.
Rather than symbolically analyzing the space of possible user
changes, we tested
$\solveOne{\varSubst}{\varLoc}{\varNum+d}{\varTrace}$
with two concrete values, namely, $d=1$ and $d=100$.
Of the \numInsideFragment{} pre-equations in the fragment,
\numSolutionForOne{} were solvable for $d=1$
(\ie{} a green highlight)
and the remaining
\numNoSolutionForOne{} (\pctNoSolutionForOne{} of all unique pre-equations)
were not (\ie{} a red highlight).
Note that simply computing an update does not necessarily
mean that the change is acceptable to the user.

Of the \numSolutionForOne{} pre-equations solvable for $d=1$,
\numSolutionForOneHundred{} (\pctSolutionForOneHundred{} of all
unique pre-equations) were also solvable with $d=100$.
The remaining
\numNoSolutionForOneHundred{} (\pctNoSolutionForOneHundred{} of
all pre-equations) were not.
Upon inspection, several of these equations are of the form
$\myEquation{\varNum+d}{f(\cos{\varLoc})}$, where $f$ is some
function of $\cos{\varLoc}$. Because
the cosine function is bounded to the range $[-1,1]$, the equation
does not always have a solution.
Indeed, there is a mismatch between the interpretation of user updates
in the Cartesian plane and attributes like rotation that have more
natural representations in other coordinate systems.
In our experience, we have found that manipulating rotation angles
in \sns{} often works better with explicit sliders or using separate
built-in rotation zones in our implementation, which we have not
described in the paper.

\subsubsection{Performance}

In our experience, \sns{} is responsive for many, but not all,
of our examples.
We have not attempted to measure the observed frame rate of
\sns{}, which depends on several factors beyond our implementation.
We have, however, measured the performance of four critical
aspects of our implementation: parsing and evaluating a program,
preparing for a user action, and
solving a pre-equation.
We performed our experiments on an Intel Core i7 (four cores, 2.6-GHz)
running Mac OS X 10.9.5.
For ``Parse,'' ``Eval,'' and ``Prepare,''
we tested the operation five times on every example using Firefox 45 and
five times on every example using Chrome 49.
For ``Solve,'' we tested the operation on Chrome 49 twice per pre-equation
across all examples.
The ``Min'' and ``Max'' columns report the minimum and maximum
times across all runs;
``Med'' and ``Avg'' report the median and average across all runs.
Detailed statistics by example may be found in~\theAppendixCite{}.

\vspace{0.3em}
\begin{center}
\begin{tabular}{l r r r r}
\hline
\textbf{Operation} & \textbf{Min} & \textbf{Med} & \textbf{Avg} & \textbf{Max}
\\\hline
{Parse} & \minParseTime{} & \medParseTime{} & \avgParseTime{} & \maxParseTime{} \\
{Eval} & \minEvalTime{} & \medEvalTime{} & \avgEvalTime{} & \maxEvalTime{} \\
{Prepare} & \minPrepareTime{} & \medPrepareTime{} & \avgPrepareTime{} & \maxPrepareTime{} \\
{Solve} & \minSolveTime{} & \medSolveTime{} & \avgSolveTime{} & \maxSolveTime{}
\\\hline
\end{tabular}
\end{center}
\vspace{0.3em}

As the user drags the mouse during direct manipulation,
\sns{} repeatedly
solves the trace equations for the zone being manipulated and
re-evaluates the program to immediately display the interaction
results. The average time to ``Solve'' each trace equation is
negligible, \avgSolveTime{} on average,
because our solver uses a simple, syntax-directed procedure. Re-evaluation
takes longer, \avgEvalTime{} on
average. Our implementation re-runs the entire program even though much
of the output may not change. In the future, it would be
useful to optimize the implementation to recompute only the parts of
the program needed (\eg{}~\cite{incrementalcomputation}).

The slowest operations reported above, ``Parse'' and ``Prepare,'' are not run during
direct manipulation. ``Prepare'' encapsulates the
computation of both shape assignments and triggers for all zones. We only
perform this computation when the program is run initially and after the user
finishes dragging a zone. Some of
the data structures and
algorithms we use for computing candidate location
assignments and choosing from among them are rather naive
and can be optimized in the future.

\section{Examples}
\label{sec:examples}

We have used \sns{} to implement a variety of designs.
In this section, we will highlight observations that pertain
specifically to the combination of programmatically defined graphics
and direct manipulation. The implementations resemble typical programs
in other functional languages, but for the domain of SVG.

\begin{figure*}[t]
\centering
\includegraphics[scale=0.30]{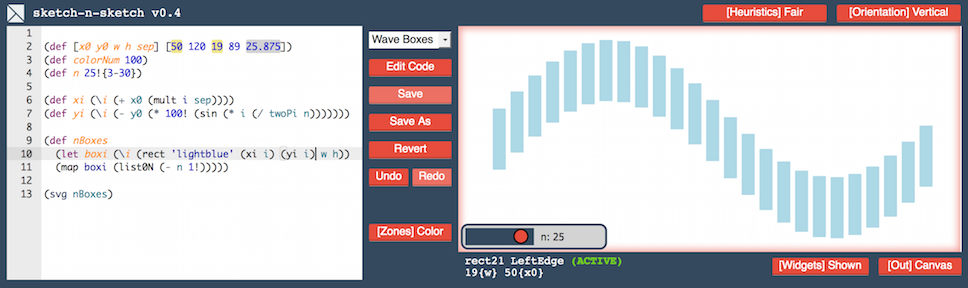}
\hspace{0.05in}
\includegraphics[scale=0.21]{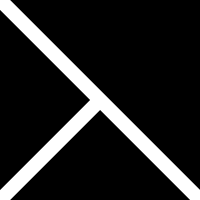}
\hspace{0.05in}
\includegraphics[scale=0.33]{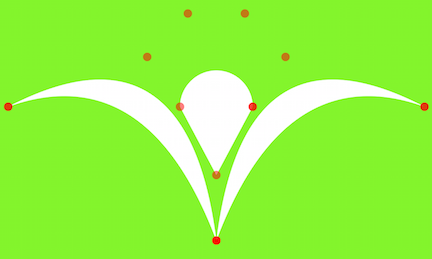}
\hspace{0.05in}
\includegraphics[scale=0.15]{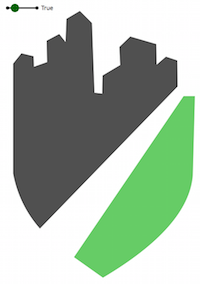}
\hspace{0.05in}
\includegraphics[scale=0.18]{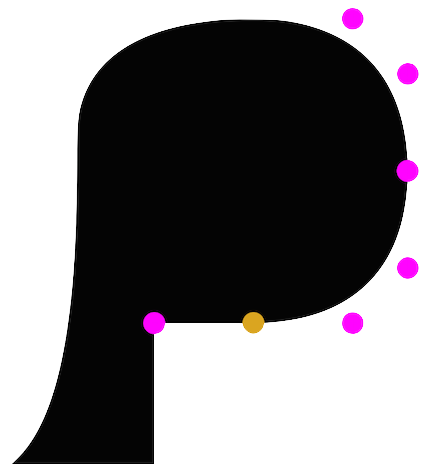}
\hspace{0.05in}
\includegraphics[scale=0.30]{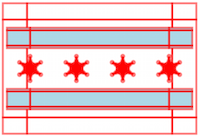}
\hspace{0.05in}
\includegraphics[scale=0.24]{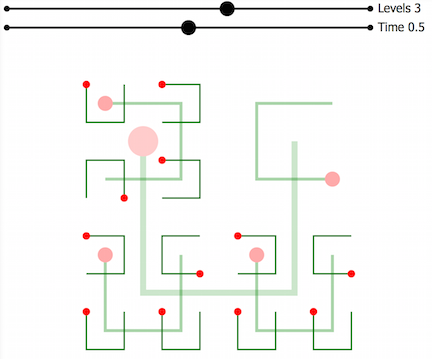}
\caption{Examples (left to right):
Sine Wave of Boxes,
\sns{} Logo,
Chicago Botanic Garden Logo,
Active Transportation Alliance Logo,
Icon from Lillicon~\cite{Lillicon},
City of Chicago Flag,
Hilbert Curve Animation
}
\label{fig:thumbnails}
\end{figure*}

\subsection{Programmatic Abstractions}

Our current implementation does not allow new shapes to be added
directly using the GUI. Nevertheless, we have used \sns{} to
effectively program and manipulate several designs that would be
difficult to edit or maintain using existing direct manipulation
tools such as \illustrator{} and \powerpoint{}.
\autoref{fig:thumbnails} provides thumbnails for some of the examples
we will discuss.

\parahead{Variables as Abstractions}

\sns{} does not attempt to infer any abstractions. It only propagates
abstractions that result from shared constants in the program.
Therefore, our \little{} programs are structured to use variables
(bound to constants) to encode explicit relationships between
attributes. Once these relationships have been defined,
the \sns{} editor preserves them during direct
manipulation. Many examples benefit from using variables as
abstractions, such as: our \sns{} logo, which comprises three black
polygons evenly spaced by white lines; the logo for the Chicago
Botanic Garden ({\url{www.chicagobotanic.org}}), which contains
several B\'{e}zier curves reflected across a vertical axis; the Active
Transportation Alliance logo ({\url{www.activetrans.org}}), which
uses several points along a path to depict a city skyline; and a logo
adapted from the Lillicon~\cite{Lillicon} project, where several
curves are used to define a semi-circle. For each example, a single
direct manipulation update changes all related attributes, without the
need for any secondary edits.

\parahead{Derived Shapes}

It is useful to define abstractions on top of the primitive
SVG shapes. We define an \verb+nStar+ function (and include it
in \texttt{Prelude}) that creates an \verb+n+-sided star
centered at \verb+(cx,cy)+ and rotated \verb+rot+ radians
in the clockwise direction, where the distance from the center to the outer
points is \verb+len1+ and the distance to the inner points is \verb+len2+.

\begin{Verbatim}[commandchars=\\\{\},codes={\catcode`\$=3\catcode`\^=7\catcode`\_=8},fontsize=\codeSize]
  (def nStar
    ($\lambda$(fill stroke w n len1 len2 rot cx cy) ...))
\end{Verbatim}

\noindent
We use \verb+nStar+ to implement the City of Chicago flag, which contains
four evenly-spaced six-sided stars. By directly manipulating the \zone{Point}
zones of a star in live mode, we can control the outer and
inner distances of all four stars. Modifying length parameters this way
can be surprising. For example, using negative lengths leads to
interesting patterns, even though one might not think to try them when
programming without immediate visual feedback.

\parahead{Group Box Pattern}

We occasionally find it useful to create a transparent rectangle
in the background with the width \verb+w+ and height \verb+h+ of an
entire design. Then, the \zone{BotRightCorner} zone of this box will,
predictably, be assigned the location set \set{\locName{w},
\locName{h}}. If we define all other shapes relative to
\verb+w+ and \verb+h+, we gain direct manipulation control over the
size of the entire design. In future work, it may be useful to
provide built-in support for grouping shapes.

\parahead{Dealing with Ambiguities}

We often start programming a design
with all constants unfrozen
except those that are not design parameters, such
as \verb+2+ in the expression \verb+(* 2! (pi))+.
Then, after seeing how
direct manipulation induces changes, we edit the
program to freeze some constants. Finally, to deal
with any remaining undesirable automatic choices from the heuristics,
we add range annotations to certain numbers so that we can
unambiguously and easily manipulate them with sliders instead.

We performed a preliminary study (described in \theAppendix{})
that demonstrates the existence of scenarios
(A) where using sliders is preferable to relying on heuristics for
disambiguation, and
(B) where relying on heuristics is preferable to using sliders.
A systematic user study would be a useful direction for future work.

\paraheadQuotes{Animations}

In several examples, like for the rendering of Hilbert curves, we use
sliders to control the SVG design as a function of a numeric
parameter. The effect is that we can ``animate'' the design as we
manipulate the slider. In the future, we plan to support dynamic,
time-varying animations as language and editor primitives.

\parahead{Procedural vs. Relational Constructions}

There is a tradeoff between procedural programming (in a functional
language like \little{}) and constraint-oriented programming (in a
system like SketchPad~\cite{SketchpadThesis}).
A detailed comparison of programming graphic
designs in these two styles may be an interesting avenue for future
work.

\subsection{Detailed Case Study: Ferris Wheel}

To provide a sense of when direct manipulation works smoothly,
and how to deal with situations when it does not,
we discuss one of our examples in more detail.
In this design,
we manipulate a ferris wheel comprising a number of equal-length
spokes emanating from a central hub, each of which has a passenger
car at its end.

\newcommand\verbBoxBlue[1]
  {\textcolor[rgb]{0,0,0.8}{\miniSepOne\fbox{\miniSepOne{#1}\miniSepOne}}\miniSepOne}
\setlength{\fboxsep}{1.5pt}
\setlength{\fboxrule}{0.5pt}

\newcommand{\inducesChange}[2]
  {\ensuremath{{#1}\rhd{#2}}}
\newcommand{\tiedTo}[2]{\ensuremath{\inducesChange{#2}{#1}}}

\newcommand\differentLocChoices[2]
  {\textcolor[rgb]{0,0,0}
    {\miniSepOne\fbox{\miniSepOne
      {\textcolor[rgb]{1,0,0}{\locName{#1}} \miniSepOne vs. \miniSepOne
      {\textcolor[rgb]{0,0,0.8}{\locName{#2}}}}}\miniSepOne}}

\newcommand{\myfrac}[2]
  {({#1}/{#2})}

\begin{figure*}
\begin{center}
\textbf{(A)}
  Initial \texttt{ferrisWheel}\texttt{.little} program in \texttt{black}
  and manual code edits in \verbBoxBlue{\texttt{boxed blue}}.\\

\begin{Verbatim}[commandchars=\\\{\},
                 codes={\catcode`\$=3\catcode`\^=7\catcode`\_=8} ,
                 fontsize=\codeSize,
                 numbers=left,numbersep=-7pt]
   (def [cx cy spokeLen rCenter wCar rCap] [220 300 80 20 30 7])
   (def [numSpokes rotAngle] [5\verbBoxBlue{!\{3-15\}} 0\verbBoxBlue{!\{-3.14-3.14\}}])
  
   (def ferrisWheel
     (let rim      [(ring 'darkgray' 6 cx cy spokeLen)]
     (let center   [(circle 'black' cx cy rCenter)]
     (let frame    [(nStar 'transparent' 'darkgray' 3 numSpokes spokeLen 0 rotAngle cx cy)]
     (let spokePts (nPointsOnCircle numSpokes rotAngle cx cy spokeLen)
     (let cars     (mapi ($\lambda$[i [x y]] (squareCenter \verbBoxBlue{(if (= 0 i) 'pink'} 'lightgray'\verbBoxBlue{)} x y wCar)) spokePts)
     (let hubcaps  (map ($\lambda$[x y] (circle 'black' x y rCap)) spokePts)
       (concat [rim cars center frame hubcaps]) )))))))
  
   (svg ferrisWheel)
\end{Verbatim}

\textbf{(B)} Traces for the \expStr{x} and \expStr{y} attributes of
the five \expStr{rect} \verb+cars+:

\footnotesize
\begin{align*}
\mmVar{CAR}_x(i) & \defeq
\mmVar{HUBCAP}_x(i) - \myfrac{\locName{wCar}}{2}
\hspace{0.20in}
\mmVar{HUBCAP}_x(i) \defeq
\locName{cx}
+ \locName{spokeLen}*\cos(\myfrac{\pi}{2}
                          - \locName{rotAngle}
                          + 2*{\pi}*\myfrac{i}{\locName{numSpokes}})
\\
\mmVar{CAR}_y(i) & \defeq
\mmVar{HUBCAP}_y(i) - \myfrac{\locName{wCar}}{2}
\hspace{0.20in}
\mmVar{HUBCAP}_y(i) \defeq
\locName{cy}
- \locName{spokeLen}*\sin(\myfrac{\pi}{2}
                          - \locName{rotAngle}
                          + 2*{\pi}*\myfrac{i}{\locName{numSpokes}})
\end{align*}

\end{center}
\caption{
  Ferris Wheel Example in \sns{}
}
\label{fig:little-ferris}
\end{figure*}

\parahead{Phase 1: Initial Development}

\begin{wrapfigure}{r}{0pt}
\includegraphics[scale=0.21]{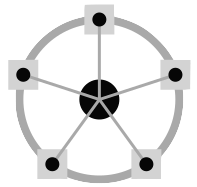}
\end{wrapfigure}
In \autoref{fig:little-ferris}A, we define a \little{} program that
embodies our design; for now, ignore the parts typeset in boxed blue.
We define several parameters on lines 1 and 2:
the center \verb+(cx,cy)+ of the wheel;
the number \verb+numSpokes+ and length \verb+spokeLen+ of the spokes;
the radius \verb+rCenter+ for the central disc;
the width \verb+wCar+ of each passenger car;
the radius \verb+rCap+ of each car's hubcap; and
the rotation \verb+rotAngle+ for the entire design.
We draw several components of the wheel on lines 4 through 11
using circles and rectangles, and we draw the spokes in terms
of the \verb+nStar+ function described earlier.
The \verb+cars+ are defined so that they remain vertical even
when the wheel is rotated, in order to accurately
portray the physical characteristics of a ferris wheel in motion.
The visual rendering of the output is shown above.

\parahead{Phase 2: Direct and Programmatic Edits}

\begin{wrapfigure}{r}{0pt}
\includegraphics[scale=0.18]{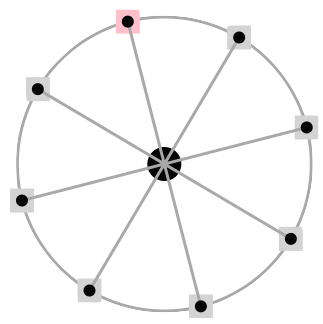}
\end{wrapfigure}
Suppose we wish to edit the program so that the output resembles
the picture on the right.
In particular, we will adjust
the size and location of the wheel,
the size of the passenger cars,
the number of spokes,
the rotation angle,
and the color of the first car.
These changes will require a combination of programmatic
and direct manipulation edits.

First, we want to change the size and location of the wheel.
When we hover over the \zone{Interior} of the \verb+rim+, \sns{}
shows a caption to indicate
that \locName{cx} and \locName{cy} will be updated.
When we hover over the \zone{Edge} of the \verb+rim+, we see
that \locName{spokeLen} will be updated.
In other words,
\sns{} has chosen the following assignments:
\begin{center}
\codeSize
\begin{tabular}{ccc}
  (\texttt{rim} , \zone{Interior}) &$\mapsto$&
  $\aSubstTwo{\expStr{cx}}{\locName{cx}}{\expStr{cy}}{\locName{cy}}$
\\
  (\texttt{rim} , \zone{Edge}) &$\mapsto$&
  $\aSubstOne{\expStr{r}}{\locName{spokeLen}}$
\end{tabular}
\end{center}

\noindent
These are, in fact, the only choices that could have been made,
because the traces for the relevant attributes were atomic locations.
Dragging these zones makes it easy to adjust the location and size
of the overall design.

Next, suppose we want to change the size of the passenger \verb+cars+.
The \expStr{width} of each rectangle is defined by a single location,
\locName{wCar}.
Therefore, the assignment
maps the \zone{RightEdge} of every car to \locName{wCar}:

\begin{center}
\codeSize
\begin{tabular}{ccc}
  (\texttt{cars}$_i$ , \zone{RightEdge}) &$\mapsto$&
  $\aSubstOne{\expStr{width}}{\locName{wCar}}$
\end{tabular}
\end{center}

\noindent
Dragging any of these \zone{RightEdge} zones allows us to
easily change the \expStr{width} of all \verb+cars+.

Now, suppose we want to change the number of spokes and the
rotation angle.
When hovering over the \zone{Interior} of several cars,
we see that, based on the heuristics,
\sns{} has chosen to vary $\locName{numSpokes}$
and $\locName{rotAngle}$ for several cars.

\begin{center}
\codeSize
\begin{tabular}{ccc}
  (\texttt{cars}$_0$ , \zone{Interior}) &$\mapsto$&
  $\aSubstOne{\expStr{x,y}}{\locName{wCar}}$
  \\
  (\texttt{cars}$_1$ , \zone{Interior}) &$\mapsto$&
  $\aSubstOne{\expStr{x,y}}{\locName{numSpokes}}$
  \\
  (\texttt{cars}$_2$ , \zone{Interior}) &$\mapsto$&
  $\aSubstOne{\expStr{x,y}}{\locName{rotAngle}}$
  \\
  (\texttt{cars}$_3$ , \zone{Interior}) &$\mapsto$&
  $\aSubstOne{\expStr{x,y}}{\locName{spokeLen}}$
  \\
  (\texttt{cars}$_4$ , \zone{Interior}) &$\mapsto$&
  $\aSubstOne{\expStr{x,y}}{\locName{numSpokes}}$
\end{tabular}
\end{center}

\noindent
Dragging some of the \verb+cars+ has strange effects.
To understand why, consider the traces of their \expStr{x} and
\expStr{y} attributes, shown in \autoref{fig:little-ferris}B;
we have simplified the traces slightly (using constant folding) and displayed
them using infix notation to improve readability.
The sines and cosines that appear in the traces come from the
\verb+nPointsOnCircle+ library function, which we use to position
the \verb+cars+ at the end of each spoke.
If we drag \verb+cars+$_1$ or \verb+cars+$_4$, the updated value
for \locName{numSpokes} is approximately \verb+0.3+,
which has the unintended effect of
changing the number of spokes. In fact, this is an example
where condition (c) of the definition of plausible updates is not
satisfied; the new program does not compute an output value that is
structurally equivalent to the original.
If we drag \verb+cars+$_2$, \locName{rotAngle} is updated but the
rotation is not smooth and intuitive
(this kind of equation was discussed in \autoref{sec:interactivity-solve}).
So, we use the editor's Undo feature to restore the original values of
\verb+numSpokes+ and \verb+rotAngle+.

Because we cannot easily manipulate the \verb+numSpokes+ and
\verb+rotAngle+ parameters, we annotate them with ranges on line 2;
these changes are depicted with blue boxes.
Furthermore, we annotate them as frozen so that no direct
manipulation zones (such as the \zone{Interior} ones for cars)
change these values.
Instead, we rely on the sliders to control them.

Finally, suppose we want to change the color of the first car, which
will make it easier to observe how the wheel is rotated.
Currently, \sns{} does not infer updates that introduce new
control-flow into the program, so we edit the expression on
line 9 to choose a different color for the car with index \verb+0+.
As a result of our programmatic edits, direct manipulation,
and indirect manipulation via sliders, the output of our final
program resembles the image at the beginning of this section.
Furthermore, having identified what changes are easy to make
with direct manipulation and what changes to make via sliders,
we can quickly make subsequent changes to the design parameters.

\subsection{Helper Value Design Pattern}
\label{sec:examples-helper}

The sliders provided by \sns{}, which
we refer to as \emph{user interface widgets}, are similar to the
notions of \emph{instruments}~\cite{InstrumentCHI2000} and
\emph{surrogate objects}~\cite{SurrogateCHI2011}, both of which aim to
provide GUI-based control over attributes that are not traditionally
easy to directly manipulate~\cite{Shneiderman1983}. Next, we show how
to derive \emph{custom} user interface widgets directly in \little{}.
Our key observation is to implement ``helper'' shapes whose attributes
affect other parameters of interest.

\parahead{User-Defined Widgets}

\setlength{\intextsep}{6pt}%
\setlength{\columnsep}{10pt}%
\begin{wrapfigure}{r}{0pt}
\includegraphics[scale=0.45]{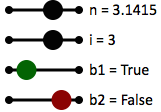}
\end{wrapfigure}

Suppose we are unhappy with the sliders built-in to \sns{}
(\autoref{sec:overview-sliders}).
We can design our own in \little{},
which are used by the program below
and depicted in the adjacent screenshot.
One slider controls a floating-point number \verb+n+,
one controls an integer \verb+i+, and
two control booleans \verb+b1+ and \verb+b2+.

\begin{Verbatim}[commandchars=\\\{\},codes={\catcode`\$=3\catcode`\^=7\catcode`\_=8},fontsize=\codeSize]
 (def [n  s1] (numSlider ... 0! 5! 'n = ' $\ttNumLoc{3.1415}{\varLoc_1}$))
 (def [i  s2] (intSlider ... 0! 5! 'i = ' $\ttNumLoc{3.1415}{\varLoc_2}$))
 (def [b1 s3] (boolSlider ... 'b1 = ' $\ttNumLoc{0.25}{\varLoc_3}$))
 (def [b2 s4] (boolSlider ... 'b2 = ' $\ttNumLoc{0.75}{\varLoc_3}$))
\end{Verbatim}

\noindent
\emph{Directly} manipulating the sliders \emph{indirectly}
manipulates the constants at locations
$\varLoc_1$, $\varLoc_2$, $\varLoc_3$, and $\varLoc_4$
(and, hence, the values bound to \verb+n+, \verb+i+, \verb+b1+, and \verb+b2+).

Both \verb+numSlider+ and \verb+intSlider+ are defined in terms
of a \verb+slider+ helper function:

\begin{Verbatim}[commandchars=\\\{\},codes={\catcode`\$=3\catcode`\^=7\catcode`\_=8},fontsize=\codeSize]
 (def slider ($\lambda$(round x0 x1 y min max s src) ...))
\end{Verbatim}

\noindent
The former returns \verb+src+ clamped to the range
$[$\verb+min+$,\ $\verb+max+$]$, if necessary;
the latter, furthermore, rounds \verb+src+ to the nearest integer.
We refer to the number supplied as the \verb+src+ parameter to
be the ``source'' number (or ``seed'') used to derive the ``target''
value, which is the first element of the pair returned by \verb+slider+.
The second element of the pair is the list of shapes
that comprise its visuals. The idea is to place a ``ball''
on the line between \verb+(x0,y)+ and \verb+(x1,y)+ at a distance
proportional to \verb+(src - min) / (max - min)+.
The editor provides a button for hiding shapes marked with a special
\expStr{HIDDEN} attribute, which we add to these helper shapes.
We employ the same approach to implement \verb+boolSlider+ for directly
manipulating booleans. In particular, a \verb+boolSlider+ is tied to a
source value between \verb+0.0+ and \verb+1.0+, where
values less than (resp. greater than) \verb+0.5+
represent \verb+true+ (resp. \verb+false+).

\parahead{Rounded Rectangles}

\begin{wrapfigure}{r}{0pt}
\includegraphics[scale=0.20]{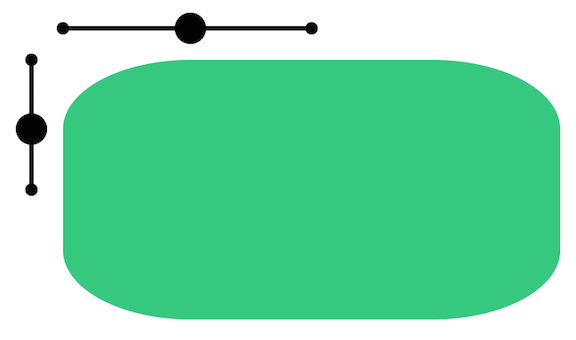}
\end{wrapfigure}

The zones supported by \sns{} control only the primary
attributes for each SVG shape kind (\eg{} \expStr{x}, \expStr{y},
\expStr{width}, and \expStr{height} for rectangles). By combining
user-defined sliders and the thin wrapper around
the full SVG specification language, it is easy to write a \little{}
function that abstracts over additional parameters, such as
\expStr{rx} and \expStr{ry} for specifying rounded corners, and
draws sliders (scaled based on the primary attributes)
next to the rectangle to control them.
\begin{Verbatim}[commandchars=\\\{\},codes={\catcode`\$=3\catcode`\^=7\catcode`\_=8},fontsize=\codeSize]
 (def roundedRect ($\lambda$(fill x y w h rx ry) ...))
\end{Verbatim}

\parahead{Tile Pattern}

Our last example demonstrates how custom UI widgets can control more
than just individual parameters.
In the screenshot below, the left (resp. right) half shows the
canvas with helper shapes displayed (resp. hidden).
We employ three new kinds of helper shapes in this design.
\begin{wrapfigure}{r}{0pt}
\includegraphics[scale=0.52]{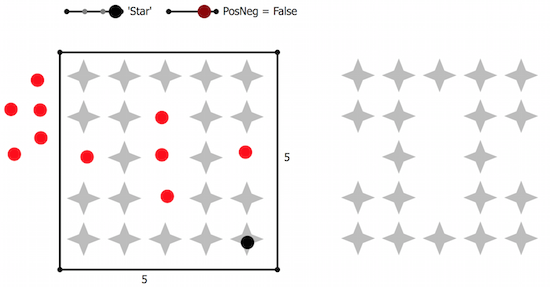}
\end{wrapfigure}
First, \verb+xySlider+ is a ``two-dimensional'' slider that allows
the control of two parameters simultaneously. In this example, we
draw the \verb+xySlider+ directly atop the grid. Dragging its
handle, the black circle in the lower-right corner, around
the grid provides an intuitive way to change the number of
rows and columns.
Next, we use \verb+enumSlider+ (drawn above the grid)
to select from a list of different shapes.
Then, we define red circles (to the left of the grid) to be
``tokens'' that denote ``selection'' when dragged
over particular tiles in the grid.
We define a helper function \verb+isCovered+
to check whether any token is currently placed over the tile centered
at \verb+(cx,cy)+.
Once we are done using these helper objects to manipulate the
grid, we use the built-in editor feature to toggle the
visibility of helper objects, leaving us with the final design
shown in the right half of the screenshot above.

\parahead{Recap: Customizing the UI}

\sns{} could provide built-in support for some of the
helper objects we described
(custom sliders and rounded rectangles).
However, no matter how many features are built-in, we believe there will
always be situations where a custom tool would be a better fit for the
task at hand.
With prodirect manipulation, the user can
push the frontier beyond what is provided.
Exploring this boundary between primitive and custom
UI widgets may be fruitful, both for designing useful
libraries as well as motivating new built-in features.

\section{Discussion}

In this paper, we presented an approach for live synchronization
of a program and changes to its output, by instrumenting program
evaluation to record run-time traces, phrasing user updates in terms
of a new framework called trace-based program synthesis, and
designing heuristics to automatically resolve ambiguities.
One may think of programs in our approach as \emph{sketches} (in the
program synthesis sense~\cite{SketchingThesis}) where the holes are
numeric constants, and the requirements for filling holes (\ie{}
changing numbers) come from the \emph{sketches} (in the drawing sense) in
the graphical user interface. Hence the name \sns{}.

\subsection{Related Work}

In a recent position paper~\cite{prodirectvision}, we provided a broad overview of relevant
\emph{program synthesis} (\eg{}~\cite{SrivastavaPOPL2010,KuncakPLDI2010,KuncakOOPSLA2013}),
\emph{programming by example} (\eg{}~\cite{GulwaniPOPL2011,GulwaniPLDI2015,Mayer2015}),
and \emph{bidirectional programming} techniques
(\eg{}~\cite{DagstuhlBX}).
Here, we focus our discussion on projects related to vector graphics.

Several projects use programming languages,
direct manipulation interfaces~\cite{Shneiderman1983}, or some combination
to provide expressive means for manipulating visual output.
We classify them using the following interaction modes
identified by Bret Victor in a talk on drawing tools~\cite{VictorDrawingShort}:
``Use'' for using built-in functionality through menus and buttons;
``Draw'' for directly manipulating domain objects; and
``Code'' for writing programs that manipulate domain objects.

\parahead{Dynamic Drawing (Use + Draw)}

Victor's prototype interactive drawing editor~\cite{VictorDrawingShort},
Apparatus~\cite{Apparatus},
and Programming by Manipulation~\cite{PBM} provide expressive direct
manipulation capabilities that serve as a way to build programs in
restricted, domain-specific languages.
By design, these tools tend to prohibit or discourage the user
from manipulating content via the ``indirect'' mechanism of code.

Although this choice may be desirable for many application domains and end
users, we believe there are limits to what can be accomplished
using features and transformations provided by any tool.
Therefore, our work targets users who wish to
work both via direct and programmatic manipulation (\ie{} Draw + Code).

\parahead{Programs that Generate Graphics (Code)}

Processing~\cite{Processing} is a language and environment
for generating visual output that has been popular both in classroom
and commercial settings.
Follow-on projects, such as Processing.js~\cite{ProcessingJs},
provide similar development environments for Web programming.
These systems provide immediate and interactive output, but they
do not provide ways to directly manipulate output in order to modify
the program that generated it.

\parahead{GUIs that Generate Programs (Draw + Code)}

Graphical user interfaces (GUIs) for creating visual output in
many domains often generate ``code behind'' what the user directly
manipulates.
Such tools include
PaintCode~\cite{PaintCode},
DrawScript~\cite{DrawScript},
SVG-edit~\cite{SvgEdit}, and
Adobe Fireworks
for graphic design.
Programs generated by these tools, however, are typically just as low-level as
the output itself, making them difficult to modify, maintain, and reuse.

\parahead{Constraint Programming (Draw + Code)}

Constraint-oriented programming systems, such as the classic
SketchPad~\cite{SketchpadThesis} and ThingLab~\cite{Borning1981}
systems as well as their more recent incarnations~\cite{Sketchpad14,Borning2015},
are characterized by 
(i) declarative programming models that allow programs to specify
constraints between program elements, and
(ii) constraint solvers that attempt to satisfy these constraints,
often using iterative and approximate numerical methods.
Together with full-featured GUIs, SketchPad and ThingLab provide
user experiences that tightly integrate programmatic and direct
manipulation.

Our goal is to support a similar workflow but for more
traditional, deterministic programming models, which are used
more regularly in a variety of domains.
That is, we wish to factor all constraint solving to a program synthesis
phase, rather than including it within the semantics of the
programming language itself.

\parahead{Synthesis for Vector Graphics}

The problem of \emph{beautifying} user drawings has been long-studied
in the graphics community and has recently been approached with
programming-by-example techniques~\cite{GulwaniGeo,QuickDraw2012,QuickDraw2014}.
These approaches synthesize artifacts in domain-specific representations
and languages.

In order to eliminate the need for secondary direct manipulation edits,
Lillicon~\cite{Lillicon} synthesizes different representations for
the same graphic design based on the intended edits.
In \sns{}, the user must pick a particular representation. But because
this representation is a general-purpose program, we can often build
abstractions that are preserved by prodirect manipulation, which
avoids the need for secondary edits.

\balance

\subsection{Future Work}

We mentioned several ways to build on our work throughout the paper,
including smarter heuristics and richer trace languages
that record control flow~(\eg{}~\cite{AngelicDebugging}).
We foresee several additional opportunities.

\parahead{Trace-Based Program Synthesis}

There are several ``knobs to turn'' within the
framework defined in \autoref{sec:sync}.
\vspace{0.02in} 
The current formulation synthesizes updates given a run-time trace
and a single updated value.
In other settings, it may be fruitful to consider
multiple traces, a history of user edits, and a history of previous
program updates.
Furthermore, it may be useful to rank candidate solutions according
to the (soft) constraints not changed by the user.

\parahead{Live Synchronization for Other Domains}

We plan to retarget our approach
(language instrumentation, synthesis algorithm, and
prodirect manipulation editor)
to meet the specific challenges of different domains,
such as layout in text documents,
formulas in spreadsheets,
dynamic animations in presentations and data visualizations, and
multiple rendering configurations for Web applications.

\parahead{Prodirect Manipulation}

The vision of prodirect manipulation, which we identified in
a position paper~\cite{prodirectvision}, comprises three goals:
(a) the ability to directly modify the output of a program and infer
updates in real-time to match the changes
({live synchronization});
(b) the ability to synthesize program expressions from output created
directly via the user interface; and
(c) the ability to temporarily break the relationship between program
and output so that ``larger'' changes can be made,
and then reconcile these changes with the original program
(called \emph{ad hoc synchronization}).

We addressed the first goal in this paper.
For the second, we plan to investigate ways to design direct
manipulation operations that generate programmatic relationships.
For the third goal, we plan to develop richer trace-based synthesis
algorithms to infer larger, ``structural'' program
updates, in contrast to the small, local updates we sought in this paper.
These directions of future work will help fully realize the
long-term vision of combining programming languages and direct
manipulation interfaces.

\acks
The authors would like to thank
Shan Lu,
Kathryn McKinley,
Loris D'Antoni, and
the anonymous reviewers
for valuable suggestions
that improved the final version of this paper.


\newpage
\appendix
\nobalance
\begin{center}
\fbox{\textbf{Supplementary Appendices for the PLDI 2016 Paper}}
\end{center}

\section{\littleItals{}}

The syntax of \little{} was defined in \autoref{fig:syntax}
of \autoref{sec:overview}.
Expressions $\varExp$ include booleans $\varBool$, strings $\varStr$,
floating-point numbers $\varNum$,
lists (encoded as cons cells, or pairs), functions, function application,
let-bindings, case expressions, and applications of primitive operations.
Patterns $\varPat$ are either variables or list patterns used for
deconstructing values.

\parahead{Syntactic Sugar}
The following are derived in terms of the core expression forms, as usual.
\small
\begin{eqnarray}
\expDef{\varPat}{\varExp_1}{\varExp_2}
  &\defeq&
  \expLet{\varPat}{\varExp_1}{\varExp_2}
\nonumber\\[-2pt]
\expDefRec{\varPat}{\varExp_1}{\varExp_2}
  &\defeq&
  \expLetRec{\varPat}{\varExp_1}{\varExp_2}
\nonumber\\[-2pt]
\expIte{\varExp_1}{\varExp_2}{\varExp_3}
  &\defeq&
  \expCaseTwo{\varExp_1}
             {\expBranch{\expTrue}{\varExp_2}}
             {\expBranch{\expFalse}{\varExp_3}}
\nonumber\\[-2pt]
\expFun{\parens{\threeThings{\varPat_1}{\cdots}{\varPat_m}}}{\varExp}
  &\defeq&
  \expFun{\varPat_1}{\twoThings{\cdots}{\expFun{\varPat_m}{\varExp}}}
\nonumber\\[-2pt]
\expAppMulti{\varExp_0}{\varExp_1}{\varExp_m}
  &\defeq&
  \expApp{\expApp{\expApp{\varExp_0}{\varExp_1}}{\cdots}}{\varExp_m}
\nonumber\\[-2pt]
\expList{\varExp_1}{\cdots}{\varExp_m}
  &\defeq&
  \expCons{\threeThings{\varExp_1}{\cdots}{\varExp_m}}{\expNil}
\nonumber\\[-2pt]
\expCons{\threeThings{\varExp_1}{\cdots}{\varExp_m}}{\varExp_0}
  &\defeq&
  \expCons{\varExp_1}{\expCons{\cdots}{\expCons{\varExp_m}{\varExp_0}}}
\nonumber\\[-2pt]
\expList{\varPat_1}{\cdots}{\varPat_m}
  &\defeq&
  \expCons{\threeThings{\varPat_1}{\cdots}{\varPat_m}}{\expNil}
\nonumber\\[-2pt]
\expCons{\threeThings{\varPat_1}{\cdots}{\varPat_m}}{\varPat_0}
  &\defeq&
  \expCons{\varPat_1}{\expCons{\cdots}{\expCons{\varPat_m}{\varPat_0}}}
\nonumber
\end{eqnarray}
\normalsize

\vspace{-0.15in} 

\parahead{SVG Attributes}

We represent SVG elements with three-element lists
$\expTriple{\varVal_1}{\varVal_2}{\varVal_3}$,
where the second element $\varVal_2$ is a list
$\expList{\expPair{\varValW_{k1}}{\varValW_{v1}}}{\cdots}
         {\expPair{\varValW_{km}}{\varValW_{vm}}}$
of $m$ key-value pairs.
Below,
we define $\expPair{\varValW_k}{\varValW_v} \toSvg  \mmVar{svgAttr}$
to be the translation from key-value pairs in \little{} to
attributes in the target, XML-based SVG format.

\vspace{-0.15in} 
{
\centering
\small
\begin{align*}
 \expPair{\varStr}{\varStr_1} &\ \toSvg\
   \svgRHS{\varStr=\svgStr{\varStr_1}}
\\
 \expPair{\varStr}{\varNum}   &\ \toSvg\
   \svgRHS{\varStr=\svgStr{\varNum}}
\\
 \expPair{\expStr{points}}{\mmVar{pts}} 
  &\ \toSvg\
  \svgRHS{\texttt{points}= \varStr}
  ,\hspace{0.10in}
  \mmVar{pts} \toSvgSub{\mmVar{pts}} \varStr
\\
\expPair{\expStr{fill}}{\mmVar{rgba}}
  &\ \toSvg\
  \svgRHS{\texttt{fill}=\varStr}
  ,\hspace{0.10in}
  \mmVar{rgba} \toSvgSub{\mmVar{rgba}} \varStr
\\
 \expPair{\varStr}{\varStr_1} &\ \toSvg \ \svgRHS{}
  ,\hspace{0.10in}
 \varStr \in \set{\expStr{ZONES},\expStr{HIDDEN}}
\\
\expPair{\expPair{\varNum_{x1}}{\varNum_{y1}}}{\cdots}
  &\ \toSvgSub{\mmVar{pts}}
  \svgRHS{\svgStr{\varNum_{x1},\varNum_{y1} \cdots}}
\\
\expListFour{\varNum_r}{\varNum_g}{\varNum_b}{\varNum_a}
  &\ \toSvgSub{\mmVar{rgba}}
  \svgStr{\texttt{rgba(}\varNum_r,\varNum_g,\varNum_b,\varNum_a\texttt{)}}
\end{align*}
\normalsize
}

%
We provide a thin wrapper over the target format by allowing
\little{} programs to specify arbitrary attributes using
strings (the first translation rule). In our current implementation,
we do not include units in the translation of numeric attributes
(the second rule), so numbers are interpreted as pixels.
We support specialized encodings for several SVG attributes in order
to provide direct manipulation access to their constituent numeric
values. For example, the points of a polygon or polyline may be
specified as a list (the third rule), and color attributes may be
specified as RGBA components (the fourth rule). We also encode
\texttt{path} commands (\ie{} the \expStr{d} attribute) so as to
access their data points and B\'{e}zier curve control points; we elide
the details of this encoding.
We use non-standard attributes \expStr{ZONES} and \expStr{HIDDEN} for
specific purposes in \sns{}.
So, we eliminate them when translating to SVG (the fifth rule).

\parahead{Semantics}

Compared to the constant literal range annotations
in \autoref{fig:syntax}, we can extend the syntax to allow
expressions in ranges, written $\numRange{\varExp_1}{\varExp_2}$.
The changes required to evaluate range expressions to numbers
and propagate them through the evaluation relation are straightforward,
so we elide them.
Also, compared to the version shown in \autoref{fig:syntax},
we define a more general version of the \ruleName{E-Op-Num} rule:
$$\inferrule*
  {\reducesToMulti{\varExp_0}{\varOp_m} \sepPremise
   \reducesToMulti{\varExp_1}{\labeledNum{\varNum_1}{\varTrace_1}} \sepPremise
   \cdots \sepPremise
   \reducesToMulti{\varExp_m}{\labeledNum{\varNum_m}{\varTrace_m}} \\
   \varNum = \interp{\expAppMulti{\varOp_m}{\varNum_1}{\varNum_m}}
   \sepPremise
   \varTrace = \expAppMulti{\varOp_m}{\varTrace_1}{\varTrace_m}
  }
  {\reducesToMulti
     {\expAppMulti{\varExp_0}{\varExp_1}{\varExp_m}}
     {\labeledNum{\varNum}{\varTrace}}}
$$

\section{Live Synchronization for SVG}

We provide additional discussion of some topics
from \autoref{sec:prodirect}.

\parahead{Zones and User Actions}

\autoref{fig:zones} shows the zones associated with each kind of SVG node,
as well as the attributes that are affected by directly manipulating that
kind of zone. Recall that in the definition of \helperop{MouseTrigger}
for \zone{Interior} zones of rectangles, the new value of the \expStr{x}
attribute is defined as $\varNum_x + \mmVar{dx}$. Therefore, we
say that \expStr{x} varies \emph{covariantly} with $\mmVar{dx}$
(denoted by $\dxPos$ in \autoref{fig:zones}).
Similarly, \expStr{y} varies covariantly with $\mmVar{dy}$
(written $\dyPos$).
As shown in the table, some zone attributes vary \emph{contravariantly}
(written $\dxNeg$ and $\dyNeg$) to match the expected physical behavior
in the user interface.
We refer to these changes in attribute values as \emph{offsets}.
Note that in \autoref{fig:zones} we write \expStr{w} and \expStr{h}
to abbreviate \expStr{width} and \expStr{height}.

\subsection{Mouse Triggers}

We describe how to generalize the approach we described for
computing triggers for \zone{Interior} zones.
Compared to the discussion in \autoref{sec:prodirect},
a {shape assignment} $\varZoneSubst$ maps a shape \emph{and} zone
to an attribute assignment.

\parahead{Computing Triggers}

Compared to the description in \autoref{sec:prodirect},
the $\helperop{ComputeTrigger}$ procedure is also indexed
by the shape kind (the first argument).

For each \expStr{rect} node in the output, we evaluate the helper procedure 
$
  \makeTrigger
    {\zone{Interior}}{\varSubst_0}{\varZoneSubst_0}{\mmVar{box_i}}
$,
where $\helperop{SolveOne}$ is a solver
that is given exactly one univariate equation to solve:
\begin{align*}
& \makeTrigger
    {\zone{Interior}}{\varSubst}{\varZoneSubst}{\varVal} \defeq
\\
& \hspace{0.20in}
  \lambda(\mmVar{dx},\mmVar{dy}).\ 
    \varSubst
    \substPlus
    \solveAndUpdate{\varSubst}{\varLoc_x}{\varNum_x'}{\varTrace_x}
\\
& \hspace{0.20in}
  \phantom{\lambda(\mmVar{dx},\mmVar{dy}).\ \varSubst}
    \substPlus
    \solveAndUpdate{\varSubst}{\varLoc_y}{\varNum_y'}{\varTrace_y}
\end{align*}
\vspace{-0.35in} 
\begin{align*}
\textrm{where }\ \
\numTrace{\varNum_x}{\varTrace_x} &= \attrVal{\varVal}{\expStr{x}}
&
\varLoc_x &= \varZoneSubst(\varVal)(\zone{Interior})(\expStr{x})
\\
\numTrace{\varNum_y}{\varTrace_y} &= \attrVal{\varVal}{\expStr{y}}
&
\varLoc_y &= \varZoneSubst(\varVal)(\zone{Interior})(\expStr{y})
\\
\varNum_x' &\defeq \varNum_x + \mmVar{dx}
&
\varNum_y' &\defeq \varNum_y + \mmVar{dy}
\end{align*}

\noindent
Notice how the shape assignment is indexed by value $\varVal$
\emph{and} zone \zone{Interior} in order to retrieve the
attribute assignment, which determines which locations to update
in response to the user's changes to the \expStr{x} and \expStr{y}
values.

\begin{figure}[t]
\centering

\vsepRule

\begin{tabular}{lcccc}
\hline
\texttt{rect} & \expStr{x} & \expStr{y} & \expStr{w} & \expStr{h} \\\hline
\zone{Interior} & \dxPos & \dyPos & & \\
\zone{RightEdge} & & & \dxPos & \\
\zone{BotRightCorner} & & & \dxPos & \dyPos \\
\zone{BotEdge} & & & & \dyPos \\
\zone{BotLeftCorner} & \dxPos & & \dxNeg & \dyNeg \\
\zone{LeftEdge} & \dxPos & & \dxNeg & \\
\zone{TopLeftCorner} & \dxPos & \dyPos & \dxNeg & \dyNeg \\
\zone{TopEdge} & & \dyPos & & \dyNeg \\
\zone{TopRightCorner} & & \dyPos & \dxPos & \dyNeg
\\[7pt]
\hline
\texttt{line} & \expStr{x1} & \expStr{y1} & \expStr{x2} & \expStr{y2} \\\hline
$\zone{Point}_1$ & \dxPos & \dyPos & & \\
$\zone{Point}_2$ & & & \dxPos & \dyPos \\
\zone{Edge} & \dxPos & \dyPos & \dxPos & \dyPos
\\[7pt]
\hline
\texttt{ellipse} & \expStr{cx} & \expStr{cy} & \expStr{rx} & \expStr{ry} \\\hline
\zone{Interior} & \dxPos & \dyPos & & \\
\zone{RightEdge} & & & \dxPos & \\
\zone{BotEdge} & & & & \dyPos
\\[7pt]
\hline
\texttt{circle} & \expStr{cx} & \expStr{cy} & \expStr{r} \\\hline
\zone{Interior} & \dxPos & \dyPos & \\
\zone{RightEdge} & & & \dxPos & \\
\zone{BotEdge} & & & \dyPos
\\[7pt]
\hline
\texttt{polygon}
  & \multicolumn{2}{c}{$(\expStr{x},\expStr{y})_i$}
  & \multicolumn{2}{c}{$(\expStr{x},\expStr{y})_{i+1}$} \\\hline
$\zone{Point}_i$
  & \multicolumn{2}{c}{$(\dxPos,\dyPos)$}
  & \\
$\zone{Edge}_i$
  & \multicolumn{2}{c}{$(\dxPos,\dyPos)$}
  & \multicolumn{2}{c}{$(\dxPos,\dyPos)$} \\
\zone{Interior} $\sep\sep\sep\cdots$
  & \multicolumn{2}{c}{$(\dxPos,\dyPos)$}
  & \multicolumn{2}{c}{$(\dxPos,\dyPos)$}
\\[7pt]
\hline
\texttt{polyline} & \multicolumn{4}{l}{(omitted)} \\\hline
\\
\hline
\texttt{path} & \multicolumn{4}{l}{(omitted)} \\\hline
\\
\end{tabular}

\caption{SVG Zones and Offsets}
\label{fig:zones}
\end{figure}

To generalize the definition of \helperop{ComputeTrigger},
we add versions for each different kind of shape
(the first argument). The table in \autoref{fig:zones} dictates
which attributes are updated by manipulation each zone
(\expStr{x} and \expStr{y} for \zone{Interior}) and
the {offset} to apply to their previous values
($\dxPos$ and $\dyPos$ above).
For example,
we compute triggers for the \zone{LeftEdge} zone of \expStr{rect}
as follows. Notice that \mmVar{dx} affects two attributes,
and that \mmVar{dy} does not affect any.
\begin{align*}
& \makeTrigger
    {\zone{LeftEdge}}{\varSubst}{\varZoneSubst}{\varVal} \defeq
\\
& \hspace{0.20in}
  \lambda(\mmVar{dx},\mmVar{dy}).\ 
    \varSubst
    \substPlus
    \solveAndUpdate{\varSubst}{\varLoc_x}{\varNum_x'}{\varTrace_x}
\\
& \hspace{0.20in}
  \phantom{\lambda(\mmVar{dx},\mmVar{dy}).\ \varSubst}
    \substPlus
    \solveAndUpdate{\varSubst}{\varLoc_w}{\varNum_w'}{\varTrace_w}
\end{align*}
\vspace{-0.35in} 
\begin{align*}
\textrm{where }\ \
\numTrace{\varNum_x}{\varTrace_x} &= \attrVal{\varVal}{\expStr{x}}
&
\varLoc_x &= \varZoneSubst(\varVal)(\zone{LeftEdge})(\expStr{x})
\\
\numTrace{\varNum_w}{\varTrace_w} &= \attrVal{\varVal}{\expStr{w}}
&
\varLoc_w &= \varZoneSubst(\varVal)(\zone{LeftEdge})(\expStr{w})
\\
\varNum_x' &\defeq \varNum_x + \mmVar{dx}
&
\varNum_w' &\defeq \varNum_w - \mmVar{dx}
\end{align*}
As another example,
we compute triggers for the \zone{BotLeftCorner} zone
(abbreviated to \zone{BotLeft}) as follows.
Notice that \mmVar{dx} affects the same two attributes
as \zone{LeftEdge}, and that \mmVar{dy} affects \expStr{height}.
\begin{align*}
& \makeTrigger
    {\zone{BotLeft}}{\varSubst}{\varZoneSubst}{\varVal} \defeq
\\
& \hspace{0.20in}
  \lambda(\mmVar{dx},\mmVar{dy}).\ 
    \varSubst
    \substPlus
    \solveAndUpdate{\varSubst}{\varLoc_x}{\varNum_x'}{\varTrace_x}
\\
& \hspace{0.20in}
  \phantom{\lambda(\mmVar{dx},\mmVar{dy}).\ \varSubst}
    \substPlus
    \solveAndUpdate{\varSubst}{\varLoc_w}{\varNum_w'}{\varTrace_w}
\\
& \hspace{0.20in}
  \phantom{\lambda(\mmVar{dx},\mmVar{dy}).\ \varSubst}
    \substPlus
    \solveAndUpdate{\varSubst}{\varLoc_h}{\varNum_h'}{\varTrace_h}
\end{align*}
\vspace{-0.35in} 
\begin{align*}
\textrm{where }\ \
\numTrace{\varNum_x}{\varTrace_x} &= \attrVal{\varVal}{\expStr{x}}
&
\varLoc_x &= \varZoneSubst(\varVal)(\zone{BotLeft})(\expStr{x})
\\
\numTrace{\varNum_w}{\varTrace_w} &= \attrVal{\varVal}{\expStr{w}}
&
\varLoc_w &= \varZoneSubst(\varVal)(\zone{BotLeft})(\expStr{w})
\\
\numTrace{\varNum_h}{\varTrace_h} &= \attrVal{\varVal}{\expStr{h}}
&
\varLoc_h &= \varZoneSubst(\varVal)(\zone{BotLeft})(\expStr{h})
\\
\varNum_x' &\defeq \varNum_x + \mmVar{dx}
&
\varNum_w' &\defeq \varNum_w - \mmVar{dx}
\\
&&
\varNum_h' &\defeq \varNum_h - \mmVar{dy}
\end{align*}
As a final example,
we compute triggers for the \zone{RightEdge} zone of \expStr{circle}
nodes as follows.
\begin{align*}
& \makeTrigger
    {\zone{RightEdge}}{\varSubst}{\varZoneSubst}{\varVal} \defeq
\\
& \hspace{0.20in}
  \lambda(\mmVar{dx},\mmVar{dy}).\ 
    \varSubst
    \substPlus
    \solveAndUpdate{\varSubst}{\varLoc_r}{\varNum_r'}{\varTrace_r}
\end{align*}
\vspace{-0.35in} 
\begin{align*}
\textrm{where }\ \
\numTrace{\varNum_r}{\varTrace_r} &= \attrVal{\varVal}{\expStr{r}}
\\
\varLoc_r &= \varZoneSubst(\varVal)(\zone{RightEdge})(\expStr{r})
\\
\varNum_r' &\defeq \varNum_r + \mmVar{dx}
\end{align*}
In our implementation, we generalize this approach to allow
the user to click and drag anywhere on the \zone{Edge} of a
\expStr{circle} or \expStr{ellipse} to manipulate the radius
attributes.

\parahead{``Biased'' Heuristic}

Program locations that are used in relatively few run-time traces of
the output have fewer opportunities to be assigned to a shape according
to the fair heuristic.
Therefore, we offer an alternative that
is ``biased'' towards program locations that occur in relatively
few run-time traces. In particular, we count the number of occurrences
that each location appears across \emph{all} traces of the output canvas,
which we use to compute a score for the location set as
$
  \helperop{Score}(\set{\varLoc_1, \cdots, \varLoc_n}) =
  \helperop{Count}(\varLoc_1)
  \times \cdots \times
  \helperop{Count}(\varLoc_n)
$.
Then, when choosing among multiple attribute assignments, we pick the one with
the lowest score, in order to favor locations that appear relatively
infrequently.
In our experience, these two simple heuristics work equally well
on many examples.
Next, we show an example where the biased heuristic is preferable.

Consider the variation below of the \sineWaveBoxes{} example
from \autoref{sec:overview}. Compared to the original version,
the value \verb+x0'+, computed from \verb+x0+ and two new
values \verb+a+ and \verb+b+, is used as the ``base'' x-position.

\begin{Verbatim}[commandchars=\\\{\},codes={\catcode`\$=3\catcode`\^=7\catcode`\_=8},fontsize=\codeSize]
 ...
 (def [a b] [0 0])
 (def [x0'] [(+ x0 (+ a (+ a (+ b b))))])
 ...
    (let xi (+ x0' (* i sep))
    ...
\end{Verbatim}

\noindent
According to the fair heuristic,
each of the four locations in
$\set{\locName{x0},\locName{a},\locName{b},\locName{sep}}$ get
chosen equally often (\ie{} for every fourth box in the pattern).
Because the first three of these locations all contribute to the
same initial base position, the effect is that dragging three out
of every four boxes manipulates the base position and one out of
every four manipulates the spacing between them.

The biased heuristic makes different choices for this example.
There are $n$ occurrences of both $\locName{x0}$ and $\locName{sep}$
in the traces of all attribute values of all shapes, and there
$2n$ occurrences of both $\locName{a}$ and $\locName{b}$. Therefore,
the former two locations are preferred by the biased heuristic over
the latter two. Assuming that this choice is combined with a
``rotating'' assignment, then alternating boxes are assigned
$\locName{x_0}$ and $\locName{sep}$, and no boxes are assigned
$\locName{a}$ or $\locName{b}$.

Although this example is contrived, we find biased heuristic to be useful
in some examples (such as the Floral Logo in our project repository),
where many control points are positioned and rotated relative to a common
initial position.

\subsection{Synthesis Algorithm}

In \autoref{sec:prodirect},
we described our approach for decomposing a system of constraints
(induced by manipulating SVG zones) into several univariate equations.
And in \autoref{sec:implementation},
we described the simple solvers that we currently use.
In order to provide a more detailed description of our overall
implementation, now we describe both of these pieces together
and present more details on the simple solvers.
Our algorithm rests on three design principles:
\setlength\leftmargini{24pt}
\begin{itemize}
\setlength\itemsep{-2pt}
\item[(I)] Solve only one equation at a time,
rather than\\ simultaneously solving a system of equations;
\item[(II)] Solve only univariate equations; and
\item[(III)] Solve equations only in simple, stylized forms.
\end{itemize}
\setlength{\leftmargini}{13pt} 

\noindent The first two design decisions follow the decomposition
approach described in \autoref{sec:prodirect}. The last follows the
observation that even a simple solver facilitates a variety of
interesting use cases (\autoref{sec:examples}), while being fast
enough to incorporate and easy to deploy in our Web-based setting.
As we will discuss, we infer plausible updates rather than
faithful ones.

\parahead{The Algorithm}

Suppose the user manipulates the $m$ values
$\numTrace{\varNum_1}{\varTrace_1}$ through
$\numTrace{\varNum_m}{\varTrace_m}$
to be $\varNum_1'$ through $\varNum_m'$, respectively,
which leads to the set of $m$ trace-value equations
${\set{\varNum_1'=\varTrace_1,\cdots,\varNum_m'=\varTrace_m}}$.
Furthermore, let $\varSubst_0$ be
the mapping of locations to numbers in the original program.
We compute a set of local updates using the algorithm below,
which we will walk through to explain how it encompasses
our three design decisions.
\begin{align*}
&\inferPlausible
   {\varSubst_0}
   {\set{\varNum_1'=\varTrace_1,\cdots,\varNum_m'=\varTrace_m}} = \\
&\hspace{0.20in}
 \setComp
  {\varSubst_0({\substPlus_i(\aSubst{\varLoc_i}{\varNumK_i})})}
  {(\varLoc_1,\cdots,\varLoc_m) \in \varLocs_1'\times\cdots\times\varLocs_m'} \\
&\hspace{0.60in}{\textrm{where }}
 \varNumK_i = \solve{\varSubst_0}{\varLoc_i}{\varNum_i'}{\varTrace_i}, \\
&\hspace{0.60in}\phantom{\textrm{where }}
 \varLocs_i = \locsOf{\varTrace_i}, \textrm{and }
 \varLocs_i' = \varLocs_i.
\end{align*}

\parahead{Solving Univariate Equations (I and II)}

First, we define $\varLocs_i=\locsOf{\varTrace_i}$ to be the set of
non-frozen locations (\ie{} constants) that occur in $\varTrace_i$, each
of which is a candidate ``variable'' to solve for.
Then, our equation solver, \helperop{Solve}, attempts to solve
each equation $\myEquation{\varNum_i'}{\varTrace_i}$ by treating the
location $\varLoc_i$ as the single variable, and using the substitution
$\varSubst_0$ to define the values of all other locations referred to
in $\varTrace_i$.
In general, a location may appear in multiple traces.
As a result, solutions may provide different new values for a
location, and any subsequent bindings will ``shadow'' or override
previous ones. As a result, the solutions are only plausible.

If we wanted to synthesize faithful updates, we could instead define
the sets $\varLocs_i'=\varLocs_i\setminus(\cup_{j \neq i}\
\varLocs_j)$ to be disjoint sets of locations. The resulting
equations would then be a set of independent equations.
We choose to forgo this approach, however, because it can be overly
restrictive. For example, consider a common
pattern found when programming with points along a circle of
length \verb+len+:

\begin{Verbatim}[commandchars=\\\{\},codes={\catcode`\$=3\catcode`\^=7\catcode`\_=8},fontsize=\codeSize]
   (let xi (+ cx (* len (cos (angle i))))
   (let yi (+ cy (* len (sin (angle i)))) ...))
\end{Verbatim}

\noindent
Taking the disjoint location approach, only \verb+cx+ can be
varied for the value of \verb+xi+, and only \verb+cy+ can be
varied for \verb+yi+. Manipulating these parameters via direct
manipulation is useful, but it would be nice to directly manipulate
the \verb+len+ parameter as well. As a result, in our current
implementation, we choose to forgo faithfulness in favor of admitting
more possible updates for direct manipulation.

\parahead{Solving Simple Equations (III)}

\begin{figure}
\codeSize
\textbf{(O)} Overall Solver:
\begin{align*}
&\solve{\varSubst}{\varLoc}{\varNum}{\varTrace} = \varNumK \\
&\hspace{0.20in}
 \textrm{ if } \solvea{\varSubst}{\varLoc}{\varNum}{\varTrace} = \varNumK
 \textrm{ or } \solveb{\varSubst}{\varLoc}{\varNum}{\varTrace} = \varNumK
\end{align*}
\textbf{(A)} ``Addition-Only'' Solver:
\begin{align*}
\solvea{\varSubst}{\varLoc}{\varNum}{\varTrace} &= (\varNum - s)/c,
\\[-2pt]
\textrm{where }
(c,s) &= \countPlus{\varSubst}{\varLoc}{\varTrace}
\\[4pt]
\countPlus{\varSubst}{\varLoc}{\varLoc}
  &= (1,0) \\
\countPlus{\varSubst}{\varLoc}{\varLoc'}
  &= (0,\varSubst\varLoc') \\
\countPlus{\varSubst}{\varLoc}{\expAppTwo{\op{+}}{\varTrace_1}{\varTrace_2}}
  &= (c_1+c_2,s_1+s_2) \\[-2pt]
\textrm{where }
(c_1,s_1) &= \countPlus{\varSubst}{\varLoc}{\varTrace_1} \\[-2pt]
(c_2,s_2) &= \countPlus{\varSubst}{\varLoc}{\varTrace_2}
\end{align*}
\textbf{(B)} ``Single-Occurrence'' Solver:
\begin{align}
\solveb{\varSubst}{\varLoc}{\varNum}{\varLoc} &= \varNum \nonumber\\
\solveb{\varSubst}{\varLoc}{\varNum}{\expApp{\varOp_1}{\varTrace_1}} &=
  \solveb{\varSubst}{\varLoc}{\inv{\varOp_1}{\varNum}}{\varTrace_1} \nonumber\\
\solveb{\varSubst}{\varLoc}{\varNum}{\expAppTwo{\varOp_2}{\varTrace_1}{\varTrace_2}} &=
\nonumber \\
\left \{
  \begin{tabular}{ll}
  $\solveb{\varSubst}{\varLoc}{\invL{\varOp_2}{\varNum_1}{\varNum}}{\varTrace_2}$ &
  \textrm{if } $\varSubst\varTrace_1=\varNum_1$ \\
  $\solveb{\varSubst}{\varLoc}{\invR{\varOp_2}{\varNum_2}{\varNum}}{\varTrace_1}$ &
  \textrm{if } $\varSubst\varTrace_2=\varNum_2$
  \end{tabular} \right. \span
\nonumber
\end{align}
\begin{align*}
\inv{\op{cos}}{\varNum} &= \interp{\expApp{\op{arccos}}{\varNum}} \\
\inv{\op{arccos}}{\varNum} &= \interp{\expApp{\op{cos}}{\varNum}} \\
\invL{\op{+}}{\varNum_1}{\varNum} &= \interp{\expAppTwo{\op{-}}{\varNum}{\varNum_1}} \\
\invR{\op{+}}{\varNum_2}{\varNum} &= \interp{\expAppTwo{\op{-}}{\varNum}{\varNum_2}} \\
\invL{\op{-}}{\varNum_1}{\varNum} &= \interp{\expAppTwo{\op{+}}{\varNum}{\varNum_1}} \\
\invR{\op{-}}{\varNum_2}{\varNum} &= \interp{\expAppTwo{\op{-}}{\varNum_2}{\varNum}} \\
\invL{\op{*}}{\varNum_1}{\varNum} &= \interp{\expAppTwo{\op{/}}{\varNum}{\varNum_1}} \\
\invR{\op{*}}{\varNum_2}{\varNum} &= \interp{\expAppTwo{\op{/}}{\varNum}{\varNum_2}} \\
\invL{\op{/}}{\varNum_1}{\varNum} &= \interp{\expAppTwo{\op{/}}{\varNum_1}{\varNum}} \\
\invR{\op{/}}{\varNum_2}{\varNum} &= \interp{\expAppTwo{\op{*}}{\varNum}{\varNum_2}} \\
\cdots &= \cdots
\end{align*}
\caption{Simple Value-Trace Equation Solver}
\label{fig:solve}
\end{figure}

\autoref{fig:solve} shows how
we combine two different solvers, one for ``addition-only'' equations
(where the only primitive operation is $\op{+}$) and one for
``single-occurrence'' equations (where the location $\varLoc$ being
solved for occurs exactly once). These two kinds of stylized equations
are already enough to enable program synthesis for a variety of
interesting examples (\autoref{sec:examples}). Future work, however,
may incorporate more powerful solvers while taking care
to ensure that synthesis is quick enough to incorporate into an
interactive, portable, direct manipulation editor.

Our first solver, \helperop{Solve_A}, handles traces where the only
operator used is \op{+} (\cf{} the discussion of programming with
integers.
Such equations are easy to solve: count the
number of occurrences $c$ of the unknown variable, and divide the
partial sum $s$ by $c$.
Note that this yields a number, not necessarily an integer, as output.

Our second solver, \helperop{Solve_B},
handles equations where there is \emph{exactly one}
occurrence of the unknown location variable being solved for.
For equations like these, we can define a top-down, syntax-directed
procedure that uses inverses of primitive operations.

In practice, \helperop{Solve_B} subsumes \helperop{Solve_A} on
virtually all equations encountered in our examples.

\parahead{Inverse Operations}

Our ``single-occurrence'' equation solver, \helperop{Solve_B},
recursively reduces the goal equation at each step by using inverses
of primitive operators.
We define $\helperop{Inv}(\varOp)$ to return the inverse operator
of unary operator $\varOp$.
We define $\helperop{Inv_L}(\varOp,\varNum_1)$
(resp. $\helperop{Inv_R}(\varOp,\varNum_2)$) to be
a function that denotes the inverse of binary operator $\varOp$
partially applied to the first argument $\varNum_1$
(resp. second argument $\varNum_2$).

Not all primitive operations have total inverses, so our
solver sometimes fails to compute a solution.
For example, recall \autoref{eq:box1} from \autoref{sec:overview}
and consider a situation where $\locName{sep}$ is assigned to
change (instead of $\locName{x_0}$) when the user manipulates the
\expStr{x}-position to be \verb+115+;
there is no solution for
$\solveOne{\varSubst_0}{\locName{sep}}{\mathtt{155}}
  {\expAppTwo{\op{+}}
    {\locName{x_0}}
    {\expAppTwo{\op{*}}{\varLoc_0}{\locName{sep}}}}$ because
the expression
\texttt{(+ 50 (* 0 sep))} evaluates to \verb+0+ for all choices
of \verb+sep+.

\section{Implementation}

Our implementation provides several \little{} and SVG features
features beyond those described in \autoref{sec:implementation}.

\parahead{Prelude}

We have implemented a small \texttt{Prelude} library of \little{}
functions, in approximately \preludeLoc{} lines of code, that is included
in every program. Some \texttt{Prelude} functions support general
functional programming, and some support SVG programming, in
particular.

\parahead{Thawing and Freezing Constants}

By default, all constants that are not frozen may be changed by the
synthesis algorithm. We allow the user to choose the alternative,
treating all constants as frozen except those that are explicitly
\emph{thawed} (written $\expThaw{\varNum}$) in the program.

\parahead{Programming with Integers}

The \little{} language provides a single, floating-point number type.
When programming with (non-negative) integers, however, the user may
choose to use the library functions \verb+mult+, \verb+minus+, and
\verb+div+ instead of the primitive multiplication, subtraction, and
division operators, respectively. These library functions result in
values with addition-only traces.
For example, compared to \autoref{eq:box3}, the expression
\texttt{(+ x0 (mult i sep)} leads to the addition-only trace
$
  {\expAppTwo{\op{+}}
    {\locName{x_0}}
    {\expAppTwo{\op{+}}{\locName{sep}}{\locName{sep}}}}
$ when \verb+i+ is \verb+2+.

\parahead{Color Numbers}

In addition to specifying colors as strings or RGBA numbers, we allow
colors to be specified using a non-standard \emph{color number}, where
the range 0 to 500 is interpreted as a spectrum of colors, including
grayscale. For an SVG shape with a fill color number (\ie{}
\expPair{\expStr{fill}}{\varNum}), our editor displays a slider right
next to the object that allows direct manipulation control over the
\expStr{fill} attribute.

\parahead{Layers}

Our implementation currently provides two kinds of layers. We
allow the user to toggle between displaying and hiding the zones
associated with each shape. We also allow the user to toggle between
displaying and hiding shapes that contain the
non-standard \expStr{HIDDEN} attribute. We have found hidden shapes to
be useful in several of our examples (\autoref{sec:examples-helper}).

\parahead{Exporting to SVG}

In order to facilitate interoperation with other SVG editing tools, we
provide an option to print the resulting canvas in SVG format (rather than
rendering it), which can then be copied and pasted into other tools.

\section{Examples}

Compared to our discussion in \autoref{sec:examples},
here we provide more low-level details of some of our \little{}
examples.
All of our examples are available on the
Web.\footnote{\blindUrl}

\parahead{\example{Boxes}}

\begin{wrapfigure}{r}{0pt}
\includegraphics[scale=0.30]{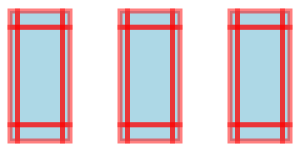}
\end{wrapfigure}

The \verb+threeBoxesInt+ program is our ``hello world'' example for
prodirect manipulation. The number of boxes and their location, width,
and height are simple parameters to change in the program. In addition,
the location, width, and height
can easily be changed in the direct manipulation editor.
When manipulated in live mode, all of the boxes are updated
together in real-time. The screenshot on the right shows the zones displayed
to the user.

\parahead{\example{Elm Logo}\protect\footnote{\url{github.com/evancz/elm-svg/blob/1.0.2/examples/Logo.elm}}}

The Elm logo is a tangram consisting of seven polygons.
We implemented this logo by massaging the definition from the SVG format
to the representation in \little{}.
This process will be automatic once we add support for importing
SVG images directly.

\begin{wrapfigure}{r}{0pt}
\includegraphics[scale=0.25]{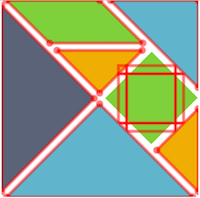}
\end{wrapfigure}

There are two noteworthy aspects of this example in \sns{}.
The first is that our definition uses the \expStr{viewBox}
attribute to define a local coordinate system for shapes within
the canvas. Because of the thin wrapper around SVG, the result in
\sns{} is that the output canvas is scaled to fit the size of the
canvas, no matter how large or small the application window it is.
As a result, even though our current implementation does not provide a way
to pan within or scale a canvas, one can use \expStr{viewBox} in order
to render the output in ``full screen'' mode.

The second interesting aspect is the square, which is rotated using
the SVG \expStr{matrix} command within the \expStr{transform} attribute.
Again, even though we currently provide no special support for these
features, the zone we display for the square (though not rotated to
match) can still be used to directly manipulate it.

Unlike the \sns{} logo,
the high-level relationships between the shapes in the
Elm logo are not captured by the definition in SVG format,
nor in its direct translation to \little{}.
As a result, directly manipulating any one of the pieces does not
affect the others, therefore breaking the intended abstraction
of the logo.

\parahead{\example{\snsItals{} Logo}}

\setlength{\intextsep}{6pt}%
\setlength{\columnsep}{10pt}%
\begin{wrapfigure}{r}{0pt}
\includegraphics[scale=0.25]{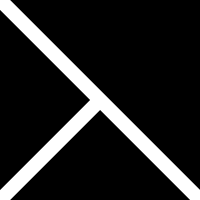}
\end{wrapfigure}

To implement our logo, which
pays homage to the Elm logo,
we use the abstraction facilities of a programming language to
declare relationships between the shapes.
The definition is
parameterized by a position \verb+(x0,y0)+ for the top-left
corner, a width \verb+w+ and height \verb+h+, and a \verb+delta+
parameter that determines the size of the gap between the three
shapes:
\begin{Verbatim}[commandchars=\\\{\},codes={\catcode`\$=3\catcode`\^=7\catcode`\_=8},fontsize=\codeSize]
  (let [x0 y0 w h delta] [50 50 200 200 10] ...)
\end{Verbatim}
\noindent
The rest of the definition (not shown) computes the three polygons
in terms of these parameters. It is, thus, simple to change any of
these values and re-run the program to generate an updated logo.

Better yet is the ability to manipulate the parameters \emph{directly}
through the canvas in live mode. For example, say that we want to stretch the
logo, that is, by changing the \verb+w+ and \verb+h+ parameters. If we
click and drag bottom-right corner (\ie{} a \zone{Point} zone) of the
bottom triangle in live mode,
the height of the logo is adjusted but not the width;
instead, the x-position of the logo is. In other words, the location
set assigned to this particular zone happens to be
\set{\locName{x_0}, \locName{h}} instead of
\set{\locName{w}, \locName{h}} as we might have liked.

We can proceed in a couple of ways. One option is to edit the code
to freeze the \verb+x0+, \verb+y0+, and \verb+delta+ values,
thereby directing \sns{} towards assigning the desired location set
and trigger to this \zone{Point} zone. With this change, directly
manipulating this corner point allows us to stretch the
logo in either direction.

Another option is to create a
transparent rectangle in the background with dimensions
\verb+w+ and \verb+h+.
The \zone{BotRightCorner} zone of this box will, predictably,
be assigned the location set \set{\locName{w}, \locName{h}}, thus
providing direct manipulation control over the desired attributes of
the logo. This second option, creating an explicit ``group box,''
is a design pattern that is often useful for mixing programmatic
and direct manipulation (\ie{} prodirect manipulation) in the
current version of \sns{}. In future work, it may be useful to provide
some built-in support for grouping shapes.

\parahead{\example{Chicago Flag}}

Like with the \sns{} logo, we define a transparent group box
(visible when displaying zones, as in the screenshot) to give direct
manipulation control over the width and height of the flag.
Unlike that
example, however, there is no way to produce the same exact result by manipulating
only one of the polygons.
\begin{wrapfigure}{r}{0pt}
\includegraphics[scale=0.40]{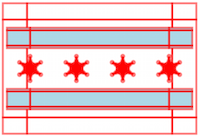}
\end{wrapfigure}
If the user changes, say, the bottom stripe by
moving the mouse cursor a given distance, the overall dimensions of the flag
will change, but \emph{not} by the amount the cursor has moved. As a result,
the relationship between stretching one of the stripes and the overall
flag is not a smooth, intuitive one. Using a group box, however, provides
the simple and expected behavior.

\parahead{\example{Chicago Botanic Garden Logo}\protect\footnote{\url{www.chicagobotanic.org}}}

\begin{wrapfigure}{r}{0pt}
\includegraphics[scale=0.35]{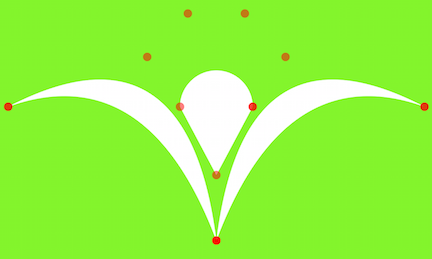}
\end{wrapfigure}

The symmetric design of this logo
uses curves, defined with B\'{e}zier \verb+path+ commands.
By programming in \little{}, we can define the coordinates
and control points such that they are reflected across
a vertical axis running down the middle of the logo.
Then, in live mode, direct manipulation of any position or control point
(the ``floating'' \zone{Point} zones in the screenshot)
in either half is immediately reflected in the other half.

\parahead{\example{Active Trans Logo}\protect\footnote{\url{www.activetrans.org}}}

The logo of the Active Transportation Alliance contains two paths,
each of which has a single curved edge and some number of straight edges.
In our current implementation, \sns{} does not provide a GUI-based way
to create shapes or add extra points to existing shapes. Therefore,
these two paths must be generated using \little{} code, at least initially.

\begin{wrapfigure}{r}{0pt}
\includegraphics[scale=0.25]{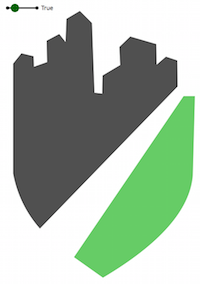}
\end{wrapfigure}

Nevertheless, we found that we can quickly and easily begin implementing
this logo as follows. First, we implement a \verb+makePath+ function
that stitches together a path based on a list of points and a single
B\'{e}zier control point. Next, we define two intially-empty lists,
\verb+grayPts+ and \verb+greenPts+, that will store the points of each
path. Then, we use \verb+makePath+ to construct two paths out of these
lists.

\begin{Verbatim}[commandchars=\\\{\},codes={\catcode`\$=3\catcode`\^=7\catcode`\_=8},fontsize=\codeSize]
  (let makePath ($\lambda$(color pts [xCtrl yCtrl]) ...)
  (let [grayPts greenPts] [ [] [] ]
  (let [p1 p2] [(makePath ...) (makePath ...)]
    ...)))
\end{Verbatim}

\noindent
Now the task is to define the list of points for each path.
We would like to do this visually by directly manipulating points
into the desired positions, but we need some points to begin with.
As is, \verb+grayPts+ and \verb+greenPts+ are empty, so there are
no shapes to render.

One option is to use a text editor to populate the list with dummy
points, but this could be tedious for a large number of points,
especially because they should be reasonably spaced out so that
they can be manipulated in the visual editor. Instead, we wrote
a \little{} function to generate such a list of points and
evaluated it using the Elm REPL (read-eval-print loop). We
then copied this list into our program, rendered it, and proceeded
to directly manipulate the points. Our helper function essentially
created a ``ball of clay'' that we massaged into the desired shapes.
In future work, the visual editor might provide support for generating
complex shapes using templates such as this one.

Once we settled on the desired shapes of our paths, we returned to
the program to introduce structure that relates the topmost points of
the top shape (corresponding to the city skyline). As a result,
dragging any one of these points up or down in live mode affects
all of the others. So, if the skyline grows taller (which has been
known to happen in Chicago), we can easily adapt the logo to match.

Lastly, we include a button in our development
and use it to toggle between a ``positive'' version, where the
shapes are colored and the background is white, and a ``negative''
version, where the shapes are white and the background is colored.
These two versions of the logo are easy to develop
in tandem using \sns{}.

\parahead{\example{User-Defined Widgets}}

\begin{figure*}[t]
\centering

\begin{minipage}{0.64\textwidth}
\centering
\begin{Verbatim}[commandchars=\\\{\},codes={\catcode`\$=3\catcode`\^=7\catcode`\_=8},fontsize=\codeSize]
; slider : Bool -> Int -> Int -> Int -> Num -> Num -> Str -> Num
;       -> [Num (List Svg)]
(def slider ($\lambda$(roundInt x0 x1 y minVal maxVal caption srcVal)
  (let preVal (clamp minVal maxVal srcVal)
  (let targetVal (if roundInt (round preVal) preVal)
  (let shapes
    (let ball
      (let [xDiff valDiff] [(- x1 x0) (- maxVal minVal)]
      (let xBall (+ x0 (* xDiff (/ (- srcVal minVal) valDiff)))
      (if (= preVal srcVal) (circle 'black' xBall y 10!)
                            (circle 'red' xBall y 10!))))
    [ (line 'black' 3! x0 y x1 y)
      (text (+ x1 10) (+ y 5) (+ caption (toString targetVal)))
      (circle 'black' x0 y 4!)
      (circle 'black' x1 y 4!)
      ball ])
  [targetVal (ghosts shapes)])))))

(def [numSlider intSlider] [(slider false) (slider true)])
\end{Verbatim}
\caption{Horizontal Slider in \little{}.}
\label{fig:slider}
\end{minipage}
\hspace{0.30in}
\begin{minipage}{0.27\textwidth}
\centering
\includegraphics[scale=0.30]{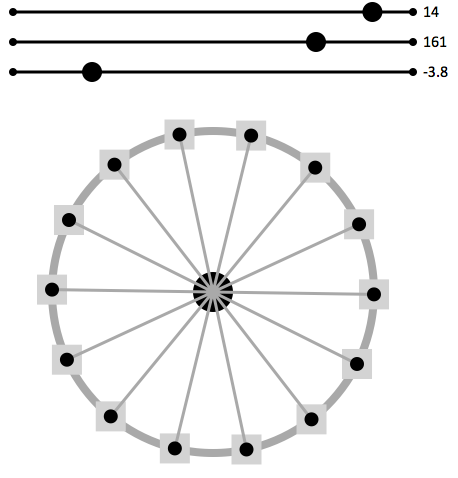}
\caption{Ferris Wheel Example. Generated using a combination of
programmatic, direct, and user-defined indirect manipulation.}
\label{fig:ferris}
\end{minipage}

\end{figure*}

\autoref{fig:slider} shows the definition of \verb+slider+
described in \autoref{sec:examples}. The \verb+ghosts+ function
adds the \expStr{HIDDEN} attribute to a list of shapes so that
they can hidden by toggling a button.

\begin{Verbatim}[commandchars=\\\{\},codes={\catcode`\$=3\catcode`\^=7\catcode`\_=8},fontsize=\codeSize]
  (def ghosts
    (map ($\lambda$shape (consAttr shape ['HIDDEN' '']))))
\end{Verbatim}

\parahead{\example{Ferris Wheel}}

We designed a ferris wheel that consists
of some number of equal-length spokes emanating from a central
hub, each of which has a passenger car at its end. Furthermore,
we wanted the ability to rotate the wheel while
keeping the passenger cars vertical, in order to accurately
portray the physical characteristics of a ferris wheel in motion.
It is hard to imagine how one could develop these relationships
in a modular way using tools like \illustrator{} or \powerpoint{}.

In \sns{}, we combine programmatic, direct manipulation,
and indirect manipulation (via user-defined sliders) to develop
our design in a way that is highly-reusable and easy to edit.
First, we write a function

\begin{Verbatim}[commandchars=\\\{\},codes={\catcode`\$=3\catcode`\^=7\catcode`\_=8},fontsize=\codeSize]
  (def ferrisWheel
    ($\lambda$(numSpokes spokeLen rotAngle
      $\phantom{\lambda}$sizeCar radiusCenter cx cy) ...))
\end{Verbatim}

\noindent
that, given several parameters, draws the desired circles, lines,
and rectangles. The function is straightforward to write, making use
of a \texttt{Prelude} function

\begin{Verbatim}[commandchars=\\\{\},codes={\catcode`\$=3\catcode`\^=7\catcode`\_=8},fontsize=\codeSize]
  (def nPointsOnCircle ($\lambda$(n rot cx cy r) ...))
\end{Verbatim}

\noindent
that generates a list of \verb+n+ points evenly spaced around a circle
of \verb+r+ radius centered at \verb+(cx,cy)+.
A drawing that results from \verb+ferrisWheel+ is shown in the bottom
of \autoref{fig:ferris}.

We can directly manipulate several parameters of the ferris wheel:
we can adjust the location \verb+(cx,cy)+ of the wheel by dragging
the \zone{Interior} zone of the central hub;
we can adjust \verb+radiusCenter+ to change the size of the central hub by
manipulating its \zone{Edge} zone; and
we can adjust the width \verb+sizeCar+ of all passenger cars by
manipulating any one of their \zone{Edge} zones.
While this workflow in \sns{} is already unique and quite useful,
it would be nice to also have a way to adjust \verb+numSpokes+
and \verb+rotAngle+ in the visual editor. However, no zones are connected
to these parameters.

Therefore, we add \verb+slider+s
to control the parameters
\verb+numSpokes+, \verb+rotAngle+, and \verb+spokeLen+ from the GUI editor.

\begin{Verbatim}[commandchars=\\\{\},codes={\catcode`\$=3\catcode`\^=7\catcode`\_=8},fontsize=\codeSize]
  (let [num s1] (intSlider ... $\ttNumLoc{5}{\varLoc_1}$)
  (let [len s2] (intSlider ... $\ttNumLoc{80}{\varLoc_2}$)
  (let [rot s3] (numSlider ... $\ttNumLoc{0}{\varLoc_3}$)

  (let wheel (ferrisWheel num len rot ...)
  (let sliders
    (let show true
    (if show (concat [s1 s2 s3]) []))

  (svg (append sliders wheel)))))))
\end{Verbatim}

\noindent
The resulting canvas is shown in \autoref{fig:ferris}.
With this setup, we can easily tweak any of the parameters
to \verb+ferrisWheel+ in live mode without having to modify the
program. If we wanted to change something about the ferris wheel
abstraction, of course, we could easily switch to programmatic
manipulation as needed.

To wrap up, we note how easy it is to export our ferris wheel
design once we have finished modifying it.
We can set the \verb+show+ parameter to \verb+false+ in order to hide
the sliders from the output. From there, we use the export
facility in \sns{} to generate the raw SVG for our design, which
we can copy and paste into other SVG editors.

\parahead{Procedural vs. Relational Constructions}

As mentioned in \autoref{sec:examples},
a detailed comparison of programming graphic
designs in these two styles may be an interesting avenue for future
work. Here, we merely note one relational specification that can be
easily rephrased procedurally, namely, that of an equilateral triangle. We can use
the \verb+nStar+ function defined earlier to derive an equilateral
triangle:

\begin{Verbatim}[commandchars=\\\{\},codes={\catcode`\$=3\catcode`\^=7\catcode`\_=8},fontsize=\codeSize]
  (def tri ($\lambda$(c x y sideLen rot)
    (let len1 (* sideLen (/ 2! 3!))
    (let len2 (* sideLen (/ 1! 3!))
    (nStar c 'none' 0 3! len1 len2 rot x y)))))
\end{Verbatim}

\newpage
\section{User Study}
\label{sec:user-study}

\newcommand{\ciInterval}[2]{($#1$, $#2$)}
\newcommand{\ci}[2]{$95\%$ CI: \ciInterval{#1}{#2}}
\newcommand{\pci}[2]{(\ci{#1}{#2})}
\newcommand{\pciShort}[2]{\ciInterval{#1}{#2}}

In this paper, we presented several ways of mitigating ambiguities
to enable live synchronization:
freezing constants, using
heuristics for automatic disambiguation, and using sliders. We
performed a user study to evaluate the relative strengths and
weaknesses of these techniques. We also sought to find out whether the
combination of programmatic and direct manipulation is a compelling
idea.

\parahead{Hypotheses} Our aim was to test the following hypotheses:

\begin{enumerate}

\item Simple heuristics for automatic disambiguation are sometimes preferable
to using sliders.

\item Direct manipulation (either with sliders or heuristics) is usually
preferable to making all edits programmatically.

\item Adding variables and other programmatic constructs to full-featured
direct manipulation tools is desired.

\end{enumerate}

\subsection{Procedures}

\sns{} is not a full-featured direct manipulation tool (\eg{}
\illustrator{}), nor is it a polished language implementation with
rich libraries and tools (\eg{} Processing). As a result, asking users
to learn to use \sns{} (i) would be time-consuming, (ii) would not allow
apples-to-apples comparisons with the aforementioned tools, and
(iii) would not be likely to isolate the interesting questions at
the boundary between programmatic and direct manipulation.
Instead, we designed a series of videos showing expert \sns{} users
(the authors) working with the tool. We asked participants to
watch the videos and then answer survey questions based on their
observations.
For reference, the survey questions as administered and a link to
our videos can be found in
\theAppendix{}.

\parahead{Video 0: Background (3 minutes)}
We described the relative strengths and weaknesses of programming and
direct manipulation for creating graphic designs.
We then asked background questions about programming and
graphic design experience.

\parahead{Video 1: Intro to \snsItals{} (9 minutes)}
We introduced \little{}, basic SVG features,
unambiguous direct manipulation updates,
and freezing constants.

\parahead{Video 2: Examples (2 minutes)}
We showed two examples where variables relate different
attributes and unambiguous direct manipulation updates
preserve these relationships.

\parahead{Video 3: Heuristics (6 minutes)}
We described the heuristics for automatic disambiguation.

\parahead{Video 4: Sliders (3 minutes)}
We demonstrated sliders as a way to control
otherwise hard-to-manipulate parameters.

\parahead{Video 5: Side-by-Side Comparisons (22 minutes)}
This section was designed to answer the question,
``Are direct manipulation heuristics better or worse than simply
providing a slider for every constant in the program?''

\begin{figure}[t]
\centering
\begin{tabular}{ c | c | c }
\textbf{Before} &
\textbf{After} &
\textbf{Histograms} \\[2pt]\hline
\raisebox{-.5\height}{\includegraphics[scale=0.21]{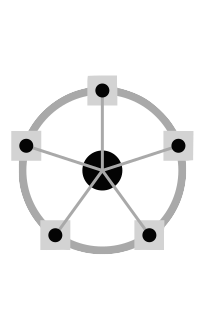}} &
\raisebox{-.5\height}{\includegraphics[scale=0.19]{ferris-task-after.png}} &
\raisebox{-.5\height}{\includegraphics[scale=0.45]{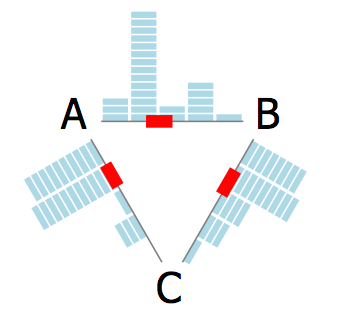}}
\\ \hline
\raisebox{-.5\height}{\includegraphics[scale=0.12]{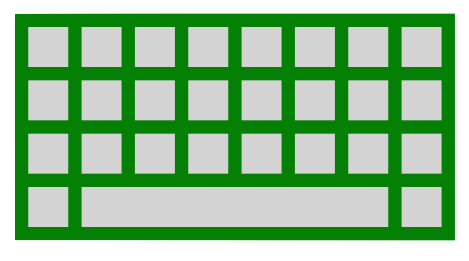}} &
\raisebox{-.5\height}{\includegraphics[scale=0.12]{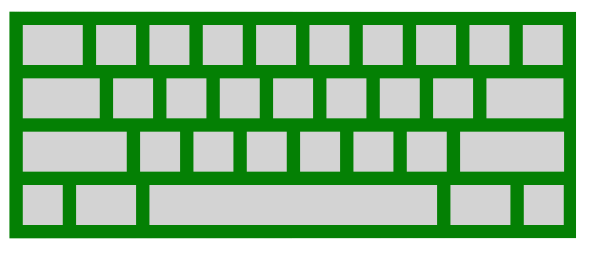}} &
\raisebox{-.5\height}{\includegraphics[scale=0.45]{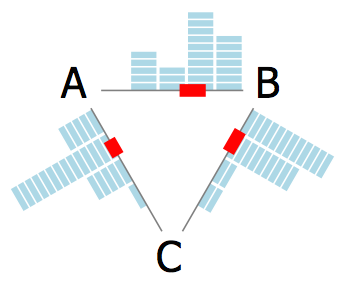}}
\\ \hline
\raisebox{-.5\height}{\includegraphics[scale=0.28]{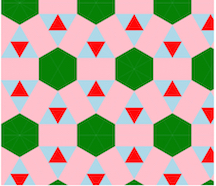}} &
\raisebox{-.5\height}{\includegraphics[scale=0.28]{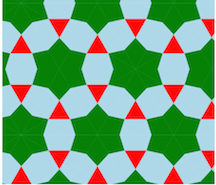}} &
\raisebox{-.5\height}{\includegraphics[scale=0.45]{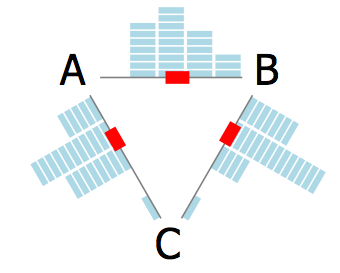}}
\end{tabular}
\caption{
The goal of each task (Ferris Wheel, Keyboard, and Tessellation) was
to convert the ``Before'' design into the ``After'' design.
For each task, users evaluated three ways of dealing with ambiguities
during the editing process:
(A) Sliders; (B) Heuristics; (C) Programmatic Manipulation Only.
The ``Histograms'' show the results.
}
\label{fig:study-tasks}
\end{figure}

We demonstrated a series of three tasks (Ferris Wheel, Keyboard, and
Tessellation), each starting with an
initial design (the ``Before'' column in \autoref{fig:study-tasks})
with the goal of editing it to realize a target design
(the ``After'' column). We performed each task twice: (A)
using only sliders when needing to break ambiguities (heuristics were
disabled) and (B) relying on the heuristics and freezing constants to
break ambiguities (sliders were not allowed). In both interaction
modes, unambiguous direct manipulation updates were allowed and
several programmatic edits were required.

After showing a task performed both ways,
we asked participants to rate the relative
effectiveness of interaction modes (A) and (B). We also asked
participants to rate each compared to a third mode: (C) using only
programmatic edits (no direct manipulation updates or sliders). Our
videos did not explicitly demonstrate mode (C).
To conclude the study, our survey asked about
final impressions of \sns{} and prodirect manipulation.


\parahead{Participants}

We sought users with programming experience, because the current
version of \sns{} requires it. We advertised our study to
undergraduate, Masters, and PhD students in the Computer Science
Department at the home institution of the authors. We held three
separate, in-person sessions where we showed the videos and
administered anonymous surveys. A total of \numUsers{} students
attended the sessions and completed surveys. Each person was paid \$20
for their participation, except one person who refused payment.
The study was reviewed and approved by the
Institutional Review Board (IRB) at our home institution.

\subsection{Results}

Our participants had significant programming background, with
$64\%$ reporting at least $3$ years of experience.
We also found that participants, on average,
generate $18\%$ of their graphic design work programmatically.

The side-by-side comparisons for the Ferris Wheel, Keyboard, and Tessellation
tasks comprised the primary evaluative aspect of our study.
For each pair of
interaction modes (M1) and (M2), we provided a five-option, balanced
rating scale which we interpret as a number in the range $[-2,2]$,
where $-2$ and $-1$ represent strong and weak preference,
respectively, for (M1) and $1$ and $2$ represent weak and strong
preference, respectively, for (M2).
The ``Histograms'' column of \autoref{fig:study-tasks} shows
the survey results.
Each edge between modes (M1) and (M2) of the triangle displays a histogram
of the relative preferences between (M1) and (M2).
We calculated the means along with $95\%$ \mbox{bootstrap-t} confidence
intervals,\endnote{A. Canty and B. Ripley, ``Package Boot,''
\url{https://cran.r-project.org/web/packages/boot/boot.pdf}}\endnote{A. C. Davison and D. Kuonen,
``An Introduction to the Bootstrap with Applications in R,''
2002}\endnote{A. C. Davison and D. V. Hinkley,
``Bootstrap Methods and Their Application,''
1997}
which are displayed with red lines along the edges
of the triangle.

\parahead{Hypothesis 1}
This table summarizes the mean preference ratings, along with
$95\%$ confidence intervals,
between sliders (A) and heuristics (B) for each of the three tasks.
\begin{wrapfigure}{r}{0pt}
\begin{tabular}{c | c}
  & (A) vs. (B) \\\hline
F & $-0.52$ \pciShort{-0.92}{0.01} \\
K & $\phantom{-}0.76$ \pciShort{\phantom{-}0.26}{1.18} \\
T & $\phantom{-}0.20$ \pciShort{-0.20}{0.64}
\end{tabular}
\end{wrapfigure}
Neither sliders nor heuristics were preferred for the Ferris Wheel task (F),
heuristics were weakly preferred over sliders for the Keyboard task (K),
and neither was preferred for the Tessellation task (T).
These data suggest that even simple heuristics can provide an
advantage over sliders.
Therefore, we conclude that
developing smarter heuristics in future work may provide an even more
desirable workflow.

\parahead{Hypothesis 2} Compared to manual code edits (C), both
sliders (A) and heuristics (B) were, on average, preferred on every task.
The following summarizes the mean preference ratings and
$95\%$ confidence intervals:
\begin{center}
\begin{tabular}{c | c | c}
  & (C) vs. (A) & (C) vs. (B) \\\hline
F & $1.12$ \pciShort{0.59}{1.47} & $0.80$ \pciShort{0.25}{1.23} \\
K & $0.92$ \pciShort{0.59}{1.21} & $1.24$ \pciShort{0.73}{1.57} \\
T & $0.76$ \pciShort{0.34}{1.10} & $1.00$ \pciShort{0.53}{1.32}
\end{tabular}
\end{center}

\noindent
We deliberately did not present tasks that could
be best accomplished through manual code edits, so this weak preference cannot be
generalized to all graphics manipulation tasks.
However, the data does suggest
that the ways in which \sns{} provides interactive, direct manipulation are
sometimes preferable to purely programmatic manipulation.

\parahead{Hypothesis 3} Participants reported that, on average, $50\%$
of designs they have created using direct manipulation tools would
have benefited from programmatic manipulation. Our participants had
significant programming background,
so this percentage is likely to be
higher than if participants had been drawn from a wider variety of users.
Nevertheless,
this finding supports the need for combining programmatic and direct
manipulation at least among expert programmers.

\parahead{Threats to Validity}

Because we showed each task twice, insights from the first
observation may have affected the second. To mitigate learning
effects, we performed similar edits and narration at the same pace in
the second version, even when they were redundant when viewed
after the first.

Choices in program representation have a significant impact on what
kinds of direct manipulation may or may not be possible. In an attempt
not to bias our initial programs (used to generate the ``Before''
designs) towards either mode (A) or (B), we wrote
near-final drafts of the programs before deciding exactly what
manipulation tasks to perform. We also chose the specific
tasks by taking inspiration from outside references, such as photos.

Finally, the participants made judgments about the relative merits of sliders
and heuristics based on observations, not by using the tool directly.
We plan to run ``hands-on'' experiments in the future, which will
first require achieving more feature-parity with existing tools.

\begingroup
\parindent 0pt
\parskip 6pt
\def\enotesize{\footnotesize}
\theendnotes
\endgroup

\newpage
\section{User Study Survey}

This section provides the survey instrument that we
administered (reformatted for this paper).
For multiple-choice
questions, we display the number of participants (out of $25$)
that chose each option in parentheses.
Our videos can be found on our project Web page.



\small

\newcommand{\nextQ}{\vspace{3pt}\noindent}

\newcommand{\percentagebar}{
\begin{tikzpicture}
\draw (0,0) -- (200 pt,0);
\foreach \x in {0,10,20,30,40,50,60,70,80,90,100}
    \draw (\x * 2.0 pt,3pt) -- (\x * 2.0 pt,-3pt) node[anchor=north] {\small \x \%};
\end{tikzpicture}
}

\newcommand{\surveyCount}[2]{\ensuremath{({#1})}}

\newcommand{\sidebysidetaskquestions}[9]{
\subsection*{Side By Side Comparison: #1}
  \def\taskName{#1}

\vspace{3pt}
\textbf{Interaction A:} Sliders and unambiguous direct manipulation.\\
\textbf{Interaction B:} Direct manipulation with heuristics and freezing.
\vspace{5pt}

\nextQ
Which interaction worked better for the #1 task?

\begin{itemize}
  \setlength{\itemsep}{-1pt}
  \item Interaction A worked much better \surveyCount{#2}{}
  \item Interaction A worked a little better \surveyCount{#3}{}
  \item They are about the same \surveyCount{#4}{}
  \item Interaction B worked a little better \surveyCount{#5}{}
  \item Interaction B worked much better \surveyCount{#6}{}
\end{itemize}

\nextQ
\textit{The next two questions ask you to consider how each interaction would compare to just making all edits in the code manually.}
\vspace{5pt}

\nextQ
For the #1 task, how good was interaction A?

\begin{itemize}
  \setlength{\itemsep}{-1pt}
  \item Manual code edits only would have worked much better \surveyCount{#7}{}
  \item Manual code edits only would have worked a little better \surveyCount{#8}{}
  \item About the same as manual code edits \surveyCount{#9}{}
  \relay
}
\newcommand{\relay}[7]{
  \item Interaction A worked a little better \surveyCount{#1}{}
  \item Interaction A worked much better \surveyCount{#2}{}
\end{itemize}

\nextQ
For the \taskName{} task, how good was interaction B?

\begin{itemize}
  \setlength{\itemsep}{-1pt}
  \item Manual code edits only would have worked much better \surveyCount{#3}{}
  \item Manual code edits only would have worked a little better \surveyCount{#4}{}
  \item About the same as manual code edits \surveyCount{#5}{}
  \item Interaction B worked a little better \surveyCount{#6}{}
  \item Interaction B worked much better \surveyCount{#7}{}
\end{itemize}
}


\subsection*{Background Questions}

\nextQ
How often do you use graphic design applications?\\
(Examples: Adobe Illustrator, Microsoft PowerPoint, GIMP)

\begin{itemize}
  \setlength{\itemsep}{-1pt}
  \item Less than once a year \surveyCount{0}{}
  \item A few times a year \surveyCount{9}{}
  \item A few times a month \surveyCount{11}{}
  \item A few times a week \surveyCount{5}{}
  \item Every day or almost every day \surveyCount{0}{}
\end{itemize}

\nextQ
How many years of programming experience do you have?

\begin{itemize}
  \setlength{\itemsep}{-1pt}
  \item Less than 1 \surveyCount{3}{}
  \item 1-2 \surveyCount{6}{}
  \item 3-5 \surveyCount{8}{}
  \item 6-10 \surveyCount{8}{}
  \item 11-20 \surveyCount{0}{}
  \item More than 20 \surveyCount{0}{}
\end{itemize}

\nextQ
What percent of your graphic design work is programmatically generated?
(Place a mark on the line.)
\\

\percentagebar

\sidebysidetaskquestions{Ferris Wheel}
  {3}{14}{2}{5}{1}
  {0}{3}{1}{11}{10}
  {1}{3}{4}{9}{8}

\newpage

\sidebysidetaskquestions{Keyboard}
  {0}{5}{3}{10}{7}
  {0}{1}{5}{14}{5}
  {0}{2}{2}{9}{12}

\sidebysidetaskquestions{Tessellation}
  {0}{7}{9}{6}{3}
  {1}{0}{8}{11}{5}
  {1}{0}{4}{13}{7}

\subsection*{Additional Questions}

\nextQ
Thinking of all the graphics you have created with direct manipulation tools, what percentage of those graphics would have been easier to create if your direct manipulation tool also allowed programmatic manipulation?
(Place a mark on the line.)
\\

\percentagebar

\nextQ
Do you plan to try using Sketch-n-Sketch to create graphics?

\begin{itemize}
  \setlength{\itemsep}{-1pt}
  \item Certainly not \surveyCount{0}{}
  \item Probably not \surveyCount{2}{}
  \item Maybe \surveyCount{8}{}
  \item Likely \surveyCount{12}{}
  \item Certainly \surveyCount{3}{}
\end{itemize}

\nextQ
What improvements or new features would make Sketch-n-Sketch better?
\\

\nextQ
Are there any other comments about Sketch-n-Sketch or the idea of prodirect manipulation that you would like to share?
\\

\nextQ
Today you saw a demonstration of prodirect manipulation for graphics editing. Are there other applications where you would like to see prodirect manipulation?
\\

\normalsize

\newpage
\section{Measurements}

The following measurements, summarized and discussed in
\autoref{sec:implementation}, were collected on
\verb+v0.4.2+ of our implementation.

\begin{figure*}
\tiny
\begin{tabular}{c c c c c c}
\hline
 &  &  & \multicolumn{3}{c}{\textbf{Number of zones with}} \\
\textbf{Example Name} & \textbf{Shape Count} & \textbf{Zone Count} & \multicolumn{3}{c}{\textbf{$0$, $1$, $>1$ loclist choices}} \\
\cline{4-6}
 &  &  & $\bm{0}$ & $\bm{1}$ & $\bm{>1}$ \textbf{(avg)} \\
\hline \hline
Wave Boxes & 12 & 108 & 0 & 36 & 72 (2.67) \\
\hline
Wave Boxes Grid & 78 & 702 & 18 & 246 & 438 (10.71) \\
\hline
Logo & 4 & 30 & 12 & 9 & 9 (3.22) \\
\hline
Botanic Garden Logo & 4 & 25 & 1 & 23 & 1 (2.00) \\
\hline
Active Trans Logo & 9 & 40 & 8 & 32 & 0 \\
\hline
Sailboat & 36 & 132 & 15 & 75 & 42 (4.07) \\
\hline
Chicago Flag & 7 & 127 & 4 & 73 & 50 (5.84) \\
\hline
Sliders & 24 & 34 & 30 & 4 & 0 \\
\hline
Buttons & 6 & 7 & 6 & 1 & 0 \\
\hline
Widgets & 24 & 33 & 29 & 4 & 0 \\
\hline
xySlider & 13 & 19 & 18 & 1 & 0 \\
\hline
Tile Pattern & 49 & 123 & 104 & 19 & 0 \\
\hline
Color Picker & 25 & 36 & 32 & 4 & 0 \\
\hline
Ferris Wheel & 13 & 80 & 11 & 49 & 20 (2.25) \\
\hline
Ferris Task Before & 13 & 80 & 0 & 49 & 31 (3.45) \\
\hline
Ferris Task After & 19 & 125 & 0 & 76 & 49 (3.47) \\
\hline
Ferris Wheel Slideshow & 14 & 18 & 16 & 2 & 0 \\
\hline
Survey Results & 207 & 752 & 188 & 564 & 0 \\
\hline
Hilbert Curve Animation & 16 & 21 & 16 & 5 & 0 \\
\hline
Bar Graph & 48 & 60 & 25 & 30 & 5 (3.00) \\
\hline
Pie Chart & 41 & 105 & 91 & 5 & 9 (6.00) \\
\hline
Solar System & 23 & 41 & 8 & 25 & 8 (7.00) \\
\hline
Clique & 37 & 105 & 0 & 105 & 0 \\
\hline
Eye Icon & 4 & 21 & 7 & 12 & 2 (2.00) \\
\hline
Wikimedia Logo & 6 & 20 & 10 & 10 & 0 \\
\hline
Haskell.org Logo & 4 & 22 & 2 & 8 & 12 (2.50) \\
\hline
Cover Logo & 16 & 159 & 9 & 150 & 0 \\
\hline
POP-PL Logo & 8 & 36 & 22 & 12 & 2 (2.00) \\
\hline
Lillicon P & 8 & 30 & 7 & 21 & 2 (2.00) \\
\hline
Lillicon P, v2 & 3 & 15 & 2 & 13 & 0 \\
\hline
Keyboard & 40 & 360 & 0 & 61 & 299 (6.17) \\
\hline
Keyboard Task Before & 28 & 252 & 0 & 1 & 251 (5.82) \\
\hline
Keyboard Task After & 33 & 297 & 0 & 7 & 290 (5.81) \\
\hline
Tessellation Task Before & 1364 & 4992 & 0 & 1536 & 3456 (2.56) \\
\hline
Tessellation Task After & 980 & 2496 & 0 & 384 & 2112 (2.45) \\
\hline
Floral Logo 1 & 20 & 140 & 0 & 0 & 140 (7.86) \\
\hline
Floral Logo 2 & 18 & 126 & 0 & 0 & 126 (8.52) \\
\hline
Spiral Spiral-Graph & 140 & 420 & 0 & 280 & 140 (12.00) \\
\hline
Rounded Rect & 12 & 28 & 8 & 14 & 6 (2.33) \\
\hline
Thaw/Freeze & 1 & 9 & 5 & 4 & 0 \\
\hline
3 Boxes & 3 & 27 & 0 & 19 & 8 (2.00) \\
\hline
N Boxes Sli & 3 & 27 & 0 & 19 & 8 (2.00) \\
\hline
N Boxes & 10 & 49 & 5 & 27 & 17 (3.65) \\
\hline
Elm Logo & 7 & 53 & 0 & 53 & 0 \\
\hline
Logo 2 & 4 & 46 & 15 & 22 & 9 (2.22) \\
\hline
Logo Sizes & 12 & 138 & 61 & 52 & 25 (2.08) \\
\hline
Rings & 5 & 10 & 0 & 6 & 4 (2.50) \\
\hline
Polygons & 5 & 73 & 0 & 39 & 34 (5.00) \\
\hline
Stars & 5 & 105 & 7 & 48 & 50 (2.88) \\
\hline
Triangles & 2 & 26 & 0 & 0 & 26 (2.46) \\
\hline
US-13 Flag & 22 & 354 & 14 & 203 & 137 (3.90) \\
\hline
US-50 Flag & 59 & 181 & 14 & 110 & 57 (3.75) \\
\hline
French Sudan Flag & 12 & 59 & 11 & 31 & 17 (3.76) \\
\hline
Frank Lloyd Wright & 31 & 119 & 42 & 9 & 68 (5.06) \\
\hline
Bezier Curves & 22 & 42 & 11 & 19 & 12 (7.50) \\
\hline
Fractal Tree & 27 & 60 & 18 & 4 & 38 (13.74) \\
\hline
Stick Figures & 54 & 193 & 53 & 56 & 84 (3.92) \\
\hline
Matrix Transformations & 5 & 20 & 0 & 0 & 20 (3.20) \\
\hline
Misc Shapes & 6 & 30 & 0 & 30 & 0 \\
\hline
Interface Buttons & 21 & 70 & 14 & 32 & 24 (2.21) \\
\hline
Paths 1 & 3 & 8 & 0 & 8 & 0 \\
\hline
Paths 2 & 9 & 36 & 0 & 36 & 0 \\
\hline
Paths 3 & 3 & 13 & 0 & 13 & 0 \\
\hline
Paths 4 & 1 & 6 & 0 & 6 & 0 \\
\hline
Paths 5 & 4 & 12 & 0 & 12 & 0 \\
\hline
Sample Rotations & 2 & 15 & 0 & 14 & 1 (9.00) \\
\hline
Grid Tile & 24 & 79 & 22 & 14 & 43 (2.98) \\
\hline
Zones & 4 & 29 & 0 & 24 & 5 (5.00) \\
\hline
\hline
\textbf{Totals} & 3772 & 14106 & 991 & 4856 & 8259 (3.83) \\
\hline \hline
\end{tabular}
\end{figure*}

\clearpage

\begin{figure*}
\tiny
\begin{tabular}{c c c c c}
\hline
\textbf{Example Name} & \textbf{\# Output Locs} & \textbf{Unfrozen} & \textbf{Unassigned} & \textbf{Assigned (avg times) (avg rate)} \\
\hline \hline
Wave Boxes & 7 & 6 & 0 & 6 (40.0) (67\%) \\
\hline
Wave Boxes Grid & 29 & 24 & 1 & 23 (64.5) (34\%) \\
\hline
Logo & 6 & 6 & 1 & 5 (7.6) (36\%) \\
\hline
Botanic Garden Logo & 22 & 19 & 7 & 12 (3.2) (98\%) \\
\hline
Active Trans Logo & 71 & 44 & 1 & 43 (1.5) (100\%) \\
\hline
Sailboat & 52 & 38 & 22 & 16 (14.3) (85\%) \\
\hline
Chicago Flag & 19 & 8 & 1 & 7 (24.4) (23\%) \\
\hline
Sliders & 32 & 12 & 8 & 4 (1.0) (100\%) \\
\hline
Buttons & 9 & 3 & 2 & 1 (1.0) (100\%) \\
\hline
Widgets & 18 & 4 & 0 & 4 (1.0) (100\%) \\
\hline
xySlider & 18 & 7 & 5 & 2 (1.0) (100\%) \\
\hline
Tile Pattern & 49 & 27 & 0 & 27 (1.1) (100\%) \\
\hline
Color Picker & 27 & 4 & 0 & 4 (1.0) (100\%) \\
\hline
Ferris Wheel & 11 & 9 & 2 & 7 (11.0) (53\%) \\
\hline
Ferris Task Before & 11 & 11 & 2 & 9 (9.6) (42\%) \\
\hline
Ferris Task After & 11 & 11 & 2 & 9 (14.9) (42\%) \\
\hline
Ferris Wheel Slideshow & 22 & 12 & 10 & 2 (1.0) (100\%) \\
\hline
Survey Results & 94 & 64 & 63 & 1 (564.0) (100\%) \\
\hline
Hilbert Curve Animation & 44 & 34 & 27 & 7 (1.0) (100\%) \\
\hline
Bar Graph & 40 & 5 & 0 & 5 (9.0) (46\%) \\
\hline
Pie Chart & 46 & 13 & 7 & 6 (2.3) (24\%) \\
\hline
Solar System & 38 & 16 & 0 & 16 (2.8) (74\%) \\
\hline
Clique & 14 & 14 & 1 & 13 (20.5) (100\%) \\
\hline
Eye Icon & 40 & 22 & 15 & 7 (2.0) (90\%) \\
\hline
Wikimedia Logo & 19 & 8 & 4 & 4 (3.0) (85\%) \\
\hline
Haskell.org Logo & 58 & 23 & 16 & 7 (3.1) (59\%) \\
\hline
Cover Logo & 15 & 2 & 0 & 2 (75.0) (100\%) \\
\hline
POP-PL Logo & 25 & 13 & 0 & 13 (1.8) (96\%) \\
\hline
Lillicon P & 18 & 15 & 1 & 14 (3.3) (98\%) \\
\hline
Lillicon P, v2 & 19 & 17 & 4 & 13 (2.0) (100\%) \\
\hline
Keyboard & 45 & 45 & 11 & 34 (18.1) (34\%) \\
\hline
Keyboard Task Before & 44 & 44 & 2 & 42 (13.3) (43\%) \\
\hline
Keyboard Task After & 46 & 46 & 2 & 44 (14.5) (38\%) \\
\hline
Tessellation Task Before & 89 & 77 & 72 & 5 (1510.2) (57\%) \\
\hline
Tessellation Task After & 87 & 76 & 71 & 5 (768.0) (60\%) \\
\hline
Floral Logo 1 & 33 & 32 & 1 & 31 (4.7) (16\%) \\
\hline
Floral Logo 2 & 32 & 29 & 1 & 28 (4.7) (16\%) \\
\hline
Spiral Spiral-Graph & 14 & 14 & 0 & 14 (30.9) (18\%) \\
\hline
Rounded Rect & 16 & 11 & 4 & 7 (5.9) (85\%) \\
\hline
Thaw/Freeze & 4 & 1 & 0 & 1 (4.0) (100\%) \\
\hline
3 Boxes & 5 & 5 & 0 & 5 (12.0) (83\%) \\
\hline
N Boxes Sli & 5 & 5 & 0 & 5 (12.0) (83\%) \\
\hline
N Boxes & 22 & 17 & 6 & 11 (7.8) (76\%) \\
\hline
Elm Logo & 43 & 43 & 1 & 42 (4.1) (100\%) \\
\hline
Logo 2 & 9 & 7 & 1 & 6 (8.2) (66\%) \\
\hline
Logo Sizes & 15 & 11 & 1 & 10 (11.7) (48\%) \\
\hline
Rings & 6 & 6 & 1 & 5 (3.0) (77\%) \\
\hline
Polygons & 29 & 26 & 2 & 24 (4.1) (28\%) \\
\hline
Stars & 8 & 6 & 1 & 5 (22.6) (25\%) \\
\hline
Triangles & 14 & 9 & 1 & 8 (3.8) (29\%) \\
\hline
US-13 Flag & 27 & 7 & 1 & 6 (76.5) (26\%) \\
\hline
US-50 Flag & 26 & 5 & 0 & 5 (54.8) (66\%) \\
\hline
French Sudan Flag & 21 & 4 & 0 & 4 (24.0) (55\%) \\
\hline
Frank Lloyd Wright & 40 & 31 & 6 & 25 (5.8) (60\%) \\
\hline
Bezier Curves & 23 & 12 & 1 & 11 (5.2) (57\%) \\
\hline
Fractal Tree & 37 & 22 & 4 & 18 (3.5) (18\%) \\
\hline
Stick Figures & 85 & 68 & 13 & 55 (6.3) (81\%) \\
\hline
Matrix Transformations & 57 & 7 & 1 & 6 (3.3) (43\%) \\
\hline
Misc Shapes & 28 & 28 & 3 & 25 (3.3) (100\%) \\
\hline
Interface Buttons & 35 & 18 & 12 & 6 (13.3) (58\%) \\
\hline
Paths 1 & 19 & 19 & 3 & 16 (1.0) (100\%) \\
\hline
Paths 2 & 72 & 72 & 0 & 72 (1.0) (100\%) \\
\hline
Paths 3 & 26 & 26 & 0 & 26 (1.0) (100\%) \\
\hline
Paths 4 & 24 & 24 & 12 & 12 (1.0) (100\%) \\
\hline
Paths 5 & 48 & 48 & 24 & 24 (1.0) (100\%) \\
\hline
Sample Rotations & 12 & 12 & 2 & 10 (2.8) (97\%) \\
\hline
Grid Tile & 14 & 8 & 2 & 6 (19.0) (62\%) \\
\hline
Zones & 31 & 28 & 1 & 27 (2.0) (87\%) \\
\hline
\hline
\textbf{Totals} & 2075 & 1440 & 465 & 975 (21.1) (69\%) \\
\hline \hline
\end{tabular}
\end{figure*}

\clearpage

\begin{figure*}
\small
\begin{tabular}{l c}
\hline
\textbf{\# Equations} & 28222 \\
\textbf{\# (example, loc, tr) Equations} & 4574 \\
\textbf{Mean Trace Size (tree nodes)} & 141.30 \\
\textbf{Max Solve Time} & 0.014 \\
\textbf{Median Solve Time} & $< 0.001$ (Calculated: 0.0000) \\
\textbf{Mean Solve Time} & $< 0.001$ (Calculated: 0.0001) \\
\textbf{\# Traces in Solve A fragment} & 778 \\
\textbf{\hspace{2em}Solved} &  \\
\textbf{\hspace{4em}$+= 1$} & 778 \\
\textbf{\hspace{4em}$+= 100$} & 778 \\
\textbf{\hspace{2em}Not Solved} &  \\
\textbf{\hspace{4em}$+= 1$} & 0 \\
\textbf{\hspace{4em}$+= 100$} & 0 \\
\textbf{\# Traces in Solve B fragment} & 3655 \\
\textbf{\hspace{2em}Solved} &  \\
\textbf{\hspace{4em}$+= 1$} & 3461 \\
\textbf{\hspace{4em}$+= 100$} & 3023 \\
\textbf{\hspace{2em}Not Solved} &  \\
\textbf{\hspace{4em}$+= 1$} & 194 \\
\textbf{\hspace{4em}$+= 100$} & 632 \\
\textbf{\# Traces in either fragment} & 3656 \\
\textbf{\hspace{2em}Solved} &  \\
\textbf{\hspace{4em}$+= 1$} & 3462 \\
\textbf{\hspace{4em}$+= 100$} & 3024 \\
\textbf{\hspace{2em}Not Solved} &  \\
\textbf{\hspace{4em}$+= 1$} & 194 \\
\textbf{\hspace{4em}$+= 100$} & 632 \\
\textbf{\# Traces in no fragment} & 918 \\
\hline
\end{tabular}
\end{figure*}

\clearpage

\begin{figure*}
\scriptsize
\begin{tabular}{c c c c c c c}
\hline
\textbf{Example Name} & \textbf{Parse} & \textbf{Eval} & \textbf{Unparse} & \textbf{valToInde...} & \textbf{prepareLi...} & \textbf{``Run Code''} \\
\hline \hline
Wave Boxes & FF: 0.016 & FF: 0.001 & FF: 0.001 & FF: 0.000 & FF: 0.027 & FF: 0.050 \\
LOC: 7 & Ch: 0.020 & Ch: 0.003 & Ch: 0.001 & Ch: 0.000 & Ch: 0.061 & Ch: 0.090 \\
\hline
Wave Boxes Grid & FF: 0.100 & FF: 0.012 & FF: 0.004 & FF: 0.001 & FF: 3.372 & FF: 3.510 \\
LOC: 42 & Ch: 0.131 & Ch: 0.025 & Ch: 0.004 & Ch: 0.002 & Ch: 6.174 & Ch: 6.365 \\
\hline
Logo & FF: 0.027 & FF: 0.007 & FF: 0.001 & FF: 0.000 & FF: 0.006 & FF: 0.052 \\
LOC: 18 & Ch: 0.035 & Ch: 0.026 & Ch: 0.002 & Ch: 0.000 & Ch: 0.012 & Ch: 0.106 \\
\hline
Botanic Garden Logo & FF: 0.060 & FF: 0.001 & FF: 0.001 & FF: 0.002 & FF: 0.005 & FF: 0.074 \\
LOC: 28 & Ch: 0.072 & Ch: 0.002 & Ch: 0.003 & Ch: 0.001 & Ch: 0.005 & Ch: 0.089 \\
\hline
Active Trans Logo & FF: 0.067 & FF: 0.004 & FF: 0.004 & FF: 0.001 & FF: 0.010 & FF: 0.095 \\
LOC: 33 & Ch: 0.089 & Ch: 0.008 & Ch: 0.002 & Ch: 0.002 & Ch: 0.019 & Ch: 0.131 \\
\hline
Sailboat & FF: 0.089 & FF: 0.009 & FF: 0.004 & FF: 0.004 & FF: 0.030 & FF: 0.151 \\
LOC: 51 & Ch: 0.115 & Ch: 0.012 & Ch: 0.004 & Ch: 0.007 & Ch: 0.053 & Ch: 0.210 \\
\hline
Chicago Flag & FF: 0.032 & FF: 0.011 & FF: 0.002 & FF: 0.000 & FF: 0.046 & FF: 0.104 \\
LOC: 16 & Ch: 0.040 & Ch: 0.025 & Ch: 0.001 & Ch: 0.000 & Ch: 0.087 & Ch: 0.178 \\
\hline
Sliders & FF: 0.083 & FF: 0.001 & FF: 0.003 & FF: 0.000 & FF: 0.002 & FF: 0.093 \\
LOC: 38 & Ch: 0.105 & Ch: 0.004 & Ch: 0.003 & Ch: 0.001 & Ch: 0.004 & Ch: 0.122 \\
\hline
Buttons & FF: 0.033 & FF: 0.001 & FF: 0.003 & FF: 0.000 & FF: 0.002 & FF: 0.043 \\
LOC: 16 & Ch: 0.039 & Ch: 0.000 & Ch: 0.002 & Ch: 0.000 & Ch: 0.002 & Ch: 0.046 \\
\hline
Widgets & FF: 0.015 & FF: 0.003 & FF: 0.000 & FF: 0.000 & FF: 0.002 & FF: 0.025 \\
LOC: 5 & Ch: 0.018 & Ch: 0.008 & Ch: 0.001 & Ch: 0.000 & Ch: 0.005 & Ch: 0.045 \\
\hline
xySlider & FF: 0.078 & FF: 0.002 & FF: 0.003 & FF: 0.000 & FF: 0.003 & FF: 0.092 \\
LOC: 33 & Ch: 0.098 & Ch: 0.002 & Ch: 0.003 & Ch: 0.000 & Ch: 0.003 & Ch: 0.112 \\
\hline
Tile Pattern & FF: 0.121 & FF: 0.033 & FF: 0.004 & FF: 0.000 & FF: 0.004 & FF: 0.202 \\
LOC: 74 & Ch: 0.154 & Ch: 0.050 & Ch: 0.005 & Ch: 0.001 & Ch: 0.009 & Ch: 0.272 \\
\hline
Color Picker & FF: 0.025 & FF: 0.003 & FF: 0.001 & FF: 0.000 & FF: 0.003 & FF: 0.039 \\
LOC: 8 & Ch: 0.028 & Ch: 0.006 & Ch: 0.001 & Ch: 0.001 & Ch: 0.003 & Ch: 0.046 \\
\hline
Ferris Wheel & FF: 0.038 & FF: 0.004 & FF: 0.004 & FF: 0.000 & FF: 0.010 & FF: 0.066 \\
LOC: 12 & Ch: 0.044 & Ch: 0.008 & Ch: 0.002 & Ch: 0.000 & Ch: 0.016 & Ch: 0.083 \\
\hline
Ferris Task Before & FF: 0.030 & FF: 0.005 & FF: 0.001 & FF: 0.000 & FF: 0.018 & FF: 0.062 \\
LOC: 16 & Ch: 0.038 & Ch: 0.009 & Ch: 0.001 & Ch: 0.000 & Ch: 0.028 & Ch: 0.090 \\
\hline
Ferris Task After & FF: 0.031 & FF: 0.008 & FF: 0.002 & FF: 0.000 & FF: 0.023 & FF: 0.073 \\
LOC: 16 & Ch: 0.037 & Ch: 0.013 & Ch: 0.002 & Ch: 0.001 & Ch: 0.041 & Ch: 0.109 \\
\hline
Ferris Wheel Slideshow & FF: 0.298 & FF: 0.028 & FF: 0.011 & FF: 0.000 & FF: 0.003 & FF: 0.372 \\
LOC: 252 & Ch: 0.375 & Ch: 0.063 & Ch: 0.012 & Ch: 0.001 & Ch: 0.004 & Ch: 0.524 \\
\hline
Survey Results & FF: 0.225 & FF: 0.068 & FF: 0.008 & FF: 0.004 & FF: 0.045 & FF: 0.415 \\
LOC: 121 & Ch: 0.270 & Ch: 0.157 & Ch: 0.010 & Ch: 0.006 & Ch: 0.082 & Ch: 0.691 \\
\hline
Hilbert Curve Animation & FF: 0.409 & FF: 0.006 & FF: 0.010 & FF: 0.000 & FF: 0.006 & FF: 0.443 \\
LOC: 289 & Ch: 0.503 & Ch: 0.010 & Ch: 0.011 & Ch: 0.000 & Ch: 0.006 & Ch: 0.542 \\
\hline
Bar Graph & FF: 0.111 & FF: 0.003 & FF: 0.003 & FF: 0.000 & FF: 0.005 & FF: 0.130 \\
LOC: 43 & Ch: 0.145 & Ch: 0.008 & Ch: 0.005 & Ch: 0.001 & Ch: 0.011 & Ch: 0.182 \\
\hline
Pie Chart & FF: 0.089 & FF: 0.004 & FF: 0.003 & FF: 0.001 & FF: 0.007 & FF: 0.112 \\
LOC: 36 & Ch: 0.112 & Ch: 0.015 & Ch: 0.006 & Ch: 0.002 & Ch: 0.015 & Ch: 0.170 \\
\hline
Solar System & FF: 0.065 & FF: 0.003 & FF: 0.003 & FF: 0.000 & FF: 0.008 & FF: 0.087 \\
LOC: 31 & Ch: 0.084 & Ch: 0.006 & Ch: 0.002 & Ch: 0.000 & Ch: 0.014 & Ch: 0.114 \\
\hline
Clique & FF: 0.027 & FF: 0.006 & FF: 0.001 & FF: 0.000 & FF: 0.023 & FF: 0.066 \\
LOC: 13 & Ch: 0.032 & Ch: 0.011 & Ch: 0.001 & Ch: 0.001 & Ch: 0.042 & Ch: 0.103 \\
\hline
Eye Icon & FF: 0.049 & FF: 0.001 & FF: 0.003 & FF: 0.001 & FF: 0.005 & FF: 0.065 \\
LOC: 37 & Ch: 0.062 & Ch: 0.001 & Ch: 0.002 & Ch: 0.003 & Ch: 0.006 & Ch: 0.079 \\
\hline
\end{tabular}
\end{figure*}

\clearpage

\begin{figure*}
\scriptsize
\begin{tabular}{c c c c c c c}
\hline
\textbf{Example Name} & \textbf{Parse} & \textbf{Eval} & \textbf{Unparse} & \textbf{valToInde...} & \textbf{prepareLi...} & \textbf{``Run Code''} \\
\hline \hline
Wikimedia Logo & FF: 0.039 & FF: 0.001 & FF: 0.001 & FF: 0.001 & FF: 0.003 & FF: 0.048 \\
LOC: 19 & Ch: 0.052 & Ch: 0.001 & Ch: 0.001 & Ch: 0.001 & Ch: 0.003 & Ch: 0.061 \\
\hline
Haskell.org Logo & FF: 0.093 & FF: 0.001 & FF: 0.002 & FF: 0.001 & FF: 0.003 & FF: 0.103 \\
LOC: 59 & Ch: 0.119 & Ch: 0.002 & Ch: 0.003 & Ch: 0.001 & Ch: 0.006 & Ch: 0.137 \\
\hline
Cover Logo & FF: 0.072 & FF: 0.003 & FF: 0.003 & FF: 0.000 & FF: 0.010 & FF: 0.094 \\
LOC: 39 & Ch: 0.102 & Ch: 0.006 & Ch: 0.005 & Ch: 0.000 & Ch: 0.019 & Ch: 0.144 \\
\hline
POP-PL Logo & FF: 0.071 & FF: 0.001 & FF: 0.003 & FF: 0.000 & FF: 0.003 & FF: 0.081 \\
LOC: 57 & Ch: 0.089 & Ch: 0.002 & Ch: 0.003 & Ch: 0.001 & Ch: 0.007 & Ch: 0.108 \\
\hline
Lillicon P & FF: 0.039 & FF: 0.000 & FF: 0.002 & FF: 0.000 & FF: 0.003 & FF: 0.048 \\
LOC: 28 & Ch: 0.048 & Ch: 0.002 & Ch: 0.002 & Ch: 0.000 & Ch: 0.007 & Ch: 0.064 \\
\hline
Lillicon P, v2 & FF: 0.026 & FF: 0.000 & FF: 0.001 & FF: 0.000 & FF: 0.002 & FF: 0.033 \\
LOC: 19 & Ch: 0.031 & Ch: 0.001 & Ch: 0.001 & Ch: 0.000 & Ch: 0.005 & Ch: 0.042 \\
\hline
Keyboard & FF: 0.110 & FF: 0.006 & FF: 0.004 & FF: 0.000 & FF: 0.873 & FF: 1.005 \\
LOC: 71 & Ch: 0.140 & Ch: 0.017 & Ch: 0.004 & Ch: 0.001 & Ch: 1.319 & Ch: 1.498 \\
\hline
Keyboard Task Before & FF: 0.067 & FF: 0.006 & FF: 0.002 & FF: 0.000 & FF: 1.077 & FF: 1.163 \\
LOC: 48 & Ch: 0.077 & Ch: 0.007 & Ch: 0.002 & Ch: 0.000 & Ch: 1.843 & Ch: 1.939 \\
\hline
Keyboard Task After & FF: 0.069 & FF: 0.005 & FF: 0.002 & FF: 0.000 & FF: 1.340 & FF: 1.422 \\
LOC: 51 & Ch: 0.083 & Ch: 0.010 & Ch: 0.003 & Ch: 0.001 & Ch: 2.127 & Ch: 2.235 \\
\hline
Tessellation Task Before & FF: 0.116 & FF: 0.002 & FF: 0.004 & FF: 0.096 & FF: 0.675 & FF: 0.995 \\
LOC: 80 & Ch: 0.142 & Ch: 0.003 & Ch: 0.004 & Ch: 0.164 & Ch: 1.319 & Ch: 1.803 \\
\hline
Tessellation Task After & FF: 0.098 & FF: 0.002 & FF: 0.002 & FF: 0.072 & FF: 0.357 & FF: 0.607 \\
LOC: 73 & Ch: 0.127 & Ch: 0.004 & Ch: 0.005 & Ch: 0.115 & Ch: 0.671 & Ch: 1.044 \\
\hline
Floral Logo 1 & FF: 0.073 & FF: 0.012 & FF: 0.004 & FF: 0.004 & FF: 0.158 & FF: 0.275 \\
LOC: 45 & Ch: 0.091 & Ch: 0.028 & Ch: 0.004 & Ch: 0.005 & Ch: 0.379 & Ch: 0.542 \\
\hline
Floral Logo 2 & FF: 0.095 & FF: 0.013 & FF: 0.005 & FF: 0.004 & FF: 0.152 & FF: 0.293 \\
LOC: 49 & Ch: 0.108 & Ch: 0.025 & Ch: 0.004 & Ch: 0.005 & Ch: 0.345 & Ch: 0.521 \\
\hline
Spiral Spiral-Graph & FF: 0.034 & FF: 0.030 & FF: 0.002 & FF: 0.002 & FF: 0.257 & FF: 0.351 \\
LOC: 16 & Ch: 0.041 & Ch: 0.060 & Ch: 0.002 & Ch: 0.003 & Ch: 0.571 & Ch: 0.745 \\
\hline
Rounded Rect & FF: 0.026 & FF: 0.003 & FF: 0.001 & FF: 0.001 & FF: 0.004 & FF: 0.039 \\
LOC: 14 & Ch: 0.032 & Ch: 0.004 & Ch: 0.001 & Ch: 0.000 & Ch: 0.005 & Ch: 0.048 \\
\hline
Thaw/Freeze & FF: 0.015 & FF: 0.000 & FF: 0.001 & FF: 0.000 & FF: 0.003 & FF: 0.021 \\
LOC: 2 & Ch: 0.011 & Ch: 0.000 & Ch: 0.003 & Ch: 0.000 & Ch: 0.003 & Ch: 0.021 \\
\hline
3 Boxes & FF: 0.018 & FF: 0.003 & FF: 0.002 & FF: 0.001 & FF: 0.011 & FF: 0.037 \\
LOC: 7 & Ch: 0.016 & Ch: 0.001 & Ch: 0.001 & Ch: 0.000 & Ch: 0.007 & Ch: 0.028 \\
\hline
N Boxes Sli & FF: 0.016 & FF: 0.001 & FF: 0.001 & FF: 0.001 & FF: 0.005 & FF: 0.027 \\
LOC: 7 & Ch: 0.018 & Ch: 0.001 & Ch: 0.001 & Ch: 0.000 & Ch: 0.009 & Ch: 0.033 \\
\hline
N Boxes & FF: 0.042 & FF: 0.003 & FF: 0.002 & FF: 0.001 & FF: 0.012 & FF: 0.067 \\
LOC: 24 & Ch: 0.053 & Ch: 0.006 & Ch: 0.002 & Ch: 0.000 & Ch: 0.022 & Ch: 0.091 \\
\hline
Elm Logo & FF: 0.029 & FF: 0.000 & FF: 0.001 & FF: 0.000 & FF: 0.018 & FF: 0.051 \\
LOC: 12 & Ch: 0.036 & Ch: 0.002 & Ch: 0.001 & Ch: 0.000 & Ch: 0.034 & Ch: 0.077 \\
\hline
Logo 2 & FF: 0.049 & FF: 0.009 & FF: 0.001 & FF: 0.000 & FF: 0.004 & FF: 0.073 \\
LOC: 18 & Ch: 0.060 & Ch: 0.026 & Ch: 0.003 & Ch: 0.000 & Ch: 0.010 & Ch: 0.129 \\
\hline
Logo Sizes & FF: 0.072 & FF: 0.027 & FF: 0.003 & FF: 0.001 & FF: 0.021 & FF: 0.151 \\
LOC: 33 & Ch: 0.084 & Ch: 0.059 & Ch: 0.002 & Ch: 0.000 & Ch: 0.031 & Ch: 0.244 \\
\hline
Rings & FF: 0.029 & FF: 0.002 & FF: 0.002 & FF: 0.000 & FF: 0.002 & FF: 0.039 \\
LOC: 12 & Ch: 0.034 & Ch: 0.004 & Ch: 0.001 & Ch: 0.000 & Ch: 0.002 & Ch: 0.047 \\
\hline
Polygons & FF: 0.031 & FF: 0.003 & FF: 0.003 & FF: 0.000 & FF: 0.027 & FF: 0.068 \\
LOC: 15 & Ch: 0.039 & Ch: 0.008 & Ch: 0.001 & Ch: 0.000 & Ch: 0.065 & Ch: 0.126 \\
\hline
\end{tabular}
\end{figure*}

\clearpage

\begin{figure*}
\scriptsize
\begin{tabular}{c c c c c c c}
\hline
\textbf{Example Name} & \textbf{Parse} & \textbf{Eval} & \textbf{Unparse} & \textbf{valToInde...} & \textbf{prepareLi...} & \textbf{``Run Code''} \\
\hline \hline
Stars & FF: 0.043 & FF: 0.009 & FF: 0.002 & FF: 0.000 & FF: 0.021 & FF: 0.085 \\
LOC: 18 & Ch: 0.049 & Ch: 0.023 & Ch: 0.002 & Ch: 0.000 & Ch: 0.040 & Ch: 0.139 \\
\hline
Triangles & FF: 0.017 & FF: 0.002 & FF: 0.001 & FF: 0.000 & FF: 0.005 & FF: 0.034 \\
LOC: 14 & Ch: 0.018 & Ch: 0.006 & Ch: 0.000 & Ch: 0.000 & Ch: 0.014 & Ch: 0.046 \\
\hline
US-13 Flag & FF: 0.048 & FF: 0.017 & FF: 0.003 & FF: 0.000 & FF: 0.107 & FF: 0.198 \\
LOC: 24 & Ch: 0.060 & Ch: 0.059 & Ch: 0.003 & Ch: 0.001 & Ch: 0.224 & Ch: 0.409 \\
\hline
US-50 Flag & FF: 0.050 & FF: 0.007 & FF: 0.002 & FF: 0.001 & FF: 0.032 & FF: 0.107 \\
LOC: 17 & Ch: 0.055 & Ch: 0.016 & Ch: 0.002 & Ch: 0.001 & Ch: 0.053 & Ch: 0.146 \\
\hline
French Sudan Flag & FF: 0.055 & FF: 0.002 & FF: 0.002 & FF: 0.000 & FF: 0.010 & FF: 0.073 \\
LOC: 27 & Ch: 0.067 & Ch: 0.003 & Ch: 0.002 & Ch: 0.000 & Ch: 0.016 & Ch: 0.092 \\
\hline
Frank Lloyd Wright & FF: 0.075 & FF: 0.004 & FF: 0.003 & FF: 0.000 & FF: 0.080 & FF: 0.168 \\
LOC: 40 & Ch: 0.101 & Ch: 0.012 & Ch: 0.004 & Ch: 0.001 & Ch: 0.157 & Ch: 0.293 \\
\hline
Bezier Curves & FF: 0.209 & FF: 0.006 & FF: 0.006 & FF: 0.001 & FF: 0.012 & FF: 0.242 \\
LOC: 124 & Ch: 0.270 & Ch: 0.012 & Ch: 0.007 & Ch: 0.002 & Ch: 0.021 & Ch: 0.326 \\
\hline
Fractal Tree & FF: 0.095 & FF: 0.015 & FF: 0.005 & FF: 0.001 & FF: 0.075 & FF: 0.218 \\
LOC: 38 & Ch: 0.115 & Ch: 0.023 & Ch: 0.004 & Ch: 0.001 & Ch: 0.145 & Ch: 0.310 \\
\hline
Stick Figures & FF: 0.120 & FF: 0.012 & FF: 0.003 & FF: 0.000 & FF: 0.273 & FF: 0.422 \\
LOC: 60 & Ch: 0.139 & Ch: 0.025 & Ch: 0.006 & Ch: 0.001 & Ch: 0.490 & Ch: 0.693 \\
\hline
Matrix Transformations & FF: 0.091 & FF: 0.043 & FF: 0.004 & FF: 0.001 & FF: 0.288 & FF: 0.471 \\
LOC: 30 & Ch: 0.116 & Ch: 0.126 & Ch: 0.004 & Ch: 0.001 & Ch: 0.730 & Ch: 1.105 \\
\hline
Misc Shapes & FF: 0.018 & FF: 0.000 & FF: 0.000 & FF: 0.000 & FF: 0.010 & FF: 0.033 \\
LOC: 8 & Ch: 0.020 & Ch: 0.001 & Ch: 0.001 & Ch: 0.000 & Ch: 0.012 & Ch: 0.036 \\
\hline
Interface Buttons & FF: 0.225 & FF: 0.007 & FF: 0.011 & FF: 0.003 & FF: 0.010 & FF: 0.267 \\
LOC: 79 & Ch: 0.276 & Ch: 0.012 & Ch: 0.012 & Ch: 0.000 & Ch: 0.017 & Ch: 0.332 \\
\hline
Paths 1 & FF: 0.012 & FF: 0.000 & FF: 0.000 & FF: 0.002 & FF: 0.003 & FF: 0.021 \\
LOC: 5 & Ch: 0.013 & Ch: 0.000 & Ch: 0.000 & Ch: 0.001 & Ch: 0.003 & Ch: 0.021 \\
\hline
Paths 2 & FF: 0.027 & FF: 0.000 & FF: 0.001 & FF: 0.001 & FF: 0.012 & FF: 0.043 \\
LOC: 11 & Ch: 0.034 & Ch: 0.000 & Ch: 0.001 & Ch: 0.002 & Ch: 0.022 & Ch: 0.063 \\
\hline
Paths 3 & FF: 0.012 & FF: 0.000 & FF: 0.001 & FF: 0.001 & FF: 0.004 & FF: 0.020 \\
LOC: 5 & Ch: 0.014 & Ch: 0.000 & Ch: 0.001 & Ch: 0.000 & Ch: 0.005 & Ch: 0.023 \\
\hline
Paths 4 & FF: 0.013 & FF: 0.001 & FF: 0.001 & FF: 0.000 & FF: 0.003 & FF: 0.020 \\
LOC: 11 & Ch: 0.015 & Ch: 0.000 & Ch: 0.000 & Ch: 0.001 & Ch: 0.004 & Ch: 0.024 \\
\hline
Paths 5 & FF: 0.018 & FF: 0.001 & FF: 0.001 & FF: 0.001 & FF: 0.003 & FF: 0.028 \\
LOC: 10 & Ch: 0.025 & Ch: 0.000 & Ch: 0.000 & Ch: 0.001 & Ch: 0.005 & Ch: 0.036 \\
\hline
Sample Rotations & FF: 0.034 & FF: 0.004 & FF: 0.003 & FF: 0.000 & FF: 0.004 & FF: 0.050 \\
LOC: 16 & Ch: 0.044 & Ch: 0.003 & Ch: 0.001 & Ch: 0.000 & Ch: 0.005 & Ch: 0.058 \\
\hline
Grid Tile & FF: 0.050 & FF: 0.010 & FF: 0.005 & FF: 0.000 & FF: 0.020 & FF: 0.097 \\
LOC: 23 & Ch: 0.053 & Ch: 0.010 & Ch: 0.001 & Ch: 0.000 & Ch: 0.023 & Ch: 0.102 \\
\hline
Zones & FF: 0.043 & FF: 0.005 & FF: 0.004 & FF: 0.001 & FF: 0.021 & FF: 0.081 \\
LOC: 14 & Ch: 0.039 & Ch: 0.003 & Ch: 0.001 & Ch: 0.001 & Ch: 0.015 & Ch: 0.063 \\
\hline
\hline
 & \textbf{FF: 0.069} & \textbf{FF: 0.007} & \textbf{FF: 0.003} & \textbf{FF: 0.003} & \textbf{FF: 0.142} & \textbf{FF: 0.238} \\
 & \textbf{Ch: 0.085} & \textbf{Ch: 0.016} & \textbf{Ch: 0.003} & \textbf{Ch: 0.005} & \textbf{Ch: 0.257} & \textbf{Ch: 0.390} \\
\textbf{Summary} & \textbf{Min 0.009} & \textbf{Min 0.000} & \textbf{Min 0.000} & \textbf{Min 0.000} & \textbf{Min 0.001} & \textbf{Min 0.017} \\
 & \textbf{Med 0.053} & \textbf{Med 0.005} & \textbf{Med 0.002} & \textbf{Med 0.000} & \textbf{Med 0.013} & \textbf{Med 0.098} \\
 & \textbf{Mean 0.077} & \textbf{Mean 0.012} & \textbf{Mean 0.003} & \textbf{Mean 0.004} & \textbf{Mean 0.200} & \textbf{Mean 0.314} \\
 & \textbf{Max 0.520} & \textbf{Max 0.165} & \textbf{Max 0.019} & \textbf{Max 0.180} & \textbf{Max 6.789} & \textbf{Max 6.980} \\
\hline \hline
\end{tabular}
\end{figure*}

\clearpage

\end{document}